\documentclass[draft,tightenlines,nofootinbib,preprint,aps,eqsecnum,amsmath,amssymb]{revtex4}

\newcommand{\beq}{\begin{equation}}
\newcommand{\eeq}{\end{equation}}
\newcommand{\bea}{\begin{eqnarray}}
\newcommand{\eea}{\end{eqnarray}}
\newcommand{\cir}{{\buildrel \circ \over =}}

\begin{document}

\title{Towards Relativistic Atomic Physics. II. Collective and Relative  Relativistic
Variables for a System of Charged Particles plus the
Electro-Magnetic Field}
\author{David Alba}
\affiliation{Dipartimento di Fisica\\
Universita' di Firenze\\
Polo Scientifico, via Sansone 1\\
50019 Sesto Fiorentino, Italy\\
E-mail ALBA@FI.INFN.IT}
\author{Horace W. Crater}
\affiliation{The University of Tennessee Space Institute \\
Tullahoma, TN 37388 USA \\
E-mail: hcrater@utsi.edu}
\author{Luca Lusanna}
\affiliation{Sezione INFN di Firenze\\
Polo Scientifico\\
Via Sansone 1\\
50019 Sesto Fiorentino (FI), Italy\\
E-mail: lusanna@fi.infn.it}

\begin{abstract}

In this second paper we complete the classical description of an
isolated system of "charged positive-energy particles, with
Grassmann-valued electric charges and mutual Coulomb interaction,
plus a transverse electro-magnetic field" in the rest-frame instant
form of dynamics.\medskip

In particular we show how to determine a collective variable
associated with the internal 3-center of mass on the instantaneous
3-spaces, to be eliminated with the constraints ${\vec {\cal
K}}_{(int)} \approx 0$. Here ${\vec {\cal K}}_{(int)}$ is the
Lorentz boost generator in the unfaithful internal realization of
the Poincare' group and its vanishing is the gauge fixing to the
rest-frame conditions ${\vec {\cal P}}_{(int)} \approx 0$. We show
how to find this collective variable for the following isolated
systems: a) charged particles with a Coulomb plus Darwin mutual
interaction; b) transverse radiation field; c) charged particles
with a mutual Coulomb interaction plus a transverse electro-magnetic
field.
\medskip

Then we define the Dixon multipolar expansion for the open particle
subsystem. We also define the relativistic electric dipole
approximation of atomic physics in the rest-frame instant form and
we find the a possible relativistic generalization of the electric
dipole representation.

\end{abstract}

\maketitle

\medskip

\today

\vfill\eject

\section{Introduction}

In Ref.\cite{1} we gave the description of the isolated system "N
charged scalar positive-energy particles with Coulomb mutual
interaction plus the electro-magnetic field in the radiation gauge"
in the rest-frame instant form of dynamics. Grassmann-valued
electric charges were used to regularize the Coulomb self-energies.
In that paper there was the determination of the regularized
Lienard-Wiechert electro-magnetic fields in the absence of an
incoming radiation field and of the complete potential acting among
Coulomb-dressed charged particles. This effective potential turned
out to be the Coulomb potential plus the full relativistic
expression of the Darwin potential, which at the lowest order in
$1/c^{2}$ reduces at the known form of the Darwin potential, till
now obtainable  only from the Bethe-Salpeter approach to
relativistic bound states in QFT. This full Darwin plus Coulomb
potential can be regarded as the classical analogue of the complete
transverse as well longitudinal effects of the single photon
exchange.\medskip

In Ref.\cite{2}, quoted as I with its formulas denoted by (I.2.5),
we extended the approach of Ref.\cite{1}. In particular in I we
relaxed the assumption of no ingoing radiation field and we were
able to show that, at least at the classical level, there exists a
canonical transformation from an initial situation with a free IN
transverse radiation field and decoupled particles with a Coulomb
plus Darwin interaction, into an interpolating non-radiation
transverse electro-magnetic field interacting with charged particles
having only a Coulomb mutual interaction. At later times we can
again canonically transform the system to a free OUT radiation field
plus charged particles with Coulomb plus Darwin mutual interaction.

\medskip

These results can be obtained within the formalism of the rest-frame
instant form of dynamics starting from the approach based on
parametrized Minkowski theories of relativistic mechanics \cite{3}.
The first part of paper I  contains a complete updated review of
this formalism with a resolution of all the ambiguities connected
with the relativistic center-of-mass notion. As shown in paper I, in
the inertial rest frame centered on the Fokker-Pryce covariant
non-canonical 4-center of inertia $Y^{\mu}(\tau )$ of the isolated
system, the instantaneous 3-spaces are the Wigner space-like
hyper-planes orthogonal to the conserved 4-momentum $P^{\mu} = Mc\,
h^{\mu} = Mc\, (\sqrt{1 + {\vec h}^2}; \vec h)$ of the isolated
system, whose invariant mass is $M$. On the Wigner hyper-plane,
described by the embedding $z_W^{\mu}(\tau ,\vec \sigma ) =
Y^{\mu}(\tau ) + \epsilon^{\mu}_r(\vec h)\, \sigma^r$ \footnote{As
shown in I, $(\tau ,\vec \sigma )$ are observer-dependent 4-radar
coordinates with $T = \tau /c$ being the Lorentz-scalar rest time
and $\epsilon^{\mu}_r(\vec h) = (- h_r; \delta^i_r - {{h^i\,
h_r}\over {1 + \sqrt{1 + {\vec h}^2}}})$ are 3 space-like 4-vectors
forming together with $h^{\mu} = \epsilon^{\mu}_{\tau}(\vec h)$ the
columns of the standard Wigner boost sending $P^{\mu}$ to its
rest-frame form $Mc\, (1; \vec 0)$.} the particles are identified by
(Wigner spin-1) 3-vectors ${\vec \eta}_i(\tau )$ determined by the
intersection of the particle world-line $x^{\mu}_i(\tau ) =
Y^{\mu}(\tau ) + \epsilon^{\mu}_r(\vec h)\, \eta_i^r(\tau )$ with
the Wigner hyperplane.\medskip

In the rest-frame instant form every isolated system is described as
an external decoupled canonical non-covariant 4-center of mass
${\tilde x}^{\mu}(\tau ) = ({\tilde x}^o(\tau ); {\vec x}_{NW} -
{{\vec P}\over {P^o}}\, {\tilde x}^o(\tau ))$ parametrized (like
also $Y^{\mu}(\tau )$) in terms of $\tau$ and of the 6 non-evolving
Jacobi data $\vec z$, $\vec h$ \footnote{${\vec x}_{NW} = \vec z /
Mc$ is the classical Newton-Wigner position of the 3-center of mass
and we have ${\tilde x}^{\mu}(\tau ) = \Big({\tilde x}^o(\tau );
{\vec x}_{NW} + {{{\vec P}}\over {P^o}}\, {\tilde x}^o(\tau
)\Big)$.}. This decoupled non-covariant free "particle" carries the
internal mass $M$ and the rest spin ${\vec {\bar S}}$ of the
isolated system. In the case of charged particles plus the
electro-magnetic field they  are functions of the 3-coordinates
${\vec \eta}_i(\tau )$, their conjugate momenta ${\vec
\kappa}_i(\tau )$ and of the transverse electro-magnetic fields
${\vec A}_{\perp}(\tau ,\vec \sigma )$, ${\vec \pi}_{\perp}(\tau
,\vec \sigma ) =  {\vec E}_{\perp}(\tau ,\vec \sigma )$. There is an
external realization of the Poincare' group, whose generators depend
upon $\vec z$, $\vec h$, $M$ and ${\vec {\bar S}}$.\medskip

Inside the Wigner hyper-plane there is an internal realization of
the Poincare' group, whose generators ${\cal P}_{(int)} = Mc$,
${\vec {\cal P}}_{(int)}$, ${\vec {\cal J}}_{(int)} = {\vec {\cal
S}}$, ${\vec {\cal K}}_{(int)}$, are evaluated from the
energy-momentum tensor of the isolated system obtained from the
action of the parametrized Minkowski theory describing the isolated
system in arbitrary non-inertial frames. This realization is
unfaithful because the rest-frame instant form centered on the
Fokker-Pryce inertial observer implies (see paper I) the following
second class constraints

\beq
 {\vec {\cal P}}_{(int)} \approx 0,\qquad
 {\vec {\cal K}}_{(int)} \approx 0.
 \label{1.1}
 \eeq

While the first 3 constraints are the rest-frame conditions (each
Wigner hyper-plane is a rest frame), the other 3 constraints
eliminate the internal 3-center of mass avoiding a double counting
of the center of mass. Therefore Eqs.(\ref{1.1}) imply that the
dynamics inside the Wigner hyper-planes depends only on relative
variables satisfying Hamilton equations having $Mc$ as Hamiltonian.
All the interaction potentials appear only in $Mc$ and ${\vec {\cal
K}}_{(int)}$ and not in ${\vec {\cal J}}_{(int)} = {\vec {\bar S}}$,
since we are in an instant form of dynamics.
\medskip

This complete understanding of the relativistic kinematics of the
rest-frame instant form of relativistic mechanics was obtained only
recently in Ref.\cite{4} and in paper I. The important point is to
have a Lagrangian description of the isolated system. This will
allow the determination of its energy-momentum tensor. Only in this
way  can we find the interaction-dependent internal generators $Mc$
and ${\vec {\cal K}}_{(int)}$. In most of the other approaches to
relativistic mechanics the Lagrangian formulation is unknown, so
that we can find only $Mc$ but not ${\vec {\cal K}}_{(int)}$.
Therefore we lack the interaction-dependent equations which
eliminate the internal 3-center mass, the gauge variable conjugate
to the rest-frame conditions and we are  able to reconstruct neither
the orbits nor the particle world-lines in the rest frame.
\medskip

In Ref.\cite{4}, even in absence of a Lagrangian, we were able to
guess the form of ${\vec {\cal K}}_{(int)}$, associated with a given
class of invariant masses for a two-body system, so that the
internal generators satisfy the Poincare' algebra in the rest frame.
In terms of the 3-vectors ${\vec \eta}_i(\tau )$, ${\vec
\kappa}_i(\tau )$, $i=1,2$, these generators are ($\vec \rho = {\vec
\eta}_1 - {\vec \eta}_2$)\footnote{Since we have $\{ \eta^r_i(\tau
), \kappa_{js}(\tau ) \} = \delta_{ij}\, \delta^r_s$, we use the
following vector notation: ${\vec \eta}_i(\tau) =
\Big(\eta^r_i(\tau)\Big)$, ${\vec \kappa}_i(\tau) =
\Big(\kappa_{ir}(\tau ) = - \kappa^r_i(\tau)\Big)$.}

\bea
 Mc &=& \sqrt{m_{1}^{2}\, c^2 + {\vec \kappa }_{1}^{2}(\tau ) +
 \Phi ({\vec \rho }^2(\tau ))} + \sqrt{m_{2}^{2}\, c^2 + {\vec
 \kappa }_{2}^{2}(\tau ) + \Phi ({\vec \rho }^2(\tau ))},\nonumber \\
 {\vec {\cal P}}_{(int)} &=& {\vec \kappa}_1(\tau ) + {\vec
 \kappa}_2(\tau ) \approx 0,\nonumber \\
 {\vec {\cal J}}_{(int)} &=& {\vec {\bar S}} = {\vec \eta}_1(\tau )
 \times {\vec \kappa}_1(\tau ) + {\vec \eta}_2(\tau )
 \times {\vec \kappa}_2(\tau ),\nonumber \\
 {\vec {\cal K}}_{(int)} &=& - {\vec \eta }_{1}(\tau )\, \sqrt{m_{1}^{2}\, c^2
 + {\vec \kappa }_{1}^{2}(\tau ) + \Phi ({\vec \rho }^2(\tau ))} -
 {\vec \eta }_{2}(\tau )\, \sqrt{m_{2}^{2}\, c^2 + {
\vec \kappa }_{2}^{2}(\tau ) + \Phi ({\vec \rho }^2(\tau ))} \approx
0,\nonumber \\
 &&{}
 \label{1.2}
 \eea

\noindent where $\Phi ({\vec \rho}^2(\tau ) )$ is an arbitrary
potential \footnote{The internal boost ${\vec {\cal K}}_{(int)}$
satisfying $\{ {\cal P}^i, {\cal K}^j \} = \delta _{ij}\, M$ in the
case of the pure Coulomb interaction $Mc = \sqrt{m_{1}^{2}\, c^2
+\vec{\kappa}^{2}} + \sqrt{m_{2}^{2}\, c^2 + \vec{\kappa}^{2}} +
\frac{Q_{1}Q_{2}}{4\pi |\vec{\eta}_{1}- \vec{\eta}_{2}|}$ are not
known. Only in the case of Coulomb plus Darwin potential, see next
Section, is the internal boost  known.}.\medskip

If $\vec \pi (\tau ) = {1\over 2}\, [{\vec \kappa}_1(\tau ) - {\vec
\kappa}_2(\tau )]$ is the momentum conjugate to ${\vec \rho}(\tau
)$, we have

\beq
 Mc \approx {\cal M}_{(int)}\, c = \sqrt{m_{1}^{2}\, c^2 +
 {\vec \pi}^{2}(\tau ) + \Phi ({\vec \rho }^2(\tau ))} +
 \sqrt{m_{2}^{2}\, c^2 + {\vec \pi}^{2}(\tau ) +
 \Phi ({\vec \rho }^2(\tau ))}.
 \label{1.3}
 \eeq

In Ref.\cite{4}, after having solved the equations ${\vec {\cal
K}}_{(int)} \approx 0$ and eliminated the internal 3-center of mass,
we obtained the following orbit and world-line reconstruction

\bea
 {\vec \eta}_i(\tau ) &\approx& {1\over 2}\, \Big((-1)^{i+1} - {{m_1^2 -
 m_2^2}\over {{\cal M}^2_{(int)}[{\vec \rho}^2(\tau ),
 {\vec \pi}^2(\tau )]}}\Big)\, \vec \rho
 (\tau ),\nonumber \\
 &&{}\nonumber \\
 &&{}\nonumber \\
 x^{\mu}_i(\tau ) &\approx& Y^{\mu}(\tau ) + {1\over 2}\,
 \Big((-1)^{i+1} - {{m_1^2 - m_2^2}\over {{\cal M}^2
 _{(int)}[{\vec \rho}^2(\tau ), {\vec \pi}^2(\tau )]}}\Big)\,\,
 \epsilon^{\mu}_r(\vec h)\, \rho^r(\tau ).
 \label{1.4}
 \eea

The relative variables ${\vec \rho}(\tau )$ and ${\vec \pi}(\tau )$
are to be found as solutions of the Hamilton equations with
Hamiltonian ${\cal M}_{(int)}\, c$.

\bigskip

In this paper we complete the classical description of the system of
charged particles plus a transverse electro-magnetic field.\medskip

In Section II we define the internal collective and relative
variables for the isolated system of N charged particles interacting
through the Coulomb plus Darwin potential of Ref.\cite{1}. Then we
will perform the elimination of the internal 3-center of mass and
the orbit reconstruction in the N=2 case.\medskip

In Section III we define the internal collective and relative
variables for the free radiation field. Here we rely on the results
of Refs. \cite{5} and \cite{6}, where for the first time there was
the study of collective and relative variables for the Klein-Gordon
field in the massive and massless cases respectively. The adaptation
of this results to the rest-frame instant form is done in Appendix
A. This allows one  to eliminate  the internal 3-center of mass of
the transverse radiation field.

\medskip

Then in Section IV we will use the results of Sections II and III
and the canonical transformation of paper I to find the collective
and relative variables of the isolated system "N charged particles
with mutual Coulomb interaction plus the transverse electro-magnetic
field" defined in paper I and to perform the elimination of the
internal 3-center of mass also in this case.

\bigskip

In Section V we review the multipolar expansion of the
energy-momentum tensor of the particle open subsystem and how to get
its (non-canonical) pole-dipole approximation, in which the
subsystem is viewed as an effective world-line (the 4-center of
motion) carrying the effective spin of the subsystem. This
pole-dipole structure must not be confused with the pole-dipole
structure carried by the non-covariant (Newton-Wigner) external
center of mass of the isolated system "charged particles with mutual
Coulomb interaction plus transverse electro-magnetic field".

\bigskip

While Section VI is devoted to finding the relativistic version of
the electric dipole approximation of atomic physics \cite{7,8,9}, in
Section VII we face the problem of defining a relativistic extension
of the electric dipole representation of the isolated system. We
study various point canonical transformations suggested by standard
atomic physics and we find that all of them generate singular terms
when the electro-magnetic field is considered dynamical and not an
external prescribed field: they are induced by the dipole
approximation. We then propose a relativistic representation in
which the singular terms are replaced by contact interactions among
the particles: the price for this regularization is that in the
semi-relativistic limit it is different from the standard electric
dipole representation. In Appendix B there is the identification of
the generating functions of these point canonical transformations by
the introduction of relativistic Lagrangians for the rest-frame
instant form description of the isolated system.

\bigskip

In the final Section there are some concluding remarks and an
introduction to the subsequent paper III \cite{10}.

\vfill\eject

\section{Relativistic Mechanics: the Isolated System of N Charged
Particles Interacting through the Coulomb plus Darwin Potential}

Let us consider an isolated system of N charged positive-energy
scalar particles either free or interacting through the Coulomb +
Darwin potential (I-4.5) (see I and Ref.\cite{1}).\medskip

On the Wigner hyper-planes $\Sigma_{\tau}$ each particle is
described by two canonically conjugate Wigner spin-1 3-vectors: the
3-position ${\vec \eta}_i(\tau )$ and the 3-momentum ${\vec
\kappa}_i(\tau )$, $i = 1,.., N$, restricted by the rest-frame
conditions ${\vec {\cal P}}_{(int)} \approx 0$ (vanishing of the
internal 3-momentum) and by the conditions eliminating the internal
3-center of mass (vanishing of the internal Lorentz boosts ${\vec
{\cal K}}_{(int)} \approx 0$).\bigskip

Therefore we must introduce collective and relative variables on the
Wigner hyper-planes and eliminate the collective ones. We consider
mostly the $N = 2$ case, but everything is well defined for
arbitrary N. There are various possibilities and we must find the
one which allows one to make the elimination explicitly.\medskip

\subsection{Collective and Relative Variables for N Particles}

In previous papers \cite{11} the problem of replacing the
3-coordinates ${\vec \eta}_i(\tau )$, ${\vec \kappa}_i(\tau)$ inside
a Wigner hyper-plane with internal collective and relative canonical
variables in the rest-frame instant form was solved in two different
ways:\medskip

a) We may introduce naive collective variables ${\vec \eta}_+ =
{1\over N}\, \sum_i^{1..N}\, {\vec \eta}_i$ (independent of the
particle masses), ${\vec \kappa}_+ = {\vec {\cal P}}_{(int)} =
\sum_i^{1..N}\, {\vec \kappa}_i \approx 0$ and then completing them
with either naive relative variables ${\vec \rho}_a = \sqrt{N}\,
\sum_{i=1}^N\, \gamma_{ai}\, {\vec \eta}_i$, ${\vec \pi}_a = {1\over
{\sqrt{N}}}\, \sum_{i=1}^N\, \gamma_{ai}\, {\vec \kappa}_i$,
$a=1,..,N-1$, or with the relative variables in the so-called spin
bases \cite{11,12} \footnote{ Both sets can be used to find the
expression of the Dixon multipoles (see Section V) for a
two-particle open subsystem in terms of canonical c.o.m and relative
canonical variables.}. This naive canonical basis is obtained with a
linear canonical transformation {\it point} both in the positions
and in the momenta.\medskip

b) We may find the canonical basis whose collective variables are
the internal 3-center of mass ${\vec q}_+$ \footnote{ Due to the
rest-frame condition ${\vec {\cal P}}_{(int)} \approx 0$, we have
${\vec q}_+ \approx {\vec R}_+ \approx {\vec y}_+$, where ${\vec
q}_+$ is the internal canonical 3-center of mass (the internal
Newton-Wigner position), ${\vec y}_+$ is the internal Fokker-Pryce
3-center of inertia and ${\vec R}_+$ is the internal M$\o$ller
3-center of energy \cite{4}. These global quantities are defined in
terms of the internal Poincare' generators \cite{11}.} and ${\vec
\kappa}_+ = {\vec {\cal P}}_{(int)} = \sum_i^{1..N}\, {\vec
\kappa}_i \approx 0$. To these collective variables are then
associated relative variables ${\vec \rho}_{qa}$, ${\vec \pi}_{qa}$,
$a=1,..,N-1$. However this non-linear canonical transformation
depends upon the interactions present among the particles (through
the internal Poincare' generators), so that it is known {\it only
for free particles} \cite{4,11}: even in this case it is {\it point
only in the momenta}. The advantage of this canonical transformation
would be to allow one to write the invariant mass in the form $Mc =
\sqrt{{\cal M}^2\, c^2 + {\vec \kappa}^2_+} \approx {\cal M}$
\footnote{For two free particles we have $Mc = \sqrt{m_1^2\, c^2 +
{\vec \kappa}_1^2} + \sqrt{m_2^2\, c^2 + {\vec \kappa}^2_2} =
\sqrt{{\cal M}^2\, c^2 + {\vec \kappa}^2_+} \approx {\cal M}\, c =
\sqrt{m_1^2\, c^2 + {\vec \pi}^2_q} + \sqrt{m_2^2\, c^2 + {\vec
\pi}_q^2}$.}, explicitly showing that in the rest-frame the internal
mass depends only on relative variables.

\bigskip

Since it is more convenient to use the naive linear canonical
transformation we will use the following collective and relative
variables which, written in terms of the masses of the particles,
make it easier to evaluate the non-relativistic limit ($m =
\sum_{i=1}^N\, m_i$)

\bea
 {\vec \eta}_+ &=& \sum_{i=1}^N\, {{m_i}\over m}\, {\vec
 \eta}_i,\qquad {\vec \kappa}_+ = {\vec {\cal P}}_{(int)} =
 \sum_{i=1}^N\, {\vec \kappa}_i,\nonumber \\
 {\vec \rho}_a &=& \sqrt{N}\, \sum_{i=1}^N\, \gamma_{ai}\, {\vec
 \eta}_i,\qquad {\vec \pi}_a = {1\over {\sqrt{N}}}\, \sum_{i=1}^N\, \Gamma_{ai}\,
 {\vec \kappa}_i,\qquad a = 1,..,N-1,\nonumber \\
 &&{}\nonumber \\
 {\vec \eta}_i &=& {\vec \eta}_+ + {1\over {\sqrt{N}}}\,
 \sum_{a-1}^{N-1}\, \Gamma_{ai}\, {\vec \rho}_a,\nonumber \\
 {\vec \kappa}_i &=& {{m_i}\over m}\, {\vec \kappa}_+ + \sqrt{N}\,
 \sum_{a=1}^{N-1}\, \gamma_{ai}\, {\vec \pi}_a,
 \label{2.1}
 \eea

\noindent with the following canonicity conditions
\footnote{Eqs.(\ref{2.1}) describe a family of canonical
transformations, because the $\gamma_{ai}$'s depend on
${\frac{1}{2}}(N-1)(N-2)$ free independent parameters.}

\bea
 &&\sum_{i=1}^{N}\, \gamma _{ai} = 0,\qquad  \sum_{i=1}^{N}\, \gamma
_{ai}\, \gamma _{bi} = \delta _{ab},\qquad  \sum_{a=1}^{N-1}\,
\gamma _{ai}\, \gamma _{aj} = \delta _{ij} - {\frac{1}{N}},
\nonumber \\
 &&{}\nonumber \\
 &&\Gamma_{ai} = \gamma_{ai} - \sum_{k=1}^N\,
{\frac{{m_k}}{ m}}\, \gamma_{ak},\qquad \gamma_{ai} = \Gamma_{ai} -
{\frac{1}{ N}}\, \sum_{k=1}^N\, \Gamma_{ak},\nonumber \\
 &&{}\nonumber \\
 &&\sum_{i=1}^N\, {\frac{{m_i}}{m}}\, \Gamma_{ai} = 0,\qquad
\sum_{i=1}^N\, \gamma_{ai}\, \Gamma_{bi} = \delta_{ab},\qquad
\sum_{a=1}^{N-1}\, \gamma_{ai}\, \Gamma_{aj} =  \delta_{ij} -
{\frac{{m_i}}{ m}}.\nonumber \\
 &&{}
 \label{2.2}
 \eea

\medskip

For $N=2$ we have $\gamma_{11} = - \gamma_{12} = {\frac{1}{
\sqrt{2}}}$, $\Gamma_{11} = \sqrt{2}\, {\frac{{m_2}}{ m}}$,
$\Gamma_{12} = - \sqrt{2}\, {\frac{{m_1}}{m}}$.\medskip

Therefore in the two-body case, by introducing the notation ${\vec
\eta}_{12} = {\vec \eta}_+$, ${\vec \kappa}_{12} = {\vec \kappa}_+ =
{\vec {\cal P}}_{(int)}$, we have the following collective and
relative variables

\bea
 {\vec \eta}_{12} &=& {\frac{{m_1}}{m}}\, {\vec \eta}_1 +
{\frac{{m_2}}{m}}\, {\vec\eta}_2,  \qquad
{\vec \rho}_{12} = {\vec \eta}_1 - {\vec \eta}_2,  \nonumber \\
{\vec \kappa}_{12} &=& {\vec \kappa}_1 + {\vec \kappa}_2 \approx 0,
\qquad {\vec \pi}_{12} =  {\frac{{m_2}}{m}}\, {\vec \kappa}_1 - {\
\frac{{m_1}}{m}}\, {\vec \kappa}_2,\nonumber \\
 &&{}\nonumber \\
 &&{}\nonumber \\
  {\vec \eta}_1 &=& {\vec \eta}_{12} + {\frac{{m_2}}{m}}\, {\vec \rho}_{12},
 \qquad {\vec \eta}_2 = {\vec \eta}_{12} - {\frac{{m_1}}{m}}\,
 {\vec \rho}_{12},  \nonumber \\
 {\vec \kappa}_1 &=& {\frac{{m_1}}{m}}\, {\vec \kappa}_{12} +
 {\vec \pi}_{12},  \qquad {\vec \kappa}_2 =
 {\frac{{ m_2}}{m}}\, {\vec \kappa}_{12} - {\vec \pi}_{12}.
  \label{2.3}
 \eea

We  use the notation $m = m_1 + m_2 = m_1\, (1 +
{\frac{{m_2}}{{m_1}}})$ , $\mu ={\frac{{m_1\, m_2}}{m}} = m_2\, (1 +
{\frac{{m_2}}{{m_1}}})^{-1}$. For $m_2 >> m_1$ we have ${\ \vec
\eta}_2 \approx {\vec \eta}_{12} - {\frac{{ m_1}}{{m_2}}}\, {\vec
\rho}_{12}$, ${\vec \eta}_1 \approx {\vec \eta}_{12} + {\vec
\rho}_{12}$.
\bigskip

The collective variable ${\vec \eta}_{12}(\tau )$ has to be
determined in terms of ${\vec \rho}_{12}(\tau )$ and ${\vec
\pi}_{12}(\tau )$ by means of the gauge fixings ${\vec
{\mathcal{K}}}_{(int)}\, {\buildrel {def}\over =}\, - M\, {\vec R}_+
\approx 0$.

\subsection{Two Charged Particles interacting through the Coulomb plus Darwin
Potential}

From Ref.\cite{1} we get the following internal Poincare' generators
for $N = 2$  \footnote{$V_{DARWIN} = V_D({\tilde {\vec \eta}}_1 -
{\tilde {\vec \eta}}_2, {\tilde {\vec \kappa}} _1, {\tilde {\vec
\kappa}}_2)$ is given in Eq.(6.19) of Ref.\cite{1} and in
Eq.(I.4.5). ${\vec {\mathcal{J}}}_{(int)}$ is given in Eq.(6.40) and
${\vec {\mathcal{K}}}_{(int)}$ in Eq.(6.46) of Ref.\cite{1}. }

\begin{eqnarray*}
\mathcal{E}_{(int)} &=& M\, c^2 = c\, \sum_{i=1}^2\, \sqrt{m^2_i\,
c^2 + { \tilde {\vec \kappa}}^2_i} + {\frac{{Q_1\, Q_2}}{{4\pi\,
|{\tilde { \vec \eta }}_1 - {\tilde {\vec \eta}}_2|}}} +
V_{DARWIN}({\tilde {\vec \eta}}_1(\tau ) - {\tilde {\vec
\eta}}_2(\tau ); {\tilde {\vec \kappa}}_i(\tau )),  \nonumber \\
{\vec {\mathcal{P}}}_{(int)} &=& {\tilde {\vec \kappa}}_1 + {\tilde
{\vec \kappa}}_2 \approx 0,  \nonumber \\
{\vec {\mathcal{J}}}_{(int)} &=& \sum_{i=1}^2\, {\tilde {\vec
\eta}}_i \times {\tilde {\vec \kappa}}_i,  \nonumber \\
&&{}  \nonumber \\
{\vec {\mathcal{K}}}_{(int)} &=&
 -  {\tilde {\vec \eta}}_{12}\, \Big[\sum_{i=1}^2\,
\sqrt{m_{i}^{2}\,
c^2 + \widetilde{\vec{\kappa}_{i}}^{2}} +  \nonumber \\
 &+& {\frac{{Q_1\, Q_2}}{c}}\, \Big({\frac{{\ {\tilde {\vec
\kappa}}_1 \cdot \Big[{\frac{1}{2}}\, {\vec \partial}_{{\tilde {\vec
\rho}}_{12}}\, {\tilde { \mathcal{K}}}_{12}({\tilde {\vec
\kappa}}_1, {\tilde {\vec \kappa}}_2, { \tilde {\vec \rho}}_{12}) -
2\, {\vec A}_{\perp S2}({\tilde {\vec \kappa}} _2, {\tilde {\vec
\rho}}_{12}) \Big]}}{{2\, \sqrt{m_1^2\, c^2 + \widetilde{
\vec{\kappa}_1}^2}}}} +  \nonumber \\
 &+& {\frac{{\ {\tilde {\vec \kappa}}_2 \cdot \Big[{\frac{1}{2}}\,
{\vec\partial}_{{\tilde {\vec \rho}}_{12}}\, {\tilde
{\mathcal{K}}}_{12}({\tilde { \vec \kappa}}_1, {\tilde {\vec
\kappa}}_2, {\tilde {\vec \rho}}_{12}) - 2\, { \vec A}_{\perp
S1}({\tilde {\vec \kappa}}_1, {\tilde {\vec \rho}}_{12}) \Big]
}}{{2\, \sqrt{m_2^2\, c^2 + \widetilde{\vec{\kappa}_2}^2}}}}\Big)
\Big] -  \end{eqnarray*}

\begin{eqnarray*}
 &-& {\tilde {\vec \rho}}_{12}\, \Big({\frac{{m_2}}{m}}\,
\sqrt{m_1^2\, c^2 + {\tilde {\vec \kappa}}_1^2} -
{\frac{{m_1}}{m}}\, \sqrt{m_2^2\, c^2 + {
\tilde {\vec \kappa}}_2^2} +  \nonumber \\
 &+& {\frac{{Q_1\, Q_2}}{c}}\, \Big[{\frac{{\ m_2\, {\tilde {\vec
\kappa}}_1 \cdot \Big[{\frac{1}{2}}\, {\vec \partial}_{{\tilde {\vec
\rho}}_{12}}\, { \tilde {\mathcal{K}}}_{12}({\tilde {\vec
\kappa}}_1, {\tilde {\vec \kappa}} _2, {\tilde {\vec \rho}}_{12}) -
2\, {\vec A}_{\perp S2}({\tilde {\vec \kappa }}_2, {\tilde {\vec
\rho}}_{12}) \Big] }}{{2\, m\, \sqrt{m_1^2\, c^2 + {
\tilde {\vec \kappa}}_1^2} }}} -  \nonumber \\
 &-& {\frac{{\ m_1\, {\tilde {\vec \kappa}}_2 \cdot
\Big[{\frac{1}{2}}\, { \vec \partial}_{{\tilde {\vec \rho}}_{12}}\,
{\tilde {\mathcal{K}}}_{12}({ \tilde {\vec \kappa}}_1, {\tilde {\vec
\kappa}}_2, {\tilde {\vec \rho}} _{12}) - 2\, {\vec A}_{\perp
S1}({\tilde {\vec \kappa}}_1, {\tilde {\vec \rho }}_{12}) \Big]
}}{{2\, m\, \sqrt{m_2^2\, c^2 + {\tilde {\vec \kappa}}_2^2}}}}
\Big]\Big) -  \end{eqnarray*}

\bea
 &-& {\frac{1}{{2\, c}}}\, Q_1\, Q_2\, \Big(\sqrt{m_1^2\, c^2 +
{\tilde {\vec \kappa}}_1^2}\, {\vec \partial}_{{\tilde {\vec
\kappa}}_1} + \sqrt{m_2^2\, c^2 + {\tilde {\vec \kappa}}_2^2}\,
{\vec \partial}_{{\tilde {\vec \kappa}} _2}\Big)\, {\tilde
{\mathcal{K}}}_{12}({\tilde {\vec \kappa}}_1, {\tilde {
\vec \kappa}}_2, {\tilde {\vec \rho}}_{12}) -  \nonumber \\
&-& {\frac{{Q_1\, Q_2}}{{4\pi\, c}}}\, \int d^3\sigma\,
\Big({\frac{{{ \tilde {\vec \pi}}_{\perp S1}(\vec \sigma -
{\frac{{m_2}}{m}}\, {\tilde { \vec \rho}}_{12}, {\tilde {\vec
\kappa}}_1)}}{{|\vec \sigma + {\frac{{m_1}}{m }}\, {\tilde {\vec
\rho}}_{12}|}}} + {\frac{{{\tilde {\vec \pi}}_{\perp S2}(\vec \sigma
+ {\frac{{m_1}}{m}}\, {\tilde {\vec \rho}}_{12}, {\tilde { \vec
\kappa}}_2)}}{{|\vec \sigma - {\frac{{m_2}}{m}}\, {\tilde {\vec
\rho}}_{12}|}}} \Big) -  \nonumber \\
&-& {\frac{{Q_1\, Q_2}}{c}}\, \int d^3\sigma\, \vec \sigma\,\,
\Big[{\tilde { \vec \pi}}_{\perp S1}(\vec \sigma -
{\frac{{m_2}}{m}}\, {\tilde {\vec \rho}} _{12}, {\tilde {\vec
\kappa}}_1) \cdot {\tilde {\vec \pi}}_{\perp S2}(\vec \sigma +
{\frac{{m_1}}{m}}\, {\tilde {\vec \rho}}_{12}, {\tilde {\vec \kappa}
}_2) +  \nonumber \\
&+& {\tilde {\vec B}}_{S1}(\vec \sigma - {\frac{{m_2}}{m}}\, {\tilde
{\vec \rho}}_{12}, {\tilde {\vec \kappa}}_1) \cdot {\tilde {\vec
B}}_{S2}(\vec \sigma + {\frac{{m_1}}{m}}\, {\tilde {\vec
\rho}}_{12}, {\tilde {\vec \kappa}
}_2) \Big] =  \nonumber \\
&&{}  \nonumber \\
&{\buildrel {def}\over =}& - M\, c\, \vec{R}_{+} \approx 0,
  \label{2.4}
\end{eqnarray}

\noindent where we have used Eqs. (\ref{2.3}) in the final
expression of the internal boost.\medskip

The above forms for the generators were found in Ref.\cite{1} by
first eliminating the electromagnetic degrees of freedom by forcing
them to coincide with the semiclassical phase space Lienard-Wiechert
solution by means of second class constraints (no ingoing radiation
field). Then having gone to Dirac brackets with respect to these
constraints, we found new (Coulomb-dressed) canonical variables
${\tilde {\vec\eta}}_i$ ,${\tilde {\vec \kappa}}_i$ ,for the
particles leading finally to a reduced phase space containing only
particles with mutual instantaneous action-at-a-distance
interactions.

These new canonical variables ${\tilde {\vec \eta}}_i$, ${\tilde {
\vec \kappa}}_i$, given in Eqs.(5.51) of Ref.\cite{1}, have the
following expression

\bea
 {\tilde {\vec \eta}}_i(\tau ) &=& {\vec \eta}_i(\tau ) + (-)^{i+1}\,{1\over
 2}\, Q_i\, \sum_{j\not= i}\, Q_j\, {{\partial\, {\cal K}_{12}({\vec \kappa}_1(\tau ),
 {\vec \kappa}_2(\tau ), {\vec \eta}_1(\tau ) - {\vec
 \eta}_2(\tau ))}\over {\partial\, {\vec \kappa}_j}},\nonumber \\
 {\tilde {\vec \kappa}}_i(\tau ) &=& {\vec \kappa}_i(\tau ) - (-)^{i+1}\,{1\over
 2}\, Q_i\, \sum_{j \not= i}\, Q_j\, {{\partial\, {\cal K}_{12}({\vec \kappa}_1(\tau ),
 {\vec \kappa}_2(\tau ), {\vec \eta}_1(\tau ) - {\vec
 \eta}_2(\tau ))}\over {\partial\, {\vec \eta}_j}},\nonumber \\
 &&{}\nonumber \\
 { \cal K}_{12}(\tau ) &=& - {\cal K}_{21}(\tau ) =
\int d^3\sigma\, [{\vec A}_{\perp S1} \cdot {\vec \pi}_{\perp S2} -
{\vec \pi}_{\perp S1} \cdot {\vec A}_{\perp S2}](\tau ,\vec \sigma).
 \label{2.5}
 \eea

\noindent The a-dimensional quantity ${ \cal K}_{12}(\tau )$ is
given in Eq.(5.35) of Ref. \cite{1} and in Eq.(I-3.5), while the
Lienard-Wiechert fields ${\vec A}_{\perp Si}$, ${\vec \pi}_{\perp
Si}$ are given in Eqs. (I-2.50), (I-2.51).

\bigskip

Therefore ${\vec {\mathcal{K}}}_{(int)} \approx 0$ can be solved to
get ${ \tilde {\vec \eta}}_{12} \approx {\tilde {\vec
\eta}}_{12}[{\tilde {\vec \kappa}}_1, {\tilde {\vec \kappa}}_2,
{\tilde {\vec \rho}}_{12}]$, so that, by taking into account the
rest-frame conditions ${\tilde {\vec \kappa}}_{12} \approx 0$, from
Eq.(\ref{2.4}) we get

\bea
 {\tilde {\vec \eta}}_1 \approx {\tilde {\vec \eta}}_{12}[{\tilde
{\vec \rho}}_{12}, {\tilde {\vec \pi}}_{12}] + {{m_2}\over m}\,
{\tilde {\vec \rho}}_{12},\qquad {\tilde {\vec \eta}}_2 \approx
{\tilde {\vec \eta}}_{12}[{\tilde {\vec \rho}}_{12}, {\tilde {\vec
\pi}}_{12}] - {{m_1}\over m}\, {\tilde {\vec \rho}}_{12},\qquad
{\tilde {\vec \kappa}}_1 \approx - {\tilde {\vec \kappa}}_2 \approx
{\tilde {\vec \pi}}_{12},\nonumber \\
 &&{} \label{2.6}
 \eea

\noindent with ${\tilde {\vec \rho}}_{12}(\tau )$ and ${\tilde {\vec
\pi}}_{12}(\tau )$ solutions of the Hamilton equations with $M\, c$
as Hamiltonian.\medskip

Then the inverse of the canonical transformation (\ref{2.5}) allows
one to get

\bea
 {\vec \eta}_i(\tau ) &=& {\tilde {\vec \eta}}_i(\tau ) - (-)^{i+1}\,{1\over
 2}\, Q_i\, \sum_{j \not= i}\, Q_j\, {{\partial\, {\tilde {\cal K}}
 _{12}({\tilde {\vec \kappa}}_1(\tau ),
 {\tilde {\vec \kappa}}_2(\tau ), {\tilde {\vec \eta}}_1(\tau ) - {\tilde {\vec
 \eta}}_2(\tau ))}\over {\partial\, {\vec \kappa}_j}} \approx\nonumber \\
 &\approx& {\tilde {\vec \eta}}_{12}[{\tilde
{\vec \rho}}_{12}, {\tilde {\vec \pi}}_{12}] +
(-)^{i+1}\,{{m_{i}}\over m}\, {\tilde {\vec \rho}}_{12} -
   (-)^{i}\,{1\over 2}\, Q_i\, \sum_{j \not= i}\, Q_j\, {{\partial\,
 {\tilde {\cal K}}_{12}({\tilde {\vec \pi}}_{12}(\tau ),
 - {\tilde {\vec \pi}}_{12}(\tau ), {\tilde {\vec \rho}}_{12}(\tau ))}
 \over {\partial\, {\tilde {\vec \pi}}_{12}}},\nonumber \\
 {\vec \kappa}_i(\tau ) &=& {\tilde {\vec \kappa}}_i(\tau ) + {1\over
 2}\, Q_i\, \sum_{j \not= i}\, Q_j\, {{\partial {\tilde {\cal K}}
 _{12}({\tilde {\vec \kappa}}_1(\tau ),
 {\tilde {\vec \kappa}}_2(\tau ), {\tilde {\vec \eta}}_1(\tau ) - {\tilde {\vec
 \eta}}_2(\tau ))}\over {\partial\, {\tilde {\vec \rho}}_{12}}},
 \label{2.7}
 \eea

 \noindent with ${\tilde {\cal K}}_{12}$ being the same function of
 its tilded arguments as ${\cal K}_{12}$ and its derivatives with respect
 to ${\tilde {\vec \pi}}_{12}$ are taken on the first argument.

 \bigskip

Finally, by using Eq.(I-2.4) for the embedding of the Wigner
hyper-planes in Minkowski space-time ($P^{\mu} = Mc\, h^{\mu} = Mc\,
(\sqrt{1 + {\vec h}^2}; \vec h)$ is the total 4-momentum of the
2-body system in an arbitrary inertial frame; $h^{\mu} =
\epsilon^{\mu}_{\tau}(\vec h)$ and $\epsilon^{\mu}_r(\vec h)$ are
the columns of the standard Wigner boost sending $P^{\mu}$ to the
rest-form $Mc\, (1; 0)$), we can reconstruct the relativistic orbits
and the 4-momenta ($p_i^2 = m_i^2\, c^2$) of the particles

\bea
 x^{\mu}_i(\tau ) &\approx& Y^{\mu}(\tau ) + \epsilon^{\mu}_r(\vec h)\,
 \Big( {\tilde \eta}^r_{12}[{\tilde
{\vec \rho}}_{12}, {\tilde {\vec \pi}}_{12}] +
(-)^{i+1}\,{{m_{i_1}}\over m}\, {\tilde \rho}^r_{12} -\nonumber \\
 &-&  (-)^{i}\,{1\over
 2}\, Q_i\, \sum_{j \not= i}\, Q_j\, {{\partial\, {\tilde
 {\cal K}}_{12}({\tilde {\vec \pi}}_{12}(\tau ),
 - {\tilde {\vec \pi}}_{12}(\tau ), {\tilde {\vec \rho}}_{12}(\tau ))}
 \over {\partial\, {\tilde \pi}_{12\, r}}} \Big), \nonumber \\
 &&{}\nonumber \\
 &&{}\nonumber \\
 p_i^{\mu}(\tau ) &\approx& \sqrt{m_i^2\, c^2 + {\tilde
 {\vec \pi}}^2_{12}(\tau )}\, h^{\mu} + (-)^{i+1}\,
 \epsilon^{\mu}_r(\vec h)\, {\tilde \pi}^r_{12}(\tau ).
 \label{2.8}
 \eea

\vfill\eject

\section{Collective and Relative variables for the Rest-Frame
Description of the Radiation Field in the Radiation Gauge}

Let us now consider the rest-frame instant form of a radiation field
in the radiation gauge, which was given in Eqs.(I-2.33)-(I-2.34) of
paper I. Given the internal Poincare' generators of Eqs.(I-2.35), we
must find a canonical basis spanned by collective and relative
variables allowing one to eliminate explicitly the internal 3-center
of mass as required by the constraints ${\vec {\cal P}}_{rad}
\approx 0$ and ${\vec {\cal K}}_{rad} \approx 0$. An approach to the
determination of such variables has been done in Ref.\cite{5} for
the massive Klein-Gordon field and in Ref.\cite{6} for massless
fields. The rest-frame instant form of this approach  was given in
Ref.\cite{13} for the massive Klein-Gordon field. In Appendix A
there is review and the adaptation to the rest-frame instant form of
the results for the massless scalar field. In this Section we will
use these methods for the transverse radiation field.

\subsection{The Radiation Field and its Internal Poincare' Generators}

As shown in I, Eqs. (I-2.33)-(I-2.34), we have the following
rest-frame representation of the radiation field in the radiation
gauge on the Wigner hyper-planes  \footnote{$\sigma^A =
(\sigma^{\tau} = \tau ; \sigma^r)$; $k^A= (k^{\tau} = |\vec k| =
\omega (\vec k); k^r)$, $k^2 = 0$ , with $\vec k$ Wigner spin-1
3-vector and $k^{\tau}$ Lorentz scalar; $ d\tilde k =
{\frac{{d^3k}}{{2\, \omega (\vec k)\, (2 \pi )^3}}}$, $\Omega (\vec
k) = 2\, \omega (\vec k)\, (2\pi)^3$, $[d\tilde k] = [l^{-2}]$.}\,\,

\begin{eqnarray*}
{\vec A}_{\perp rad}(\tau ,\vec \sigma )\, &{\buildrel \circ \over
{{=}}} \,& \int d\tilde k\, \sum_{\lambda =1,2}\, {\vec
\epsilon}_{\lambda}(\vec k)\, \Big[ a_{\lambda}(\vec k)\, e^{- i\,
[\omega (\vec k)\, \tau - \vec k \cdot \vec \sigma]} +
a^{*}_{\lambda}(\vec k)\, e^{i\, [\omega (\vec k)\,
\tau - \vec k \cdot \vec \sigma]}\Big],  \nonumber \\
&&{}  \nonumber \\
{\vec \pi}_{\perp rad}(\tau ,\vec \sigma )&=&  {\vec E} _{\perp
rad}(\tau ,\vec \sigma )\, {\buildrel \circ \over {{=}}} - {\frac{{
\partial}}{{\partial \tau}}}\, {\vec A}_{\perp rad}(\tau ,\vec \sigma ) =
\nonumber \\
&=& i \int d\tilde k \, \omega (\vec k)\, \sum_{\lambda =1,2}\,
{\vec \epsilon}_{\lambda}(\vec k)\, \Big[ a_{\lambda}(\vec k)\, e^{-
i\, [\omega (\vec k)\, \tau - \vec k \cdot \vec \sigma]} -
a^{*}_{\lambda}(\vec k)\, e^{i\, [\omega (\vec k)\, \tau - \vec k
\cdot \vec \sigma]}\Big],  \nonumber \\
&&{}  \nonumber \\
{\vec B}_{rad}(\tau ,\vec \sigma ) &=& \vec \partial \times {\vec
A}_{\perp
rad}(\tau ,\vec \sigma ) =  \nonumber \\
&=& i\, \int d\tilde k\, \sum_{\lambda}\, \vec k \times {\vec
\epsilon} _{\lambda}(\vec k)\, \Big[a_{\lambda}(\vec k)\, e^{- i\,
[\omega (\vec k)\, \tau - \vec k \cdot \vec \sigma ]} -
a^{*}_{\lambda}(\vec k)\, e^{i\, [\omega (\vec k)\, \tau - \vec k
\cdot \vec \sigma ]}\Big],
 \end{eqnarray*}

\bea
 a_{\lambda}(\vec k)&=& \int d^3\sigma\, {\vec
\epsilon}_{\lambda}(\vec k)\cdot \Big[ \omega (\vec k)\, {\vec
A}_{\perp rad}(\tau ,\vec \sigma ) - i\, {\vec \pi} _{\perp
rad}(\tau ,\vec \sigma ) \Big]\, e^{- i\, \vec k
\cdot \vec \sigma},  \nonumber \\
&&{}  \nonumber \\
&&{}  \nonumber \\
&&\{ A^r_{\perp rad}(\tau ,\vec \sigma ), \pi^s_{\perp rad}(\tau
,{\vec \sigma}_1)\} = - c\, P^{rs}_{\perp}(\vec \sigma )\,
\delta^3(\vec \sigma - {\vec \sigma}_1),  \nonumber \\
&&{}  \nonumber \\
\{ a_{\lambda}(\vec k), a^{*}_{\lambda^{^{\prime}}}({\vec
k}^{^{\prime}}) \} &=& - i\, 2\, \omega (\vec k)\, c\, (2\pi )^3\,
\delta_{\lambda \lambda^{^{\prime}}}\, \delta^3(\vec k - {\vec
k}^{^{\prime}}) { \buildrel {def}\over {{=}}} - i\, \Omega (\vec
k)\, c\, \delta_{\lambda
\lambda^{^{\prime}}}\, \delta^3(\vec k - {\vec k}^{^{\prime}}),  \nonumber \\
\{ a_{\lambda}(\hat k), a_{\lambda^{^{\prime}}}({\hat
k}^{^{\prime}}) \} &=& \{ a^{*}_{\lambda}(\hat k),
a^{*}_{\lambda^{^{\prime}}}({\hat k}
^{^{\prime}})\} = 0,  \nonumber \\
&&{}  \nonumber \\
\delta^{ij} &=& \sum_{\lambda =1,2}\, \epsilon^i_{\lambda}(\vec k)\,
\epsilon^j_{\lambda}(\vec k) + {\frac{{k^i\, k^j}}{{|\vec
k|^2}}},\quad\quad
\vec k \cdot {\vec \epsilon}_{\lambda}(\vec k)=0,  \nonumber \\
&&{}  \nonumber \\
&& {\vec \epsilon}_{\lambda}(\vec k) \cdot {\vec \epsilon}
_{\lambda^{^{\prime}}}(\vec k) = \delta_{\lambda
\lambda^{^{\prime}}},\qquad {\frac{{\vec k}}{{|\vec k|}}} \cdot
\Big[{\vec \epsilon}_1(\vec k) \times {\ \vec \epsilon}_2(\vec
k)\Big] = 1,
 \label{3.1}
\end{eqnarray}

\bigskip

\noindent and the following expression for the internal Poincare'
generators of the radiation field [ $ {\cal P}^A_{rad} = ({\cal
P}^{\tau}_{rad} = {\cal E}_{rad}/c = M_{rad}\, c; {\vec {\cal
P}}_{rad})$, $ {\cal J}^u_{rad} = {\frac{1}{2}}\, \epsilon^{urs}\,
{\cal J}^{rs}_{rad}$, ${\cal J}^{rs}_{rad} = \epsilon^{rsu}\, {\cal
J}^u_{rad}$]

\begin{eqnarray*}
M_{rad}\, c^2 &=& {\cal E}_{rad} = c\, {\cal P}^{\tau}_{rad}=
{\frac{1}{2}}\, \int d^3\sigma\, \Big[ {\vec \pi} _{\perp rad}^2 +
{\vec B}_{rad}^2\Big] (\tau ,\vec \sigma ) = \sum_{\lambda =1,2}\,
\int d\tilde k\, \omega (\vec k)\,
a^{*}_{\lambda}(\vec k)\, a_{\lambda}(\vec k),  \nonumber \\
&&{}  \nonumber \\
{\vec {\cal P}}_{rad}&=& {\frac{1}{c}}\,\int d^3\sigma\, \Big[ {\vec
\pi}_{\perp rad} \times {\vec B }_{rad}\Big] (\tau ,\vec \sigma ) =
{\frac{1}{c}}\, \sum_{\lambda =1,2}\, \int d\tilde k\, \vec k\,
a^{*}_{\lambda}(\vec k)\, a_{\lambda}(\vec k) \approx 0,
\end{eqnarray*}

\begin{eqnarray*}
 {\vec {\cal J}}_{rad} &=& {\vec {\bar S}}_{rad} =
 {\frac{1}{c}}\, \int d^3\sigma\, \vec \sigma
\times \Big({ \vec \pi}_{\perp rad} \times {\vec B}_{rad}\Big)(\tau
,\vec \sigma ) =
\nonumber \\
&=& {\frac{i}{c}}\, \sum_{\lambda}\, \int d\tilde k\,
a^*_{\lambda}(\vec k)\, \vec k \times {\frac{{\partial}}{{\partial\,
\vec k}}}\,
a_{\lambda}(\vec k) +  \nonumber \\
&+& {\frac{i}{2\, c}}\, \sum_{\lambda \lambda^{^{\prime}}}\, \int
d\tilde k\, \Big[ a_{\lambda}(\vec k)\,
a^{*}_{\lambda^{^{\prime}}}(\vec k) - a^{*}_{\lambda}(\vec k)\,
a_{\lambda^{^{\prime}}}(\vec k)\Big]\, {\vec
\epsilon}_{\lambda}(\vec k) \cdot \Big( \vec k\, \times
{\frac{{\partial}}{{
\partial\, \vec k}}}\Big)\, {\vec \epsilon}_{\lambda^{^{\prime}}}(\vec k) -
\nonumber \\
&-& {\frac{i}{c}}\, \sum_{\lambda \lambda^{^{\prime}}}\, \int
d\tilde k\, { \vec \epsilon} _{\lambda}(\vec k) \times {\vec
\epsilon}_{\lambda^{^{ \prime}}}(\vec k)\, a^*_{\lambda}(\vec k)\,
a_{\lambda^{^{\prime}}}(\vec k),
\end{eqnarray*}

\begin{eqnarray}
 {\cal K}^r_{rad}&=& {\cal J}^{\tau r}_{rad} = - {\frac{1}{2\, c}}\, \int d^3\sigma
\, \sigma^r\, \Big[ {\vec \pi}_{\perp rad}^2 + {\vec B}_{rad}^2\Big]
(\tau ,\vec \sigma ) =  \nonumber \\
&=& {\frac{i}{c}}\, \int d\tilde k\, a^{*}_{\lambda}(\vec k)\,
\omega (\vec k)\, {\frac{{\partial}}{{\partial k^r}}}\,
a_{\lambda}(\vec k) +  \nonumber \\
&+& {\frac{i}{2\, c}}\, \sum_{\lambda ,\lambda^{^{\prime}}=1,2}\,
\int d\tilde k\, \Big[ a_{\lambda}(\vec k)\,
a^{*}_{\lambda^{^{\prime}}}(\vec k) - a^{*}_{\lambda}(\vec k)\,
a_{\lambda^{^{\prime}}}(\vec k)\Big]\, {\vec
\epsilon}_{\lambda}(\vec k) \cdot \omega (\vec k)\, {\frac{{\partial
{\vec \epsilon}_{\lambda^{^{\prime}}}(\vec k)}}{{\partial k^r}}}
\approx 0,  \nonumber \\
&& {}  \nonumber \\
&&{}  \nonumber \\
h_{rad} &=& {\frac{{{\vec {\cal J}}_{rad} \cdot \vec k}}{{|\vec
k|}}} = {\frac{i}{c} }\, \int d\tilde k\, \Big[ a^{*}_2(\vec k)\,
a_1(\vec k) - a^{*}_1(\vec k)\, a_2(\vec k)\Big],
 \label{3.2}
\end{eqnarray}

\noindent where in the last line we defined the \textit{helicity}
\footnote{ Dimensions: $[h] = [{\vec {\cal J}}] = [m\, l^2\,
t^{-1}]$.}.

\bigskip

As shown in Appendix A based on Ref. \cite{6} \footnote{For the free
radiation field the Fourier coefficients and the modulus-phase
variables are $\tau$-independent.}, we must have $a_{\lambda}(\hat
k),\quad \vec \partial a_{\lambda}(\hat k) \in L_2(R^3, d^3k)$ for
the existence of the previous ten integrals (and of the occupation
number $N_{\lambda} = \int d\tilde k\, a^*_{\lambda}(\vec k)\,
a_{\lambda}(\vec k)$) as finite quantities. Moreover one can show
\cite{6} the existence of the following behavior: i)
$|a_{\lambda}(\hat k)|\, {\ \rightarrow}_{|\vec k| \rightarrow
\infty} \, |\vec k|^{-{\frac{3}{2}}-\rho}$ with $\rho > 0$; ii)
$|a_{\lambda}(\hat k)|\, {\rightarrow}_{|\vec k| \rightarrow 0}\,
|\vec k|^{-{\frac{3}{2}}+\epsilon}$ with $\epsilon > 0$.

\bigskip

By going to the circular basis $a_{\pm}(\vec k) = {1\over
{\sqrt{2}}}\, (a_1(\vec k) \mp\, i\, a_2(\vec k))$,
$\epsilon^{\mu}_{\pm}(\vec k) = {1\over {\sqrt{2}}}\,
(\epsilon^{\mu}_1(\vec k) \pm\, i\, \epsilon^{\mu}_2(\vec k))$, and
by introducing the modulus-phase variables ($[I_{\sigma}] = [m\,
l^5\, t^{-2}]$)

\begin{eqnarray}
I_{\sigma}(\vec k)&=& a^{*}_{\sigma}(\vec k) a_{\sigma}(\vec
k),\qquad \phi_{\sigma}(\vec k) = {\frac{1}{{2i}}} ln\,
{\frac{{a_{\sigma}(\vec k)}}{{ a^{*}_{\sigma}(\vec k)}}},\quad
\Rightarrow\quad a_{\sigma}(\vec k)= \sqrt{I_{\sigma}(\vec k)} e^{i
\phi_{\sigma}(\vec k)},  \nonumber \\
&&{}  \nonumber \\
&&\{ I_{\sigma}(\vec k), \phi_{\sigma^{^{\prime}}}({\vec
k}^{^{\prime}}) \} = \Omega (\vec k) \delta_{\sigma
\sigma^{^{\prime}}} \delta^3(\vec k-{\vec k} ^{^{\prime}}),
 \label{3.3}
\end{eqnarray}

\noindent the helicity and the Poincar\'e generators of the field
configuration become

\begin{eqnarray*}
 h_{rad}&=& {\frac{1}{c}}\, \int d\tilde k\, \Big[ I_{+}(\vec k) -
I_{-}(\vec k)\Big],  \nonumber \\
 {\cal P}^A_{rad}&=& {\frac{1}{c}}\, \sum_{\sigma =\pm} \int d\tilde k\,
k^A\, a^{*}_{\sigma}(\vec k) a_{\sigma}(\vec k) = {\frac{1}{c}}\,
\sum_{\sigma =\pm} \int d\tilde k\, k^A\, I_{\sigma}(\vec k) =
\nonumber \\
 &=& \Big( {\cal P}^{\tau}_{rad} = {\frac{{{\cal E}_{rad}}}{c}} = M_{rad}\, c;
{\vec {\cal P}}_{rad} \approx 0\Big),  \end{eqnarray*}

\bea
 {\vec {\cal J}}_{rad} &=& - {\frac{1}{c}}\, \sum_{\sigma}\, \int d\tilde
k\, I_{\sigma}(\vec k)\, \Big(\vec k \times
{\frac{{\partial}}{{\partial\, \vec k
}}}\Big)\, \phi_{\sigma}(\vec k) +  \nonumber \\
&+& {\frac{i}{2\, c}} \int d\tilde k\, \Big[ I_{+}(\vec k) -
I_{-}(\vec k) \Big] \Big[ {\vec \epsilon}_{+}(\vec k) + {\vec
\epsilon}_{-}(\vec k)\Big] \cdot \Big(\vec k \times
{\frac{{\partial}}{{\partial\, \vec k}}}\Big)\, \Big[ {\ \vec
\epsilon}_{-}(\vec k) - {\vec \epsilon}_{+}(\vec k)\Big] +
\nonumber \\
&+& {\frac{i}{c}}\, \int d\tilde k\, {\vec \epsilon}_{+}(\vec k)
\times { \vec \epsilon}_{-}(\vec k)\, \Big[ I_{+}(\vec k) -
I_{-}(\vec k)\Big],\nonumber \\
 {\cal K}^r_{rad} &=&  - {\frac{1}{c}}\, \sum_{\sigma =\pm}
\int d\tilde k\, I_{\sigma}(\vec k) \omega (\vec
k){\frac{{\partial}}{{\partial k^r}}} \phi_{\sigma}(\hat k)+  \nonumber \\
&+& {\frac{i}{2\, c}} \int d\tilde k\, \Big[ I_{+}(\vec k) -
I_{-}(\vec k) \Big] \Big[ {\vec \epsilon}_{+}(\vec k) + {\vec
\epsilon}_{-}(\vec k)\Big] \cdot \, \omega (\vec
k){\frac{{\partial}}{{\partial k^r}}} \Big[ {\vec \epsilon}
_{-}(\vec k) - {\vec \epsilon}_{+}(\vec k)\Big] \approx 0.
  \label{3.4}
\eea

As shown in Appendix A and Ref.\cite{6}, the existence of finite
values of these quantities requires \cite{6} :\medskip

i) $ |I_{\sigma}(\vec k)|\, {\rightarrow}_{|\vec k| \rightarrow
\infty}\, |\vec k|^{-3-\delta}$ with $\delta > 0$;
$|\phi_{\sigma}(\vec k)|\, {\rightarrow} _{|\vec k| \rightarrow
\infty}\, |\vec k|$ (this is due to the transformation properties
under Poincar\'e transformations; $\phi^{^{\prime}}_{\sigma}(\vec
k)=\phi_{\sigma}( {\vec {\Lambda^{-1}\, k}}) + k \cdot a +\sigma\,
\theta (\vec k,\Lambda^{-1}) $);\medskip

ii) $|I_{\sigma}(\vec k)|\, {\rightarrow}_{|\vec k| \rightarrow 0}\,
|\vec k|^{-2+\epsilon}$ with $\epsilon > 0$; $ |\phi_{\sigma}(\vec
k)|\, {\rightarrow}_{|\vec k| \rightarrow 0}\, |\vec k|^{\alpha}$
with $\alpha > -\delta$.

\bigskip

Let us replace the canonical variables $I_{\sigma}(\vec k)$, $
\phi_{\sigma}(\vec k)$ with the following new canonical basis
$\Psi_{rad}(\vec k)$, $\Phi_{rad}(\vec k)$, $\lambda_{rad}(\vec k)$,
$\rho_{rad}(\vec k)$  \footnote{ Dimensions: $[\Psi_{rad} ] =
[\lambda_{rad} ] = [I_{\sigma}] = [m\, l^5\, t^{-2}]$. }

\begin{eqnarray}
 \Psi_{rad}(\vec k)&=&\sum_{\sigma =\pm} I_{\sigma}(\vec k),\qquad
 \Phi_{rad}(\vec k) = {\frac{1}{2}} \sum_{\sigma =\pm}
 \phi_{\sigma}(\vec k),  \nonumber \\
 &&{}  \nonumber \\
 \lambda_{rad} (\vec k)&=&I_{+}(\vec k)-I_{-}(\vec k),\qquad
\rho_{rad}(\vec k) = {\frac{1}{2}} \Big[ \phi_{+}(\vec k) -
\phi_{-}(\vec k)\Big],  \nonumber \\
 &&{}  \nonumber \\
 &&\{ \Psi_{rad}(\vec k), \Phi_{rad}({\vec k}^{^{\prime}}) \} = \{
\lambda_{rad} (\vec k), \rho_{rad} ({\vec k}^{^{\prime}}) \} =
\Omega (\vec k) \delta^3(\vec k-{ \vec k}^{^{\prime}}),\nonumber \\
 &&{}\nonumber \\
 I_+(\vec k) &=& {1\over 2}\, \Big(\Psi_{rad}(\vec k) + \lambda_{rad}(\vec
k)\Big),\qquad I_{-}(\vec k) = {1\over 2}\, \Big(\Psi_{rad}(\vec k)
- \lambda_{rad}(\vec k)\Big),\nonumber \\
 \phi_+(\vec k) &=& \Phi_{rad}(\vec k) + \rho_{rad}(\vec k),\qquad
 \phi_{-}(\vec k) = \Phi_{rad}(\vec k) - \rho_{rad}(\vec k),
  \label{3.5}
\end{eqnarray}

\noindent in terms of which the helicity and the Poincar\'e
generators become

\begin{eqnarray}
 h_{rad}&=& {\frac{1}{c}}\, \int d\tilde k\, \lambda_{rad} (\vec k),
\nonumber \\
 {\cal P}^A_{rad}&=& {\frac{1}{c}}\, \int d\tilde k\, k^A\, \Psi_{rad}(\vec k) =
\Big( {\cal P}^{\tau}_{rad} = {\frac{{{\cal E}_{rad}}}{c}} =
M_{rad}\, c; {\vec {\cal P}}_{rad} \approx 0\Big),\nonumber \\
  {\vec {\cal J}}_{rad} &=& - {\frac{1}{c}}\,  \int d\tilde
 k\, \Psi_{rad}(\vec k)\, \Big(\vec k \times {\frac{{\partial}}{{\partial\,
\vec k}}}\Big)\, \Phi_{rad}(\vec k) -  \nonumber \\
 &-& {\frac{1}{c}}\,  \int d\tilde k\, \lambda_{rad}(\vec k)\, \Big(\vec
k \times {\frac{{\partial}}{{\partial\, \vec k }}}\Big)\,
\rho_{rad}(\vec k) +\nonumber \\
 &+& {\frac{i}{2\, c}} \int d\tilde k\, \lambda_{rad}(\vec k)\,
 \Big[ {\vec \epsilon}_{+}(\vec k) + {\vec \epsilon}_{-}(\vec
k)\Big] \cdot \Big(\vec k \times {\frac{{\partial}}{{\partial\, \vec
k}}}\Big)\, \Big[ {\ \vec \epsilon}_{-}(\vec k) - {\vec
\epsilon}_{+}(\vec k)\Big] +\nonumber \\
&+& {\frac{i}{c}}\, \int d\tilde k\, \lambda_{rad}(\vec k)\,
\Big[{\vec \epsilon}_{+}(\vec k)
\times { \vec \epsilon}_{-}(\vec k)\Big],\nonumber \\
 {\cal K}^r_{rad} &=&  {\frac{1}{c}}\, \int d\tilde k\, \Psi_{rad}
(\vec k) \omega (\hat k){\ \frac{{\partial}}{{\partial k^r}}}
\Phi_{rad} (\vec k) + {\frac{1 }{c}}\, \int d\tilde k\,
\lambda_{rad}(\vec k) \omega (\vec k){\frac{{
\partial}}{{\partial k^r}}} \rho_{rad}(\vec k)+  \nonumber \\
&+&{\frac{i}{2\, c}} \int d\tilde k\, \lambda_{rad}(\vec k) \Big[
{\vec \epsilon} _{+}(\vec k) + {\vec \epsilon}_{-}(\vec k)\Big]
\cdot \, \omega (\vec k){\frac{ {\partial}}{{\partial k^r}}} \Big[
{\vec \epsilon}_{-}(\vec k) - {\vec \epsilon }_{+}(\vec k)\Big]
\approx 0.
  \label{3.6}
\end{eqnarray}

One has:\hfill\break \hfill\break i) $|\Psi (\vec k)|\, {
\rightarrow }_{|\vec k| \rightarrow 0}\, |\vec k|^{-3+\epsilon}$
with $ \epsilon > 0$; $|\Phi (\vec k)|\, {\rightarrow}_{|\vec k|
\rightarrow 0}\, |\vec k|^{\alpha}$ with $\alpha > 1-\epsilon$;
$|\lambda_{rad}(\vec k)|\, { \rightarrow}_{|\vec k| \rightarrow 0}\,
|\vec k|^{-2-\delta}$ with $\delta > 0$; $|\rho_{rad}(\vec k)|\,
{\rightarrow}_{|\vec k| \rightarrow 0}\, |\vec k|^{\beta}$ with
$\beta > -\delta$;\hfill\break \hfill\break ii) $|\Psi (\vec k)|\,
{\rightarrow} _{|\vec k| \rightarrow \infty}\, |\vec k|^{-3-\gamma}$
with $\gamma > 0$; $|\Phi (\vec k)|\, {\rightarrow}_{|\vec k|
\rightarrow \infty}\, |\vec k|$ ; $|\lambda_{rad}(\vec k)|\,
{\rightarrow} _{|\vec k| \rightarrow \infty}\, |\vec k|^{-2-\chi}$
with $\chi > 0$; $ |\rho_{rad}(\vec k)|\, {\rightarrow}_{|\vec k|
\rightarrow \infty}\, |\vec k|^{\beta}$ with $\beta > \chi$.

\bigskip

\subsection{Collective and Relative Variables for the Radiation
Field}

We can now use the results of Appendix A to replace the canonical
variables $\Psi_{rad}(\vec k)$ and $\Phi_{rad}(\vec k)$ with a set
of collective and relative canonical variables $X^A_{rad}$, ${\cal
P}^A_{\phi}$, ${\bf H}_{rad}(\vec k)$, ${\bf K}_{rad}(\vec k)$. To
them must be added the remaining canonical variables
$\lambda_{rad}(\vec k)$ and $\rho_{rad}(\vec k)$ describing the
difference of the modulus-phase variables with different circular
polarization (they are already of the type of relative
variables).\bigskip

By using Eqs.(\ref{a6}), (\ref{a7}), (\ref{a9}) and the function
${\cal F}(\vec k, {\cal P}^B_{rad})$ of Eq.(\ref{a8}) with $\omega
(k) = k$ and with the functions $F^{\tau}(k)$, $F(k)$ given in
Eqs.(\ref{a11}), the collective and relative variables are

\bea
 X^A_{rad} &=& \int d\tilde k\, {{\partial\, {\cal F}(\vec k,
 {\cal P}^B_{\phi})}\over {\partial\, {\cal P}_{\phi\, A}}}\,
 \Phi_{rad}(\vec k),\nonumber \\
 {\cal P}^A_{rad}&=& {\frac{1}{c}}\, \int d\tilde k\, k^A\,
 \Psi_{rad}(\vec k) = \int d\tilde k\, k^A\, {\cal F}(\vec k,
 {\cal P}^B_{rad}),\nonumber \\
 &&{}\nonumber \\
  {\bf H}_{rad}(\vec k) &=& \int d\tilde q\, {\cal G}(\vec k, \vec
 q)\, \Big[\Psi_{rad}(\vec q) - \omega (q)\, F^{\tau}(q)\, \int d{\tilde q}_1\,
 \omega (q_1)\, \Psi_{rad}(,{\vec q}_1) +\nonumber \\
 &+& F(q)\, \vec q \cdot \int d{\tilde q}_1\, {\vec q}_1\,
 \Psi_{rad}({\vec q}_1)\Big],\nonumber \\
 {\bf K}_{rad}(\vec k) &=& {\cal D}_{\vec k}\, \Phi_{rad}(\vec
 k),\nonumber \\
 &&{}\nonumber \\
 &&\lambda_{rad}(\vec k),\qquad \rho_{rad}(\vec k),\nonumber \\
 &&{}\nonumber \\
 &&\{ X^A_{rad}, {\cal P}^B_{rad} \} = - \eta^{AB},\nonumber \\
 && \{ {\bf H}_{rad}(\vec k), {\bf K}_{rad}({\vec k}_1) \} = (2\pi )^3\,
 2\, \omega (k)\, \delta^3(\vec k - {\vec k}_1),\nonumber \\
 &&\{ \lambda_{rad}(\vec k), \rho_{rad}({\vec k}_1) \} = (2\pi )^3\,
 2\, \omega (k)\, \delta^3(\vec k - {\vec k}_1),
 \label{3.7}
 \eea

 \noindent with all the other Poisson brackets vanishing. The
 operator ${\cal D}_{\vec q}$ and its Green function are defined in
 Ref.\cite{6} and Appendix A. By construction $X^A_{rad}$ is a {\it
 4-center of phase}.
\bigskip

The helicity and Poincar\'e generators become [$D^{rs} = k^r\, {\
\frac{{\partial} }{{\partial k^s}}} - k^s\,
{\frac{{\partial}}{{\partial k^r}}}$, $D^{\tau r} = \omega (\hat
k)\, {\frac{{\partial}}{{\partial k^r}}}$]

\begin{eqnarray*}
 h_{rad}&=& {\frac{1}{c}}\, \int d\tilde k\, \lambda_{rad}(\vec k),
\nonumber \\
 &&{}\nonumber \\
 {\cal P}_{rad}^A&=& \Big({\cal P}^{\tau}_{rad} = {\frac{{{\cal E}_{rad}}}{c}} = M_{rad}\,
c;\quad {\vec {\cal P}}_{rad} \approx 0\Big),  \nonumber \\
 &&{}\nonumber \\
 {\vec {\cal J}}_{rad} &=& {\vec {\bar S}}_{rad} = {\vec X}_{rad} \times
 {\vec {\cal P}}_{rad} + {\vec S}_{rad},\nonumber \\
 &&{}\nonumber \\
 {\vec S}_{rad} &=& - {\frac{1}{c}}\, \int d\tilde k\, H_{rad}(\vec k)\,
 \Big(\vec k \times {\frac{{\partial}}{{\partial\, \vec k}}}\Big)\, K_{rad}(\vec k) -
\nonumber \\
 &-& {\frac{1}{c}}\, \int d\tilde k\,
\lambda_{rad}(\vec k)\, \Big(\vec k \times
{\frac{{\partial}}{{\partial\, \vec k}}}\Big)\, \rho_{rad}(\vec k)
 +\nonumber \\
&+& {\frac{1}{c}}\, \int d\tilde k\, \lambda_{rad}(\vec k)\, \,
\Big[{\vec \epsilon}_-(\vec k) \times {\vec \epsilon}_+(\vec k) +
{\frac{i}{2}}\, \Big({ \vec \epsilon}_-(\vec k) + {\vec
\epsilon}_+(\vec k)\Big) \cdot \vec k \times
{\frac{{\partial}}{{\partial\, \vec k}}}\, \Big({\vec \epsilon}
_-(\vec k) - {\vec \epsilon}_+(\vec k)\Big) \Big],
 \end{eqnarray*}

\bea
 {\cal K}^r_{rad} &=&  X^{\tau}_{rad}\, {\cal P}_{rad}^r -
X_{rad}^r\, M_{rad}\, c -  \nonumber \\
&-& {\frac{1}{c}}\, \int d\tilde k\, H_{rad}(\vec k)\, D^{\tau r}\,
K_{rad}(\vec k) - {\frac{1}{c}}\, \int d\tilde k\,
\lambda_{rad}(\vec k)\, D^{\tau r}\, \rho_{rad}(\vec k)-  \nonumber \\
&-&{\frac{i}{2\, c}} \int d\tilde k\, \lambda_{rad}(\vec k)\, \Big(
\Big[ {\vec \epsilon} _{+}(\vec k) + {\vec \epsilon}_{-}(\vec
k)\Big] \cdot \, D^{\tau r}\, \Big[ {\vec \epsilon}_{-}(\vec k)
-{\vec \epsilon}_{+}(\vec k)\Big]\Big) \approx 0.
 \label{3.8}
\end{eqnarray}

\bigskip

The gauge fixing ${\vec {\cal K}}_{rad} \approx 0$ to the rest-frame
conditions ${\vec {\cal P}}_{rad} \approx 0$ leads to the following
determination of the 3-center of phase ${\vec X}_{rad}$

\bea
 X^r_{rad} &\approx& - {1\over {M_{rad}\, c^2}}\, \Big[
\int d\tilde k\, H_{rad}(\vec k)\, D^{\tau r}\, K_{rad}(\vec k) +
\int d\tilde k\, \lambda_{rad}(\vec k)\, D^{\tau r}\,
\rho_{rad}(\vec k) +\nonumber \\
&+& {i\over 2}\, \int d\tilde k\, \lambda_{rad}(\vec k)\, \Big(
\Big[ {\vec \epsilon} _{+}(\vec k) + {\vec \epsilon}_{-}(\vec
k)\Big] \cdot \, D^{\tau r}\, \Big[ {\vec \epsilon}_{-}(\vec k)
-{\vec \epsilon}_{+}(\vec k)\Big]\Big)\,\, \Big].
 \label{3.9}
 \eea

\bigskip

If ${\vec A}_{\perp rad}(\tau ,\vec \sigma )$ satisfies $\int
d\tilde k\, w_{lm}(\vec k) \Big[ \Psi_{rad}(\vec k)- {\cal F}(\vec
k, {\cal P}^B_{rad})\Big] =0$ for $l \geq 2$ (see Ref.\cite{6} and
Appendix A), then we have the following expression of the fields of
the transverse radiation field

\begin{eqnarray}
{\vec A}_{\perp rad}(\tau ,\vec \sigma )&=& \int d\tilde k\,
\sum_{\sigma =\pm}\, \Big[ {\vec \epsilon}_{\sigma}(\vec k)
\sqrt{{\frac{1}{2}} \Big( {\cal F}(\vec k, {\cal P}^B_{rad}) +
\sigma\, \lambda_{rad}(\vec k)\Big) + {\cal D}_{\vec k}\,
{\bf H}_{rad}(\vec k)}  \nonumber \\
&&e^{i[ k_A\, (X^A_{rad} - \sigma^A) + \sigma\, \rho_{rad}(\vec k) +
\int d{\tilde k}_1\, d{\tilde k}_2\, {\bf K}_{rad}({\vec k}_1)\,
{\cal G}({ \vec k}_1,{\vec k}_2)\, \triangle ({\vec k}_2, \vec k)
]}\,\, + c.c. \Big],  \nonumber \\
&&{}  \nonumber \\
{\vec \pi}_{\perp rad}(\tau ,\vec \sigma ) &=& i\, \int d\tilde k\,
\omega(\vec k)\, \sum_{\sigma =\pm}\, \Big[ {\vec
\epsilon}_{\sigma}(\vec k) \sqrt{{\frac{1}{2}} \Big( {\cal F}(\vec
k, {\cal P}^B_{rad}) + \sigma\,
\lambda_{rad}(\vec k)\Big) + {\cal D}_{\vec k}\, {\bf H}_{rad}(\vec k)}  \nonumber \\
&&e^{i[ k_A\, (X^A_{rad} - \sigma^A) +\sigma\, \rho_{rad}(\vec k) +
\int d{\ \tilde k}_1\, d{\tilde k}_2\, {\bf K}_{rad}({\vec k}_1)\,
{\cal G}({ \vec k}_1 ,{\vec k}_2)\, \triangle ({\vec k}_2, \vec k)
]}\,\,
+ c.c.\Big],  \nonumber \\
&&{}  \nonumber \\
{\vec B}_{rad}(\tau ,\vec \sigma ) &=& i\, \int d\tilde k\,
\sum_{\sigma =\pm}\, \vec k \times \Big[ {\vec
\epsilon}_{\sigma}(\vec k) \sqrt{{\frac{1}{ 2}} \Big( {\cal F}(\vec
k, {\cal P}^B_{rad}) + \sigma\, \lambda_{rad}(\vec k)\Big)
+ {\cal D}_{\vec k}\, {\bf H}_{rad}(\vec k)}  \nonumber \\
&&e^{i[ k_A\, (X^A_{rad} - \sigma^A) +\sigma\, \rho_{rad}(\vec k)
+\int d{\ \tilde k}_1\, d{\tilde k}_2\, {\bf K}_{rad}({\vec k}_1)\,
{\cal G}({ \vec k}_1 ,{\vec k}_2)\, \triangle ({\vec k}_2, \vec k)
]}\,\, + c.c. \Big],
  \label{3.10}
\end{eqnarray}

\noindent so that the radiation fields have the following dependence
upon $\tau$ and $\vec \sigma$

\bea
 {\vec A}_{\perp rad}(\tau ,\vec \sigma ) &=& {\tilde {\vec
 A}}_{\perp rad}(\tau - X^{\tau}_{rad}, \vec \sigma - {\vec
 X}_{rad}),\nonumber \\
 {\vec \pi}_{\perp rad}(\tau ,\vec \sigma ) &=& {\tilde {\vec
 \pi}}_{\perp rad}(\tau - X^{\tau}_{rad}, \vec \sigma - {\vec
 X}_{rad}),\nonumber \\
 {\vec B}_{ rad}(\tau ,\vec \sigma ) &=& {\tilde {\vec
 B}}_{ rad}(\tau - X^{\tau}_{rad}, \vec \sigma - {\vec
 X}_{rad}),
 \label{3.11}
 \eea

\bigskip

The configurations of the radiation field admitting the collective
variables $X^A_{rad}$ have a "monopole" structure, carried by ${\vec
X}_{rad}$ and ${\vec p}_{rad} \approx 0$, plus the "multipoles"
$K_{rad}(\vec k)$, $H_{rad}(\vec k)$, $\lambda_{rad}(\vec k)$,
$\rho_{rad}(\vec k)$ (these last multipoles describe the helicity
structure). Eqs.(\ref{3.9}) expresses ${\vec X}_{rad}$ in terms of
$M_{rad}$ and of the higher multipoles.\medskip

As in the Klein-Gordon case we have the extra variables
$X^{\tau}_{rad}$, ${\cal P}^{\tau}_{rad} = M_{rad}\, c$ with a
similar interpretation (see Appendix A).

\bigskip

By using Eq.(I-2.25) we have the following representation of the
potential of the radiation field: $A^{\mu}_{rad}\Big(Y^{\alpha}(\tau
) + \epsilon^{\alpha}_r(\vec h)\, \sigma^r\Big) = -
\epsilon^{\mu}_r(\vec h)\, A^r_{\perp rad}(\tau ,\vec \sigma )$.

\vfill\eject

\section{N Charged Particles plus the Electro-Magnetic Field}

Let us now consider the problem of identifying the collective and
relative variable for the system of two \footnote{The case of N
particles follows the same scheme.} positive-energy charged
particles with mutual Coulomb interaction plus an arbitrary
electro-magnetic field in the radiation gauge (see Ref.\cite{1} and
I). This will allow to solve the constraints ${\vec {\cal
K}}_{(int)} \approx 0$ and to eliminate the internal 3-center of
mass like we made in the previous two Sections.

\subsection{The Internal Poincare' Generators before the Canonical
Transformation}

From Eq.(I-2.23) we have the following expression of the internal
Poincare' generators in the original canonical basis ${\vec
\eta}_i(\tau )$, ${\vec \kappa}_i(\tau )$, ${\vec A}_{\perp}(\tau
,\vec \sigma )$, ${\vec \pi}_{\perp}(\tau ,\vec \sigma )$

\begin{eqnarray*}
\mathcal{E}_{(int)} &=& \mathcal{P}^{\tau }_{(int)}\, c = M\, c^2 =
c\, \int d^3\sigma\, T^{\tau\tau}(\tau ,\vec \sigma ) =  \nonumber \\
 &&{}\nonumber \\
 &=&c\, \sum_{i=1}^2\, \Big(\sqrt{m^2_i\, c^2 + {\vec
\kappa}^2_i(\tau )} - { \ \frac{{Q_i}}{c}}\, {\frac{{{\vec
\kappa}_i(\tau ) \cdot {\vec A} _{\perp}(\tau , {\vec \eta}_i(\tau
))}}{\sqrt{m^2_i\, c^2 + {\vec \kappa}
^2_i(\tau )}}} \Big) +  \nonumber \\
 &+& \frac{Q_1\, Q_2}{4\pi\, \mid
\vec{\eta}_1(\tau ) - \vec{\eta} _2(\tau )\mid } + {\cal E}_{em},
 \nonumber \\
 &&{}\nonumber \\
 &&{\cal E}_{em} = {\cal P}^{\tau}_{em}\, c =
{\frac{1}{2}}\, \int d^{3}\sigma \, [{\vec{ \pi }}_{\perp }^{2} +
{\vec{B}}^{2}](\tau , \vec{\sigma}),
 \end{eqnarray*}

\begin{eqnarray*}
 \mathcal{P}^r_{(int)} &=& \int d^3\sigma\, T^{r\tau}(\tau ,\vec
\sigma ) = \sum_{i=1}^2\, \kappa^r_i(\tau ) +
 {\cal P}^r_{em} \approx 0,  \nonumber \\
 &&{}\nonumber \\
 &&{\vec {\cal P}}_{em} = {\frac{1}{c}}\,
\int d^{3}\sigma\, \lbrack {\vec{\pi}}_{\perp } \times
{\vec{B}}](\tau ,\vec{\sigma}),\nonumber \\
 &&{}  \nonumber \\
\mathcal{J}_{(int)}^r &=& {\bar S}^r = {\frac{1}{2}}\,
\epsilon^{ruv}\, \int d^3\sigma\, \sigma^u\, T^{v\tau}(\tau
,\vec \sigma ) =  \nonumber \\
&=&\sum_{i=1}^2\,\Big(\vec{\eta}_{i}(\tau )\times
{\vec{\kappa}}_{i}(\tau ) \Big)^{r} + {\cal J}^r_{em},
\nonumber \\
 &&{}\nonumber \\
 &&{\vec {\cal J}}_{em} =
{\frac{1}{c}}\, \int d^{3}\sigma (\vec{\sigma}\times \,\Big([{
\vec{\pi}}_{\perp }{\ \times }{\vec{B}}]\Big)^{r}(\tau
,\vec{\sigma}),
 \end{eqnarray*}

\bea
 \mathcal{K}_{(int)}^{r} &=& - \int d^3\sigma\, \sigma^r\,
T^{\tau\tau}(\tau ,\vec \sigma ) =\nonumber \\
 &&{}\nonumber \\
 &=& - \sum_{i=1}^2\, \eta^r_{i}(\tau )\, \Big( \sqrt{m^2_i\, c^2 +
{\vec \kappa}^2_i(\tau )} - {\frac{{Q_i}}{c}}\, {\frac{{{\vec
\kappa}_i(\tau ) \cdot {\vec A}_{\perp}(\tau , {\vec \eta}_i(\tau
))}}{\sqrt{m^2_i\, c^2 + {\
\vec \kappa}^2_i(\tau )}}} \Big) +  \nonumber \\
 &+& {\frac{Q_1\, Q_2}{4\pi\, c}}\, \Big[{{\eta_1^r(\tau ) +
 \eta_2^r(\tau )}\over {|{\vec \eta}_1(\tau ) - {\vec \eta}_2(\tau )|}}
 -\nonumber \\
 &-& \int {{d^3\sigma}\over {4\pi}}\, \Big({{\sigma^r - \eta_2^r(\tau )}
 \over {|\vec \sigma - {\vec \eta}_1(\tau )|\, |\vec \sigma -
 {\vec \eta}_2(\tau )|^3}} + {{\sigma^r - \eta_1^r(\tau )}
 \over {|\vec \sigma - {\vec \eta}_2(\tau )|\, |\vec \sigma -
 {\vec \eta}_1(\tau )|^3}}\Big)\Big] -  \nonumber \\
 &-& \sum_{i=1}^2\, {{Q_i}\over {4\pi\, c}}\, \int d^{3}\sigma\,
 {{{\pi}_{\perp }^{r}(\tau ,\vec{\sigma})}\over {|\vec{\sigma} -
 {\vec{\eta}}_{i}(\tau )|}} + {\cal K}^r_{em} \approx 0,\nonumber \\
 &&{}\nonumber \\
 &&{\vec {\cal K}}_{em} = - {\frac{1}{2c}}\, \int d^{3}\sigma\, \sigma ^{r}\,
({{\vec{\pi}}}_{\perp }^{2} + {{\vec{B} }} ^{2})(\tau
,\vec{\sigma}).
  \label{4.1}
\end{eqnarray}

\noindent For an electro-magnetic field not of the radiation type
the quantities ${\cal E}_{em}$, ${\vec {\cal P}}_{em}$, ${\vec {\cal
J}}_{em}$, ${\vec {\cal K}}_{em}$ are not conserved and do not
satisfy a Lorentz algebra.

\bigskip

In the canonical basis ${\vec A}_{\perp}(\tau ,\vec \sigma )$,
${\vec \pi}_{\perp}(\tau ,\vec \sigma )$, ${\vec \eta}_i(\tau )$,
${\vec \kappa}_i(\tau )$, it is not clear which are the collective
variables to be eliminated by means of the constraints ${\vec
{\mathcal{P}}}_{(int)} \approx 0$ and ${\vec {\mathcal{K}}}_{(int)}
\approx 0$, because we do not know how to define these variables for
an electro-magnetic field not of the radiation type. To treat this
type of field we now use results obtained in paper I.

\subsection{The Internal Poincare' Generators after the Canonical
Transformation}

After the canonical transformation defined in I, the system is
described by a transverse radiation field ${\vec A}_{\perp rad}(\tau
,\vec \sigma )$, ${\vec \pi}_{\perp rad}(\tau ,\vec \sigma )$, and
by Coulomb-dressed particle variables ${\hat {\vec \eta}}_i(\tau )$,
${\hat {\vec \kappa}}_i(\tau )$, see Eqs. (I-3.6) and (I-3.10).

From Eqs.(I-4.4), (I-4.1), (I-4.2), (I-4.6) [and (I-3.9) for ${\hat
T}_i$ and ${\hat {\cal K}}_{ij}$] we have that each internal
Poincare' generator becomes the direct sum of the radiation field
one of Eq.(\ref{3.2}) plus the particle one of Eq.(\ref{2.4}) [see
Eq.(6.46) of Ref.\cite{1}]

\begin{eqnarray*}
\mathcal{E}_{(int)}
 &=& {\cal P}^{\tau}_{(int)}\, c =
 M\,c^{2} = c\,\sum_{i=1}^2\,\sqrt{m_{i}^{2}\,c^{2}+{\hat{\vec{
\kappa}}}_{i}^{2}(\tau )} + {\frac{{Q_1\,Q_2}}{{4\pi \,|{\
\hat{\vec{\eta}}}_1(\tau )-{\hat{\vec{\eta}}}_2(\tau )|}}}+
\nonumber \\
 &+& V_{DARWIN}({\hat {\vec \kappa}}_1(\tau ), {\hat {\vec \kappa}}_2(\tau ),
 {\hat {\vec \eta}}_1(\tau ) - {\hat {\vec \eta}}_2(\tau )) +  \nonumber \\
 &&{}\nonumber \\
 &+& {\frac{1}{2}}\,\int d^{3}\sigma \,\Big({\vec{ \pi}}_{\perp
rad}^{2}+{\vec{B}}_{rad}^{2}\Big)(\tau ,\vec{\sigma}) =  \nonumber \\
 &&{}\nonumber \\
 &=& \Big(\mathcal{P}^{\tau}_{matter} + \mathcal{P}^{\tau}_{rad}\Big)\, c,
\end{eqnarray*}

\medskip

\begin{eqnarray*}
 {\vec {\mathcal{P}}}_{(int)} &=&
  \sum_{i=1}^2\, {\hat {\vec \kappa}}_i(\tau ) + {\frac{1}{c}}\, \int
d^3\sigma\, \Big({\vec \pi} _{\perp rad} \times {\vec B}_{rad}
\Big)(\tau ,\vec \sigma ) =  \nonumber \\
 &&{}\nonumber \\
 &=& {\vec {\mathcal{P}}}_{matter} + {\vec
{\mathcal{P}}}_{rad}\approx 0,
\end{eqnarray*}
\medskip

\begin{eqnarray*}
 {\vec {\mathcal{J}}}_{(int)} &=&
 \sum_i\, {\hat {\vec \eta}}_i \times {\hat {\vec \kappa}}_i +
{\frac{1}{c }}\, \int d^3\sigma\, \vec \sigma \times \Big({\vec
\pi}_{\perp rad} \times {\vec B} _{rad}\Big)(\tau ,\vec \sigma ) =
\nonumber \\
 &&{}\nonumber \\
 &=& {\vec {\mathcal{J}}}_{matter} + {\vec {\mathcal{J}}}_{rad},
\end{eqnarray*}

\medskip

\bea
 {\vec{\mathcal{K}}}_{(int)} &=& - \sum_{i=1}^2\, {\hat {\vec \eta}}_i\, \sqrt{m_i^2\, c^2 +
{\hat {\vec \kappa}}_i^2} -  \nonumber \\
&-& {\frac{1}{2}}\, {\frac{{Q_1\, Q_2}}{c}}\, \Big[{\hat {\vec
\eta}}_1\, { \frac{{{\hat {\vec \kappa}}_1 \cdot
\Big({\frac{1}{2}}\, {\frac{{\partial\, { \hat
{\mathcal{K}}}_{12}({\hat {\vec \kappa}}_1, {\hat {\vec \kappa}}_2,
{ \hat {\vec \rho}}_{12})}}{{\partial\, {\hat {\vec \rho}}_{12}}}} -
2\, {\vec A}_{\perp S2}({\hat {\vec \kappa}}_2, {\hat {\vec
\rho}}_{12})\Big)}}{\sqrt{
m_1^2\, c^2 + {\hat {\vec \kappa}}_1^2}}} +  \nonumber \\
&+& {\hat {\vec \eta}}_2\, {\frac{{{\hat {\vec \kappa}}_2 \cdot
\Big({\frac{1 }{2}}\, {\frac{{\partial\, {\hat
{\mathcal{K}}}_{12}({\hat {\vec \kappa}}_1, {\hat {\vec \kappa}}_2,
{\hat {\vec \rho}}_{12})}}{{\partial\, {\hat {\vec \rho}}_{12}}}} -
2\, {\vec A}_{\perp S1}({\hat {\vec \kappa}}_1, {\hat {\vec
\rho}}_{12})\Big)}}{\sqrt{m_2^2\, c^2 + {\hat {\vec \kappa}}_2^2}}}
\Big] -\nonumber \\
&-& {\frac{1}{2}}\, {\frac{{Q_1\, Q_2}}{c}}\, \Big(\sqrt{m_1^2\, c^2
+ { \hat {\vec \kappa}}_1^2}\, {\frac{{\partial}}{{\partial\, {\hat
{\vec \kappa} }_1}}} + \sqrt{m_2^2\, c^2 + {\hat {\vec
\kappa}}_2^2}\, {\frac{{\partial}}{{
\partial\, {\hat {\vec \kappa}}_2}}} \Big)\, {\hat {\mathcal{K}}}_{12}({
\hat {\vec \kappa}}_1, {\hat {\vec \kappa}}_2, {\hat {\vec
\rho}}_{12}) -\nonumber \\
&-& {\frac{{Q_1\, Q_2}}{{4\pi\, c}}}\, \int d^3\sigma\,
\Big({\frac{{{\hat { \vec \pi}}_{\perp S1}(\vec \sigma - {\hat {\vec
\eta}}_1, {\hat {\vec \kappa} }_1)}}{{|\vec \sigma - {\hat {\vec
\eta}}_2|}}} + {\frac{{{\hat {\vec \pi}} _{\perp S2}(\vec \sigma -
{\hat {\vec \eta}}_2, {\hat {\vec \kappa}}_2)}}{{
|\vec \sigma - {\hat {\vec \eta}}_1|}}} \Big) -  \nonumber \\
&-& {\frac{{Q_1\, Q_2}}{c}}\, \int d^3\sigma\, \vec \sigma\,\,
\Big({\hat { \vec \pi}}_{\perp S1}(\vec \sigma - {\hat {\vec
\eta}}_1, {\hat {\vec \kappa} }_1) \cdot {\hat {\vec \pi}}_{\perp
S2}(\vec \sigma - {\hat {\vec \eta}}_2, {
\hat {\vec \kappa}}_2) +  \nonumber \\
&+& {\hat {\vec B}}_{S1}(\vec \sigma - {\hat {\vec \eta}}_1, {\hat
{\vec \kappa}}_1) \cdot {\hat {\vec B}}_{ S2}(\vec \sigma - {\hat
{\vec \eta}}_2, {\hat {\vec \kappa}}_2) \Big) -  \nonumber \\
 &&{}\nonumber \\
 &-& {\frac{1}{{2\, c}}}\, \int d^3\sigma\, \vec \sigma\,\,
\Big({\vec \pi}^2_{\perp rad} + {\vec B}^2_{rad}\Big)(\tau ,\vec \sigma ) =  \nonumber \\
&&{}  \nonumber \\
&=& {\vec {\mathcal{K}}}_{matter} + {\vec {\mathcal{K}}}_{rad}
 =   \nonumber \\
&&{}  \nonumber \\
&&{}  \nonumber \\
&{\buildrel {def}\over {=}}&-\frac{1}{c}\mathcal{E}_{(int)}\,
\vec{R}_{+} = - {\frac{1}{c}}\, \Big(\mathcal{E}_{matter} +
\mathcal{E}_{rad}\Big)\, {\vec R}_+ \approx 0.
 \label{4.2}
\end{eqnarray}

\noindent Let us remark that in ${\vec {\mathcal{K}}}_{(int)}
\approx 0$ of Eq.(\ref{4.2}) all the particles terms depend on
${\hat {\vec \rho}}_{12}(\tau ) = {\hat { \vec \eta}}_1(\tau ) -
{\hat {\vec \eta}}_2(\tau )$ or $\vec \sigma - {\hat {\vec \eta}}
_i(\tau )$.\medskip

We also added the connection of ${\vec {\cal K}}_{(int)}$ with the
M$\o$ller internal 3-center of energy ${\vec R}_+ $ \footnote{We
have ${\vec R}_+ \approx {\vec q}_+ \approx {\vec y}_+$ due to the
rest-frame condition ${\vec {\cal P}}_{(int)} \approx 0$, where
${\vec q}_+$ is the internal 3-center of mass and ${\vec y}_+$ the
internal Fokker-Pryce 3-center of inertia.}.

\bigskip

Even if all the internal generators are the direct sum of the
generators of the two subsystems, {\it the two subsystems are
connected by the rest-frame conditions and by the vanishing of the
internal boosts}, i.e. by the necessity of eliminating the position
and momentum of the internal overall 3-center of mass.

\subsection{Semi-Relativistic Expansions of $\mathcal{E}_{(int)}$
and $ \mathcal{\vec{K}}_{(int)}$ before and after the Canonical
Transformation}

The semi-relativistic limit of the generators ${\cal E}_{(int)}$ and
${\vec {\cal K}}_{(int)}$ of Eqs.(\ref{4.1})  is

\bea
  {\cal E}_{(int)}
&\rightarrow_{c \rightarrow \infty}& (\sum_{i=1}^2\, m_i)\, c^2 +
\sum_{i=1}^2\, {\frac{{ {\vec \kappa}_i^2(\tau )}}{{2m_i}}} +
\frac{Q_1\, Q_2}{ 4\pi\, \mid \vec{\eta}_1(\tau ) - \vec{\eta}
_2(\tau )\mid } - \nonumber \\
&-& {\frac{1}{c}}\, \sum_{i=1}^2\, Q_i\, {\frac{{{\vec
\kappa}_i(\tau )}}{{m_i}}} \cdot {\vec A}_{\perp}(\tau ,{\vec
\eta}_i(\tau )) - {\frac{1}{{c^2}}}\,
\sum_{i=1}^2\, {\frac{{{\vec \kappa}_i^4(\tau )}}{{8 m_i^3}}} +  \nonumber \\
&+& {\frac{1}{{c^3}}}\, \sum_{i=1}^2\, Q_i\, {\frac{{{\vec
\kappa}_i^2(\tau )}}{{2 m_i^2}}}\, {\frac{{{\vec \kappa}_i(\tau
)}}{{m_i}}} \cdot {\vec A}
_{\perp}(\tau ,{\vec \eta}_i(\tau )) +  \nonumber \\
&+& O(c^{-4}) + {\frac{1}{2}}\, \int d^{3}\sigma \, [{\vec{\pi
}}_{\perp}^{2} + {\vec{B}}^{2}](\tau ,\vec{\sigma}),
  \nonumber \\
&&{}\nonumber \\
 &&{}\nonumber \\
 {\cal K}^r_{(int)} &\rightarrow_{c \rightarrow \infty}& c\, {\cal
 K}^r_{Galilei} +   O(c^{-1}) \approx 0,\nonumber \\
 &&{}\nonumber \\
 {\vec {\cal K}}_{Galilei} &=& - \sum_{i=1}^N\, m_1\, {\vec
 \eta}_i(\tau ) = - \Big(\sum_{i=1}^N\, m_i\Big)\, {\vec x}_{(n)} \approx 0,
 \label{4.3}
 \eea

\noindent where ${\vec x}_{(n)}$ is the non-relativistic center of
mass, which emerges as the limit of the M$\o$ller internal 3-center
of energy ${\vec R}_+ = - c\, {\vec {\cal K}}_{(int)}/{\cal
E}_{(int)}$.  See Section G of paper I for the non-relativistic
limit of the relativistic variables ${\vec \eta}_i(\tau )$, ${\vec
\kappa}_i(\tau )$ and of the relativistic external center of mass.
\bigskip

After the canonical transformation ${\cal E}_{(int)}$ is given by
Eq.(\ref{4.2}). The semi-relativistic limit of its matter part is

\bea
 {\cal P}^{\tau}_{matter} &=& \sum_{i=1}^{2}\, \Big(m_{i}\, c^{2} + \frac{{
\hat{\vec{\kappa}}}_{i}^{2}(\tau )}{2m_{i}} +
\frac{Q_{1}\,Q_{2}}{4\pi \,|{\hat{\vec{ \eta}}}_{1}(\tau ) -
{\hat{\vec{\eta}}}_{2}(\tau )|} - {\frac{1}{{c^{2}}}}\,\,{
\frac{{\hat{\vec{\kappa}}}_{i}^{4}{(\tau )}}{{8m_{i}^{3}}}}
+\nonumber \\
 &+& {{Q_1\, Q_2}\over {m_1\, m_2\, c^2}}\, {{\vec{\kappa}_{i}^{2}(\tau )
  - \Big[{\vec \kappa}_i(\tau ) \cdot {{{\hat{\vec{ \eta}}}_{1}(\tau ) -
  {\hat{\vec{\eta}}}_{2}(\tau )}\over {|{\hat{\vec{\eta}} }_{1}(\tau ) -
{\hat{\vec{\eta}}}_{2}(\tau )|}}\Big]^2 }\over {16\pi
|{\hat{\vec{\eta}}} _1(\tau ) - {\hat{\vec{\eta}}}_2(\tau )|}} \Big)
+ O(c^{-4}).
 \label{4.4}
 \eea

\medskip

By using Eqs.(I-2.51) - (I-2.53)  for the Lienard-Wiechert fields we
get ${\hat {\mathcal{K}}}_{12} = O(c^{-3})$, so that the
semi-relativistic limit of ${\vec {\cal K}}_{(int)}$ of
Eq.(\ref{4.2}) is

\begin{eqnarray}
\mathcal{\vec{K}}_{(int)} &=&- \sum_{i=1}^2\,
{\hat{\vec{\eta}}}_{i}(\tau )\, \Big(m_{i}\, c +
\frac{{\hat{\vec{\kappa}}}_{i}^{2}}{2m_{i}c}\Big) +
O(c^{-2}) -  \nonumber \\
&-& {\frac{1}{2c}}\int d^{3}\sigma \,\vec{\sigma}\,\,\Big({\vec{
\pi}}_{\perp rad}^{2}+{\vec{B}}_{rad}^{2}\Big)(\tau ,\vec{\sigma}) =  \nonumber \\
&&{}  \nonumber \\
&=&{\vec{\mathcal{K}}}_{matter}+{\vec{\mathcal{K}}}_{rad}=c\,{\vec{K}}
_{Galilei}+O({\frac{1}{c}}) \approx 0,  \nonumber \\
&&{}  \nonumber \\
&& {\vec {\mathcal{K}}}_{Galilei} = - (m_1 + m_2)\, {\vec x}_{(n)} =
- \sum_{i=1}^2\, m_i\, {\hat {\vec \eta}}_i \approx 0\qquad (in\,
the\, rest\, frame).
  \label{4.5}
\end{eqnarray}

\subsection{The Collective Variables after the Canonical Transformation}

The results of Section II and III allow us to find the collective
and relative variables of the matter and transverse radiation field
subsystems in the canonical basis ${\hat {\vec \eta}}_i(\tau )$,
${\hat {\vec \kappa}}_i(\tau )$, ${\vec A}_{\perp\, rad}(\tau ,\vec
\sigma )$, ${\vec \pi}_{\perp\, rad}(\tau ,\vec \sigma )$.\medskip

A) The field quantities ${\vec A}_{\perp\, rad}(\tau ,\vec \sigma
)$, ${\vec \pi}_{\perp\, rad}(\tau ,\vec \sigma )$ are canonically
equivalent to the canonical basis $X^{\tau}_{rad}$, ${\cal
P}^{\tau}_{rad} = M_{rad}\, c = { \frac{{{\cal E}_{rad}}}{c}}$,
${\vec X}_{rad}$, ${\vec {\cal P}}_{rad}$, $ \lambda_{rad}(\vec k)$,
$\rho_{rad}(\vec k)$, $H_{rad}(\vec k)$, $ K_{rad}(\vec k)$, given
in Eqs.(\ref{3.7}), which contains the collective variables for the
radiation field.\medskip

B) For the particle subsystem we make the canonical transformation
(\ref{2.3}) on the Coulomb-dressed particle variables

\bea
 {\hat {\vec \eta}}_{12} &=& {\frac{{m_1}}{m}}\, {\hat {\vec \eta}}_1 +
{\frac{{m_2}}{m}}\, {\hat {\vec \eta}}_2,  \qquad
 {\hat {\vec \rho}}_{12} = {\hat {\vec \eta}}_1 - {\hat {\vec \eta}}_2,  \nonumber \\
 {\hat {\vec \kappa}}_{12} &=& {\hat {\vec \kappa}}_1 +
 {\hat {\vec \kappa}}_2 \approx 0,
\qquad {\hat {\vec \pi}}_{12} =  {\frac{{m_2}}{m}}\, {\hat {\vec
\kappa}}_1 - {\ \frac{{m_1}}{m}}\, {\hat {\vec \kappa}}_2,\nonumber \\
 &&{}\nonumber \\
 &&{}\nonumber \\
  {\hat {\vec \eta}}_1 &=& {\hat {\vec \eta}}_{12} + {\frac{{m_2}}{m}}\,
  {\hat {\vec \rho}}_{12}, \qquad {\hat {\vec \eta}}_2 = {\hat
  {\vec \eta}}_{12} - {\frac{{m_1}}{m}}\,
 {\hat {\vec \rho}}_{12},  \nonumber \\
 {\hat {\vec \kappa}}_1 &=& {\frac{{m_1}}{m}}\, {\hat {\vec \kappa}}_{12} +
 {\hat {\vec \pi}}_{12},  \qquad {\hat {\vec \kappa}}_2 =
 {\frac{{ m_2}}{m}}\, {\hat {\vec \kappa}}_{12} - {\hat {\vec \pi}}_{12}.
  \label{4.6}
 \eea
\medskip

Eqs.(\ref{4.2}) imply ${\vec {\cal P}}_{matter} = {\hat {\vec
\kappa}}_{12} \approx - {\vec {\cal P}} _{rad} = - \int d^3\sigma\,
({\vec \pi}_{\perp rad} \times {\vec B} _{rad})(\tau ,\vec \sigma
)$.

\bigskip

C) Let us now put ${\hat {\vec \eta}}_3 = {\vec X}_{rad}$, ${\hat
{\vec \kappa}} _3 = {\vec {\cal P}}_{rad}$, with ${\vec X}_{rad}$
and ${\vec {\cal P}}_{rad}$ given by Eqs. (\ref{3.9}) and
(\ref{3.8}) respectively, and let us combine it with ${\hat {\vec
\eta}}_{12}$ and ${\ \hat {\vec \kappa}}_{12}$ of Eqs.(\ref{4.6}) to
get the canonical transformation

\begin{eqnarray}
 &&\begin{minipage}[t]{2cm}
\begin{tabular}{|l|l|l|} \hline ${\hat {\vec \eta}}_{12}$ & ${\hat {\vec \rho}}_{12}$
& ${\hat {\vec \eta}}_3$ \\
\hline ${\hat {\vec \kappa}}_{12}$ & ${\hat {\vec \pi}}_{12}$  &
${\hat {\vec \kappa}}_3$\\ \hline
\end{tabular}\end{minipage}
\ {\ \longrightarrow \hspace{.2cm}} \ \begin{minipage}[t]{2 cm}
\begin{tabular}{|l|l|l|} \hline ${\hat {\vec \eta}}$ & ${\hat {\vec
\rho}}_{12}$
& ${\hat {\vec \rho}}_{(12)3}$ \\
\hline ${\hat {\vec \kappa}}\, (\approx 0)$&${\hat {\vec \pi}}_{12}$
& ${\hat {\vec \pi}}_{(12)3}$ \\ \hline
\end{tabular} \end{minipage}\nonumber \\
 &&{}\nonumber \\
 &&{}\nonumber \\
 {\hat {\vec \eta}} &=& {\frac{1}{2}}\, ({\hat {\vec \eta}}_{12} +
{\hat { \vec \eta}}_3),\qquad {\hat {\vec \kappa}} = {\hat {\vec
\kappa}}_{12} + {\ \hat {\vec \kappa}}_3 = {\vec {\cal P}}_{(int)}
\approx 0,  \nonumber \\
{\hat {\vec \rho}}_{(12)3} &=& {\hat {\vec \eta}}_{12} - {\hat {\vec
\eta}} _3,\qquad {\hat {\vec \pi}}_{(12)3} = {\frac{1}{2}}\, ({\hat
{\vec \kappa}}_{12} - {\hat {\vec \kappa}}_3),  \nonumber \\
&&{}  \nonumber \\
{\hat {\vec \eta}}_{12} &=& {\hat {\vec \eta}} + {\frac{1}{2}}\,
{\hat {\vec \rho}}_{(12)3},\qquad {\hat {\vec \kappa}}_{12} =
{\frac{1}{2}}\, {\hat { \vec \kappa}} + {\hat {\vec \pi}}_{(12)3}
\approx {\hat {\vec \pi}}_{(12)3},\nonumber \\
{\hat {\vec \eta}}_3 &=& {\hat {\vec \eta}} - {\frac{1}{2}}\, {\hat
{\vec \rho}}_{(12)3},\qquad {\hat {\vec \kappa}}_3 = {\frac{1}{2}}\,
{\hat {\vec \kappa}} - {\hat {\vec \pi}}_{(12)3} \approx {\hat {\vec
\pi}}_{(12)3},
 \label{4.7}
\end{eqnarray}

\noindent so that we get

\begin{eqnarray}
{\hat {\vec \eta}}_1 &=& {\hat {\vec \eta}} + {\frac{1}{2}}\, {\hat
{\vec \rho}}_{(12)3} + {\frac{{m_2}}{m}}\, {\hat {\vec \rho}}_{12},  \nonumber \\
{\hat {\vec \eta}}_2 &=& {\hat {\vec \eta}} + {\frac{1}{2}}\, {\hat
{\vec \rho}}_{(12)3} - {\frac{{m_1}}{m}}\, {\hat {\vec \rho}}_{12},  \nonumber \\
{\vec X}_{rad} &=& {\hat {\vec \eta}}_3 = {\hat {\vec \eta}} -
{\frac{1}{2}}\, {\hat {\vec \rho}}_{(12)3},  \nonumber \\
&&{}  \nonumber \\
{\hat {\vec \kappa}}_1 &=& {\frac{{m_1}}{m}}\, ({\frac{1}{2}}\,
{\hat {\vec \kappa}} + {\hat {\vec \pi}}_{(12)3}) + {\hat {\vec
\pi}}_{12} \approx {\ \frac{{m_1}}{m}}\, {\hat {\vec \pi}}_{(12)3} +
{\hat {\vec \pi}}_{12},\nonumber \\
{\hat {\vec \kappa}}_2 &=& {\frac{{m_2}}{m}}\, ({\frac{1}{2}}\,
{\hat {\vec \kappa}} + {\hat {\vec \pi}}_{(12)3}) - {\hat {\vec
\pi}}_{12} \approx {\ \frac{{m_2}}{m}}\, {\hat {\vec \pi}}_{(12)3} -
{\hat {\vec \pi}}_{12},\nonumber \\
{\vec {\cal P}}_{rad} &=& {\hat {\vec \kappa}}_3 = {\frac{1}{2}}\,
{\hat {\vec \kappa }} - {\hat {\vec \pi}}_{(12)3} \approx - {\hat
{\vec \pi}}_{(12)3}.
 \label{4.8}
\end{eqnarray}

Therefore our final canonical basis contains the collective
variables ${\hat {\vec \eta}}$, ${\hat {\vec \kappa}} \approx 0$ and
the relative variables ${\hat {\vec \rho}}_{12}$, ${\hat {\vec
\pi}}_{12}$ (depending only upon the particles), ${\bf K}_{rad}(\vec
k)$, ${\bf H}_{rad}(\vec k)$, $\lambda_{rad}(\vec k)$,
$\rho_{rad}(\vec k)$ (depending only upon the "multipoles"the
transverse radiation field) and ${\hat {\vec \rho}}_{(12)3}$, ${\hat
{\vec \pi}}_{(12)3}$ (depending on the relative motion of the
particle collective variables with respect to the radiation field
collective variables, i.e. its "monopole" aspect).\bigskip

The overall collective variable ${\hat {\vec \eta}}$ is the natural
variable for the solution of the gauge fixings ${\vec
{\mathcal{K}}}_{(int)} \approx 0$.

\bigskip

\subsection{The  Internal 3-Center of Mass ${\hat {\vec \eta}}$ from the
vanishing of the Internal Boosts after the Canonical
Transformation.}

We have to find ${\hat {\vec \eta}}$ of Eq.(\ref{4.7}) from the
vanishing of the internal boosts ${\vec {\cal K}}_{(int)} \approx 0$
in the form (\ref{4.2}), put it into Eqs.(\ref{4.8} ) and make the
inverse canonical transformation (I-3.10)  to find the original
${\vec \eta}_i(\tau )$ and the particle world-lines $ x^{\mu}_i(\tau
)$ like we made in Eqs.(\ref{2.8}).\bigskip

In the boosts in the form (\ref{4.2}) the first four lines of the
particle terms  depend on ${\hat {\vec \rho}}_{12} = {\hat {\vec
\eta}}_1 - {\hat { \vec \eta}}_2$ and ${\hat {\vec \pi}}_{12} =
{\frac{{m_2}}{m}}\, {\hat {\vec \kappa}}_1 - {\frac{{m_1}}{m}}\,
{\hat {\vec \kappa}}_2$, but they  also get a dependence on ${\hat
{\vec \pi}}_{(12)3} = {\frac{1}{2}}\, ({\hat { \vec \kappa}}_1 +
{\hat {\vec \kappa}}_2 - {\vec {\cal P}}_{rad})$ through the
dependence on the particle momenta. In the next two particle terms
we must shift the integration variable to reabsorb the quantity
${\hat {\vec \eta}} + {\frac{1}{2}}\, {\hat {\vec \rho} }_{(12)3}$
present in $\vec \sigma - {\hat {\vec \eta}}_i$.\medskip

Finally the last line of the boosts in  Eqs.(\ref{4.2}) is just
${\vec {\cal K}}_{rad}$ of Eqs.(\ref{3.8}).

\bigskip

Therefore the internal boosts in the form (\ref{4.2}) has the
following form in the new canonical basis (${\hat {\vec
\kappa}}_i(\tau )$ are given by Eqs.(\ref{4.8}))

\begin{eqnarray*}
{\vec {\mathcal{K}}}_{(int)} &=& - \Big({\hat {\vec \eta}} +
{\frac{1}{2}}\, {\hat {\vec \rho}}_{(12)3}\Big)\, \Big(
\sum_{i=1}^2\, \sqrt{m_i^2\, c^2 + { \hat {\vec \kappa}}_i^2} +
{\frac{1}{2}}\, {\frac{{Q_1\, Q_2}}{c}}\times\nonumber \\
&&\times \Big[ {\frac{{{\hat {\vec \kappa}}_1 \cdot
\Big({\frac{1}{2}}\, { \frac{{\partial\, {\hat
{\mathcal{K}}}_{12}({\hat {\vec \kappa}}_1, {\hat { \vec \kappa}}_2,
{\hat {\vec \rho}}_{12})}}{{\partial\, {\hat {\vec \rho}} _{12}}}} -
2\, {\vec A}_{\perp S2}({\hat {\vec \kappa}}_2, {\hat {\vec \rho}}
_{12})\Big)}}{\sqrt{m_1^2\, c^2 + {\hat {\vec \kappa}}_1^2}}} +  \nonumber \\
&+& {\frac{{{\hat {\vec \kappa}}_2 \cdot \Big({\frac{1}{2}}\,
{\frac{{\partial\, {\hat {\mathcal{K}}}_{12}({\hat {\vec \kappa}}_1,
{\hat {\vec \kappa}}_2, {\hat {\vec \rho}}_{12})}}{{\partial\, {\hat
{\vec \rho}}_{12}}}} - 2\, {\vec A}_{\perp S1}({\hat {\vec
\kappa}}_1, {\hat {\vec \rho}}_{12}) \Big)}}{\sqrt{m_2^2\, c^2 +
{\hat {\vec \kappa}}_2^2}}} \Big]\,\Big) -\nonumber \\
&-& {\hat {\vec \rho}}_{12}\, \Big( {\frac{{m_2}}{m}}\,
\sqrt{m_1^2\, c^2 + { \hat {\vec \kappa}}_1^2} - {\frac{{m_1}}{m}}\,
\sqrt{m_2^2\, c^2 + {\hat {\vec \kappa}}_2^2} +  \nonumber \\
&+& {\frac{1}{2}}\, {\frac{{Q_1\, Q_2}}{c}}\,
\Big[{\frac{{m_2}}{m}}\, { \frac{{{\hat {\vec \kappa}}_1 \cdot
\Big({\frac{1}{2}}\, {\frac{{\partial\, { \hat
{\mathcal{K}}}_{12}({\hat {\vec \kappa}}_1, {\hat {\vec \kappa}}_2,
{ \hat {\vec \rho}}_{12})}}{{\partial\, {\hat {\vec \rho}}_{12}}}} -
2\, {\vec A}_{\perp S2}({\hat {\vec \kappa}}_2, {\hat {\vec
\rho}}_{12})\Big)}}{\sqrt{ m_1^2\, c^2 + {\hat {\vec \kappa}}_1^2}}}
-
 \end{eqnarray*}

\bea
 &-& {\frac{{m_1}}{m}}\, {\frac{{{\hat {\vec \kappa}}_2 \cdot \Big({\frac{1}{2
}}\, {\frac{{\partial\, {\hat {\mathcal{K}}}_{12}({\hat {\vec
\kappa}}_1, { \hat {\vec \kappa}}_2, {\hat {\vec
\rho}}_{12})}}{{\partial\, {\hat {\vec \rho}}_{12}}}} - 2\, {\vec
A}_{\perp S1}({\hat {\vec \kappa}}_1, {\hat {\vec
\rho}}_{12})\Big)}}{\sqrt{m_2^2\, c^2 + {\hat {\vec \kappa}}_2^2}}}
\Big]\Big)-  \nonumber \\
&-& {\frac{1}{2}}\, {\frac{{Q_1\, Q_2}}{c}}\, \Big(\sqrt{m_1^2\, c^2
+ { \hat {\vec \kappa}}_1^2}\, {\frac{{\partial}}{{\partial\, {\hat
{\vec \kappa} }_1}}} + \sqrt{m_2^2\, c^2 + {\hat {\vec
\kappa}}_2^2}\, {\frac{{\partial}}{{
\partial\, {\hat {\vec \kappa}}_2}}} \Big)\, {\hat {\mathcal{K}}}_{12}({
\hat {\vec \kappa}}_1, {\hat {\vec \kappa}}_2, {\hat {\vec
\rho}}_{12}) -\nonumber \\
&-& {\frac{{Q_1\, Q_2}}{{4\pi\, c}}}\, \int d^3\sigma\,
\Big({\frac{{{\hat { \vec \pi}}_{\perp S1}(\vec \sigma -
{\frac{{m_2}}{m}}{\hat {\vec \rho}} _{12}, {\hat {\vec
\kappa}}_1)}}{{|\vec \sigma + {\frac{{m_1}}{m}}\, {\hat { \vec
\rho}}_{12}|}}} + {\frac{{{\hat {\vec \pi}}_{\perp S2}(\vec \sigma +
{ \frac{{m_1}}{m}}\, {\hat {\vec \rho}}_{12}, {\hat {\vec
\kappa}}_2)}}{{|\vec \sigma - {\frac{{m_2}}{m}}\, {\hat {\vec
\rho}}_{12}|}}} \Big) -  \nonumber \\
&-& {\frac{{Q_1\, Q_2}}{c}}\, \int d^3\sigma\, \Big[\vec \sigma +
{\hat { \vec \eta}} + {\frac{1}{2}}\, {\hat {\vec
\rho}}_{(12)3}\Big]\,\, \Big({ \hat {\vec \pi}}_{\perp S1}(\vec
\sigma - {\frac{{m_2}}{m}}\, {\hat {\vec \rho}}_{12}, {\hat {\vec
\kappa}}_1) \cdot {\hat {\vec \pi}}_{\perp S2}(\vec \sigma +
{\frac{{m_1}}{m}}\, {\hat {\vec \rho}}_{12}, {\hat {\vec \kappa}}
_2) +  \nonumber \\
&+& {\hat {\vec B}}_{S1}(\vec \sigma - {\frac{{m_2}}{m}}\, {\hat
{\vec \rho}} _{12}, {\hat {\vec \kappa}}_1) \cdot {\hat {\vec B}}_{
S2}(\vec \sigma + { \frac{{m_1}}{m}}\, {\hat {\vec \rho}}_{12},
{\hat {\vec \kappa}}_2)\Big) -\nonumber \\
&-& X^{\tau}_{rad}\, {\hat {\vec \pi}}_{(12)3} - M_{rad}\, c\,
\Big({\hat { \vec \eta}} - {\frac{1}{2}}\, {\hat {\vec
\rho}}_{(12)3}\Big) + {\vec { \mathcal{D}}}({\bf H}_{rad}, {\bf
K}_{rad}, \lambda_{rad} ,\rho_{rad} ) \approx 0,\nonumber \\
&&{}
  \label{4.9}
\end{eqnarray}

\noindent where the term ${\vec { \mathcal{D}}}({\bf H}_{rad}, {\bf
K}_{rad}, \lambda_{rad} ,\rho_{rad} )$ has the following expression

\bea
  &&\mathcal{D}^r({\bf H}_{rad}, {\bf K}_{rad}, \lambda_{rad} ,\rho_{rad} ) = - {\frac{1}{c
}}\, \int d\tilde k\, H_{rad}(\vec k)\, D^{or}\, K_{rad}(\vec k) -
{\frac{1}{ c}}\, \int d\tilde k\, \lambda_{rad}(\vec k)\, D^{or}\,
\rho_{rad}(\vec k) -\nonumber \\
&-& {\frac{i}{{2c}}}\, \int d\tilde k\, \lambda_{rad}(\vec k)\,
\Big[{\vec \epsilon}_+(\vec k) + {\vec \epsilon}_-(\vec k)\Big]\,
\cdot D^{or}\, \Big[{ \vec \epsilon}_-(\vec k) - {\vec
\epsilon}_+(\vec k)\Big].
 \label{4.10}
 \eea

\medskip

The solution of Eq.(\ref{4.9}) is

\bea
 {\hat {\vec \eta}} &\approx&{\frac{1}{{\mathcal{A} + M_{rad}\,
c}}}\, \Big(- {\frac{1}{2}}\, (\mathcal{A} - M_{rad}\, c)\, {\hat
{\vec \rho}}_{(12)3} -\nonumber \\
&-& \Big[{\frac{{m_2}}{m}}\, \sqrt{m_1^2\, c^2 + ({\hat {\vec
\pi}}_{12} + { \frac{{m_1}}{m}}\, {\hat {\vec \pi}}_{(12)3})^2} -
{\frac{{m_1}}{m}}\, \sqrt{ m_2^2\, c^2 + ({\hat {\vec \pi}}_{12} -
{\frac{{m_2}}{m}}\, {\hat {\vec \pi}} _{(12)3})^2} +
\mathcal{B}\Big]\, {\hat {\vec \rho}}_{12} - {\vec {\mathcal{C
}}} +  \nonumber \\
&+& X^{\tau}_{rad}\, {\hat {\vec \pi}}_{(12)3} + {\vec
{\mathcal{D}}}\Big),
 \label{4.11}
  \eea

\noindent where we have introduced the following definitions (${\hat
{\vec \kappa}}_i(\tau )$ are given by Eqs.(\ref{4.8}))

\begin{eqnarray*}
&&\mathcal{A}({\hat {\vec \rho}}_{12}, {\hat {\vec \pi}}_{12}, {\hat
{\vec \pi}}_{(12)3}) = \sum_{i=1}^2\, \sqrt{m_i^2\, c^2 + {\hat
{\vec \kappa}}_i^2}
+ {\frac{1}{2}}\, {\frac{{Q_1\, Q_2}}{c}}\times  \nonumber \\
&&\times \Big[ {\frac{{{\hat {\vec \kappa}}_1 \cdot
\Big({\frac{1}{2}}\, { \frac{{\partial\, {\hat
{\mathcal{K}}}_{12}({\hat {\vec \kappa}}_1, {\hat { \vec \kappa}}_2,
{\hat {\vec \rho}}_{12})}}{{\partial\, {\hat {\vec \rho}} _{12}}}} -
2\, {\vec A}_{\perp S2}({\hat {\vec \kappa}}_2, {\hat {\vec \rho}}
_{12})\Big)}}{\sqrt{m_1^2\, c^2 + {\hat {\vec \kappa}}_1^2}}} +  \nonumber \\
&+& {\frac{{{\hat {\vec \kappa}}_2 \cdot \Big({\frac{1}{2}}\,
{\frac{{\partial\, {\hat {\mathcal{K}}}_{12}({\hat {\vec \kappa}}_1,
{\hat {\vec \kappa}}_2, {\hat {\vec \rho}}_{12})}}{{\partial\, {\hat
{\vec \rho}}_{12}}}} - 2\, {\vec A}_{\perp S1}({\hat {\vec
\kappa}}_1, {\hat {\vec \rho}}_{12})
\Big)}}{\sqrt{m_2^2\, c^2 + {\hat {\vec \kappa}}_2^2}}} \Big] -  \nonumber \\
&-& {\frac{{Q_1\, Q_2}}{c}}\, \int d^3\sigma\, \Big({\hat {\vec
\pi}}_{\perp S1}(\vec \sigma - {\frac{{m_2}}{m}}\, {\hat {\vec
\rho}}_{12}, {\hat {\vec \kappa}}_1) \cdot {\hat {\vec \pi}}_{\perp
S2}(\vec \sigma + {\frac{{m_1}}{m}
}\, {\hat {\vec \rho}}_{12}, {\hat {\vec \kappa}}_2) +  \nonumber \\
&+& {\hat {\vec B}}_{S1}(\vec \sigma - {\frac{{m_2}}{m}}\, {\hat
{\vec \rho}} _{12}, {\hat {\vec \kappa}}_1) \cdot {\hat {\vec B}}_{
S2}(\vec \sigma + { \frac{{m_1}}{m}}\, {\hat {\vec \rho}}_{12},
{\hat {\vec \kappa}}_2) \Big),
 \end{eqnarray*}

\begin{eqnarray*}
 &&\mathcal{B}({\hat {\vec \rho}}_{12}, {\hat {\vec \pi}}_{12}, {\hat
{\vec \pi}}_{(12)3}) = {\frac{1}{2}}\, {\frac{{Q_1\, Q_2}}{c}}\,
\Big[{\frac{{m_2} }{m}}\, {\frac{{{\hat {\vec \kappa}}_1 \cdot
\Big({\frac{1}{2}}\, {\frac{{
\partial\, {\hat {\mathcal{K}}}_{12}({\hat {\vec \kappa}}_1, {\hat {\vec
\kappa}}_2, {\hat {\vec \rho}}_{12})}}{{\partial\, {\hat {\vec
\rho}}_{12}}}} - 2\, {\vec A}_{\perp S2}({\hat {\vec \kappa}}_2,
{\hat {\vec \rho}}_{12})
\Big)}}{\sqrt{m_1^2\, c^2 + {\hat {\vec \kappa}}_1^2}}} -  \nonumber \\
&-& {\frac{{m_1}}{m}}\, {\frac{{{\hat {\vec \kappa}}_2 \cdot
\Big({\frac{1}{2 }}\, {\frac{{\partial\, {\hat
{\mathcal{K}}}_{12}({\hat {\vec \kappa}}_1, { \hat {\vec \kappa}}_2,
{\hat {\vec \rho}}_{12})}}{{\partial\, {\hat {\vec \rho}}_{12}}}} -
2\, {\vec A}_{\perp S1}({\hat {\vec \kappa}}_1, {\hat {\vec
\rho}}_{12})\Big)}}{\sqrt{m_2^2\, c^2 + {\hat {\vec \kappa}}_2^2}}}
\Big],
 \end{eqnarray*}

\bea
 &&{\vec {\mathcal{C}}}({\hat {\vec \rho}}_{12}, {\hat {\vec \pi}}_{12}, {
\hat {\vec \pi}}_{(12)3}) = - {\frac{1}{2}}\, {\frac{{Q_1\,
Q_2}}{c}}\, \Big( \sqrt{m_1^2\, c^2 + {\hat {\vec \kappa}}_1^2}\,
{\frac{{\partial}}{{\partial\, {\hat {\vec \kappa}}_1}}} +
\sqrt{m_2^2\, c^2 + {\hat {\vec \kappa }}_2^2}\,
{\frac{{\partial}}{{\partial\, {\hat {\vec \kappa}}_2}}} \Big)\, {
\hat {\mathcal{K}}}_{12}({\hat {\vec \kappa}}_1, {\hat
{\vec \kappa}}_2, {\hat {\vec \rho}}_{12}) -  \nonumber \\
&-& {\frac{{Q_1\, Q_2}}{{4\pi\, c}}}\, \int d^3\sigma\,
\Big({\frac{{{\hat { \vec \pi}}_{\perp S1}(\vec \sigma -
{\frac{{m_2}}{m}}{\hat {\vec \rho}} _{12}, {\hat {\vec
\kappa}}_1)}}{{|\vec \sigma + {\frac{{m_1}}{m}}\, {\hat { \vec
\rho}}_{12}|}}} + {\frac{{{\hat {\vec \pi}}_{\perp S2}(\vec \sigma +
{ \frac{{m_1}}{m}}\, {\hat {\vec \rho}}_{12}, {\hat {\vec
\kappa}}_2)}}{{|\vec \sigma - {\frac{{m_2}}{m}}\, {\hat {\vec
\rho}}_{12}|}}} \Big) -  \nonumber \\
&-& {\frac{{Q_1\, Q_2}}{c}}\, \int d^3\sigma\, \vec \sigma\,\,
\Big({\hat { \vec \pi}}_{\perp S1}(\vec \sigma - {\frac{{m_2}}{m}}\,
{\hat {\vec \rho}} _{12}, {\hat {\vec \kappa}}_1) \cdot {\hat {\vec
\pi}}_{\perp S2}(\vec \sigma + {\frac{{m_1}}{m}}\, {\hat {\vec
\rho}}_{12}, {\hat {\vec \kappa}}_2) +  \nonumber \\
&+& {\hat {\vec B}}_{S1}(\vec \sigma - {\frac{{m_2}}{m}}\, {\hat
{\vec \rho}} _{12}, {\hat {\vec \kappa}}_1) \cdot {\hat {\vec B}}_{
S2}(\vec \sigma + { \frac{{m_1}}{m}}\, {\hat {\vec \rho}}_{12},
{\hat {\vec \kappa}}_2) \Big).
 \label{4.12}
\end{eqnarray}
\bigskip

Since, as shown in Eqs.(\ref{4.2}), we have ${\vec
{\mathcal{K}}}_{(int)} = c\, {\vec {\mathcal{K}}} _{Galilei} +
O({\frac{1}{c}}) \approx 0$ we have ${\hat {\vec \eta}} = {\vec
x}_{(n)} + O(c^{- 2}) \approx 0$ in the non-relativistic limit,
namely we get the non-relativistic rest frame ${\vec {\mathcal{K}}}
_{Galilei} = - m\, {\vec x}_{(n)} = - \sum_{i=1}^2\, m_i\, {\hat
{\vec \eta}} _i \approx 0$.

\bigskip

By using Eqs. (\ref{4.11}), (\ref{3.8}), (\ref{4.8}) and the
rest-frame condition ${\vec {\cal P}}_{(int)} = {\hat {\vec \kappa}}
(\tau ) \approx 0$, we get the following expressions of the internal
invariant mass  and of the internal angular momentum in
Eqs.(\ref{4.2})

\bea
  \mathcal{E}_{(int)} &=& M\, c^2 \approx M_{rad}\, c^2 +  \nonumber \\
&+& c\, \sqrt{m_1^2\, c^2 + ({\hat {\vec \pi}}_{12} +
{\frac{{m_1}}{m}}\, { \hat {\vec \pi}}_{(12)3})^2} + c\,
\sqrt{m_2^2\, c^2 + ({\hat {\vec \pi}}
_{12} - {\frac{{m_2}}{m}}\, {\hat {\vec \pi}}_{(12)3})^2} +  \nonumber \\
&+& {\frac{{Q_1\, Q_2}}{{4\pi\, |{\hat {\vec \rho}}_{12}|}}} +
V_{DARWIN}({ \hat {\vec \rho}}_{12}; {\hat {\vec \pi}}_{12} +
{\frac{{m_1}}{m}}\, {\hat { \vec \pi}}_{(12)3}; - {\hat {\vec
\pi}}_{12} + {\frac{{m_2}}{m}}\, {\hat { \vec \pi}}_{(12)3}),
\nonumber \\
 &&{}\nonumber \\
 M_{rad}\, c^2 &=& {1\over 2}\, \int d^3\sigma\, \Big[{\vec \pi}_{\perp
 rad}^2 + {\vec B}^2_{rad}\Big](\tau ,\vec \sigma ),
 \label{4.13}
 \eea

\begin{eqnarray}
{\vec {\mathcal{J}}}_{(int)} &=& {\vec {\bar S}} \approx {\hat {\vec
\rho}}_{12} \times {\hat { \vec \pi}}_{12} + {\hat {\vec
\rho}}_{(12)3} \times
{\hat {\vec \pi}}_{(12)3} + {\vec S}_{rad},  \nonumber \\
&&{}  \nonumber \\
 {\vec S}_{rad} &=& - {\frac{1}{c}}\, \int d\tilde k\,
H_{rad}(\vec k)\, \vec k \times {\frac{{\partial}}{{\partial\, \vec
k}}}\, K_{rad}(\vec k) - {\frac{1}{c}}\, \int d\tilde k\,
\lambda_{rad}(\vec k)\, \vec k \times {\frac{{\partial}}{{\partial\,
\vec k}}}\, \rho_{rad}(\vec k)  +\nonumber \\
&+& {\frac{1}{c}}\, \int d\tilde k\, \lambda_{rad}(\vec k)\,
\Big[{\vec \epsilon}_-(\vec k) \times {\vec \epsilon}_+(\vec k) +
{\frac{i}{2}}\, \Big({ \vec \epsilon}_-(\vec k) + {\vec
\epsilon}_+(\vec k)\Big) \cdot \vec k \times
{\frac{{\partial}}{{\partial\, \vec k}}}\, \Big({\vec \epsilon}
_-(\vec k) - {\vec \epsilon}_+(\vec k)\Big) \Big].  \nonumber \\
&&{}  \label{4.14}
\end{eqnarray}

\bigskip

Eqs.(\ref{4.7}), (\ref{4.8}) and (\ref{4.9}) imply

\bea
 {\hat {\vec \eta}}_{12} &=&
 {\hat {\vec \eta}} + {\frac{1}{2}}\, {\hat {\vec
\rho}}_{(12)3} \approx { \frac{1}{{\mathcal{A} + M_{rad}\, c}}}\,
\Big(M_{rad}\, c\, {\hat {\vec \rho}}_{(12)3} -  \nonumber \\
&-& \Big[{\frac{{m_2}}{m}}\, \sqrt{m_1^2\, c^2 + ({\hat {\vec
\pi}}_{12} + { \frac{{m_1}}{m}}\, {\hat {\vec \pi}}_{(12)3})^2} -
{\frac{{m_1}}{m}}\, \sqrt{ m_2^2\, c^2 + ({\hat {\vec \pi}}_{12} -
{\frac{{m_2}}{m}}\, {\hat {\vec \pi}} _{(12)3})^2} +
\mathcal{B}\Big]\, {\hat {\vec \rho}}_{12} -\nonumber \\
 &-& {\vec {\mathcal{C}}} + X^{\tau}_{rad}\, {\hat {\vec \pi}}_{(12)3} + {\vec
{\mathcal{D}}}\Big),\nonumber \\
 &&{}  \nonumber \\
 &&{}\nonumber \\
{\hat {\vec \eta}}_1 &\approx& {\frac{1}{{\mathcal{A} + M_{rad}\,
c}}}\, \Big(M_{rad}\, c\, {\hat {\vec \rho}}_{(12)3} +
\Big[{\frac{{m_2}}{m}}\, \Big(\mathcal{A} + M_{rad}\, c -
\sqrt{m_1^2\, c^2 + ({\hat {\vec \pi}}_{12}
+ {\frac{{m_1}}{m}}\, {\hat {\vec \pi}}_{(12)3})^2}\Big) +  \nonumber \\
&+& {\frac{{m_1}}{m}}\, \sqrt{m_2^2\, c^2 + ({\hat {\vec \pi}}_{12}
- {\frac{ {m_2}}{m}}\, {\hat {\vec \pi}}_{(12)3})^2} +
\mathcal{B}\Big]\, {\hat {\vec \rho}}_{12} -
 {\vec {\mathcal{C}}} + X^{\tau}_{rad}\, {\hat {\vec
\pi}}_{(12)3} + {\vec {\mathcal{D}}}\Big),  \nonumber \\
 {\hat {\vec \eta}}_2 &\approx& {\frac{1}{{\mathcal{A} + M_{rad}\,
c}}}\, \Big(M_{rad}\, c\, {\hat {\vec \rho}}_{(12)3} -
\Big[{\frac{{m_2}}{m}}\, \sqrt{m_1^2\, c^2 + ({\hat {\vec \pi}}_{12}
+ {\frac{{m_1}}{m}}\, {\hat { \vec \pi}}_{(12)3})^2} +
{\frac{{m_1}}{m}}\, \Big(\mathcal{A} + M_{rad}\, c -
\nonumber \\
&-& \sqrt{m_2^2\, c^2 + ({\hat {\vec \pi}}_{12} -
{\frac{{m_2}}{m}}\, {\hat { \vec \pi}}_{(12)3})^2} \Big) +
\mathcal{B} \Big]\, {\hat {\vec \rho}}_{12} -  {\vec {\mathcal{C}}}
+ X^{\tau}_{rad}\, {\hat {\vec
\pi}}_{(12)3} + { \vec {\mathcal{D}}}\Big).\nonumber \\
 &&{}
  \label{4.15}
\end{eqnarray}

\medskip

These equations, together with the inverse canonical transformation
(I-3.10) and by using Eqs.(I-2.51), (I-2.52), (I-3.9), (\ref{3.8})
and (\ref{4.8}), allow to get the original variables ${\vec
\eta}_i(\tau )$ and the world-lines (I-2.18) of the particles only
in terms of relative variables

\begin{eqnarray*}
x^{\mu}_1(\tau ) &=& Y^{\mu}(\tau ) + \epsilon^{\mu}_r(\vec h)\,
\eta_1^r,  \nonumber \\
x^{\mu}_2(\tau ) &=& Y^{\mu}(\tau ) + \epsilon^{\mu}_r(\vec h)\,
\eta_2^r ,  \end{eqnarray*}

\begin{eqnarray*}
 \eta^r_1(\tau ) &\approx& {\hat \eta}_1^r(\tau ) -
  {{Q_1}\over c}\, {m\over {m_1}}\, {{\partial}\over {\partial\,
 {\hat \pi}_{(12)3\, r}}}\, \int d^3\sigma\, \Big[{\tilde {\vec
 \pi}}_{\perp rad}\Big(\tau - X^{\tau}_{rad}, \vec \sigma + {1\over 2}\,
 {\hat {\vec \rho}}_{(12)3}\Big) \cdot\nonumber \\
 &&\cdot  {\hat {\vec A}}_{\perp S1}\Big(\vec \sigma -
 {1\over 2}\, {\hat {\vec \rho}}_{(12)3} - {{m_2}\over m}\, {\hat {\vec \rho}}_{12},
 {{m_1}\over m}\, {\hat {\vec \pi}}_{(12)3} + {\hat {\vec \pi}}_{12}\Big)
 -\nonumber \\
 &-& {\tilde {\vec A}}_{\perp rad}\Big(\tau - X^{\tau}_{rad}, \vec \sigma + {1\over 2}\,
 {\hat {\vec \rho}}_{(12)3}\Big) \cdot\nonumber \\
 &&\cdot {\hat {\vec \pi}}_{\perp S1}\Big(\vec \sigma -
 {1\over 2}\, {\hat {\vec \rho}}_{(12)3} - {{m_2}\over m}\, {\hat {\vec \rho}}_{12},
 {{m_1}\over m}\, {\hat {\vec \pi}}_{(12)3} + {\hat {\vec \pi}}_{12}\Big)
 \Big] -\nonumber \\
 &-& {{Q_1\, Q_2}\over {2\, c}}\, {m\over {m_1}}\, {{\partial}\over
 {\partial\, {\hat \pi}_{(12)3\, r}}}\, \int d^3\sigma\,
 \Big[{\hat {\vec A}}_{\perp S1}\Big(\vec \sigma -
 {1\over 2}\, {\hat {\vec \rho}}_{(12)3} - {{m_2}\over m}\, {\hat {\vec \rho}}_{12},
 {{m_1}\over m}\, {\hat {\vec \pi}}_{(12)3} + {\hat {\vec \pi}}_{12}\Big) \cdot \nonumber \\
 &&\cdot {\hat {\vec \pi}}_{\perp S2}\Big(\vec \sigma -
 {1\over 2}\, {\hat {\vec \rho}}_{(12)3} + {{m_1}\over m}\, {\hat {\vec \rho}}_{12},
 {{m_2}\over m}\, {\hat {\vec \pi}}_{(12)3} - {\hat {\vec \pi}}_{12}\Big) -\nonumber \\
 &-& {\hat {\vec \pi}}_{\perp S1}\Big(\vec \sigma -
 {1\over 2}\, {\hat {\vec \rho}}_{(12)3} - {{m_2}\over m}\, {\hat {\vec \rho}}_{12},
 {{m_1}\over m}\, {\hat {\vec \pi}}_{(12)3} + {\hat {\vec \pi}}_{12}\Big) \cdot\nonumber \\
 &&\cdot {\hat {\vec A}}_{\perp S2}\Big(\vec \sigma -
 {1\over 2}\, {\hat {\vec \rho}}_{(12)3} + {{m_1}\over m}\, {\hat {\vec \rho}}_{12},
 {{m_2}\over m}\, {\hat {\vec \pi}}_{(12)3} - {\hat {\vec \pi}}_{12}\Big)
 \Big],
 \end{eqnarray*}

\bea
 \eta^r_2(\tau ) &\approx& {\hat \eta}_2^r(\tau ) -
  {{Q_2}\over c}\, {m\over {m_2}}\, {{\partial}\over {\partial\,
 {\hat \pi}_{(12)3\, r}}}\, \int d^3\sigma\, \Big[{\tilde {\vec
 \pi}}_{\perp rad}\Big(\tau - X^{\tau}_{rad}, \vec \sigma + {1\over 2}\,
 {\hat {\vec \rho}}_{(12)3}\Big) \cdot\nonumber \\
 &&\cdot  {\hat {\vec A}}_{\perp S2}\Big(\vec \sigma -
 {1\over 2}\, {\hat {\vec \rho}}_{(12)3} + {{m_1}\over m}\, {\hat {\vec \rho}}_{12},
 {{m_2}\over m}\, {\hat {\vec \pi}}_{(12)3} - {\hat {\vec \pi}}_{12}\Big)
 -\nonumber \\
 &-& {\tilde {\vec A}}_{\perp rad}\Big(\tau - X^{\tau}_{rad}, \vec \sigma + {1\over 2}\,
 {\hat {\vec \rho}}_{(12)3}\Big) \cdot\nonumber \\
 &&\cdot {\hat {\vec \pi}}_{\perp S2}\Big(\vec \sigma -
 {1\over 2}\, {\hat {\vec \rho}}_{(12)3} + {{m_1}\over m}\, {\hat {\vec \rho}}_{12},
 {{m_2}\over m}\, {\hat {\vec \pi}}_{(12)3} - {\hat {\vec \pi}}_{12}\Big)
 \Big] +\nonumber \\
 &+& {{Q_1\, Q_2}\over {2\, c}}\, {m\over {m_2}}\, {{\partial}\over
 {\partial\, {\hat \pi}_{(12)3\, r}}}\, \int d^3\sigma\,
 \Big[{\hat {\vec A}}_{\perp S1}\Big(\vec \sigma -
 {1\over 2}\, {\hat {\vec \rho}}_{(12)3} - {{m_2}\over m}\, {\hat {\vec \rho}}_{12},
 {{m_1}\over m}\, {\hat {\vec \pi}}_{(12)3} + {\hat {\vec \pi}}_{12}\Big) \cdot \nonumber \\
 &&\cdot {\hat {\vec \pi}}_{\perp S2}\Big(\vec \sigma -
 {1\over 2}\, {\hat {\vec \rho}}_{(12)3} + {{m_1}\over m}\, {\hat {\vec \rho}}_{12},
 {{m_2}\over m}\, {\hat {\vec \pi}}_{(12)3} - {\hat {\vec \pi}}_{12}\Big) -\nonumber \\
 &-& {\hat {\vec \pi}}_{\perp S1}\Big(\vec \sigma -
 {1\over 2}\, {\hat {\vec \rho}}_{(12)3} - {{m_2}\over m}\, {\hat {\vec \rho}}_{12},
 {{m_1}\over m}\, {\hat {\vec \pi}}_{(12)3} + {\hat {\vec \pi}}_{12}\Big) \cdot\nonumber \\
 &&\cdot {\hat {\vec A}}_{\perp S2}\Big(\vec \sigma -
 {1\over 2}\, {\hat {\vec \rho}}_{(12)3} + {{m_1}\over m}\, {\hat {\vec \rho}}_{12},
 {{m_2}\over m}\, {\hat {\vec \pi}}_{(12)3} - {\hat {\vec \pi}}_{12}\Big)
 \Big],\nonumber \\
 &&{}
   \label{4.16}
\end{eqnarray}

\subsection{The Final Relative Canonical Variables}

The final independent canonical relative variables are \hfill\break

a) ${\hat {\vec \rho}}_{12}(\tau ) $, ${\hat {\vec \pi}}_{12}(\tau
)$ (relative motion of the two particles);\medskip

b) ${ \hat {\vec \rho}}_{(12)3}(\tau )$, ${\hat {\vec
\pi}}_{(12)3}(\tau )$ (relative motion of the "particle system" with
respect to the "radiation field system");\medskip

c) $X^{\tau}_{rad}(\tau ) $, ${\cal P}^{\tau}_{rad}(\tau ) =
{\frac{1}{c}}\, {\cal E}_{rad}(\tau ) = M_{rad}(\tau )\, c$ (the
energy of the radiation field and its conjugate temporal variable,
the only surviving collective variables);\medskip

d) $ {\bf H}_{rad}(\vec k)$, ${\bf K}_{rad}(\vec k)$ (relative
multipoles of the radiation field with respect to its monopole-like
collective variables);\medskip

e) $\lambda_{rad}(\vec k)$, $\rho_{rad}(\vec k)$ (relative variables
describing the helicity degrees of freedom of the radiation field).

\bigskip

These variables satisfy Hamilton equations having ${\cal E}_{(int)}$
of Eq.(\ref{4.13}) as Hamiltonian.\medskip

There are the following constants of motion:\medskip

A) The relative 3-momentum ${\hat {\vec \pi}}_{(12)3} =
{\frac{1}{2}}\Big({ \hat {\vec \kappa}}_1 + {\hat {\vec \kappa}}_2 -
{\vec {\cal P}}_{rad}\Big) \approx - {\vec {\cal P}}_{rad} \approx
{\hat {\vec \kappa}}_1 + {\hat {\vec \kappa}}_2$: $ \{ {\hat {\vec
\pi}}_{(12)3}, \mathcal{E}_{(int)}\} = 0$.\medskip

The constant of the motion ${\hat {\vec \pi}}_{(12)3}$ connects the
particles and the radiation field. Its conjugate variable satisfies
the kinematical Hamilton equation

\begin{eqnarray}
{\frac{{d\, {\hat {\vec \rho}}_{(12)3}(\tau )}}{{d\, \tau}}} &{
\buildrel \circ \over {=}}& {\frac{{\partial\,
\mathcal{E}_{(int)}}}{{\partial\, { \hat {\vec \pi}}_{(12)3}}}} =
c\, {\frac{{m_1}}{m}}\, {\frac{{\ {\hat {\vec \pi}}_{12} +
{\frac{{m_1}}{m}}\, {\hat {\vec \pi}}_{(12)3}}}{{\sqrt{m_1^2\, c^2 +
({\hat {\vec \pi}}_{12} + {\frac{{m_1}}{m}}\, {\hat {\vec \pi}}
_{(12)3})^2} }}} +  \nonumber \\
&+& c\, {\frac{{m_2}}{m}}\, {\frac{{{\hat {\vec \pi}}_{12} -
{\frac{{m_2}}{m} }\, {\hat {\vec \pi}}_{(12)3}}}{\sqrt{m_2^2\, c^2 +
({\hat {\vec \pi}}_{12} - {\frac{{m_2}}{m}}\, {\hat {\vec
\pi}}_{(12)3})^2}}}
+  \nonumber \\
&+& {\frac{{\partial\, V_{DARWIN}({\hat {\vec \rho}}_{12}; {\hat
{\vec \pi}} _{12} + {\frac{{m_1}}{m}}\, {\hat {\vec \pi}}_{(12)3}; -
{\hat {\vec \pi}} _{12} + {\frac{{m_2}}{m}}\, {\hat {\vec
\pi}}_{(12)3}) }}{{\partial\, {\hat { \vec \pi}}_{(12)3}}}}.
  \label{4.17}
\end{eqnarray}

\bigskip

B) The energy  of the radiation field: $\{ {\cal P}^{\tau}_{rad},
\mathcal{E}_{(int)}\} = 0$. Its conjugate variable $
X^{\tau}_{rad}(\tau )$, $\{ X^{\tau}_{rad}, {\cal P}^{\tau}_{rad}\}
= -1$, satisfies

\begin{equation}
{\frac{{d\, X^{\tau}_{rad}}}{{d\, \tau}}}\, {\buildrel \circ \over
{=}}\,\, \{ X^{\tau}_{rad}, \mathcal{E}_{(int)}\} = - 1.
 \label{4.18}
\end{equation}
\medskip

Like for the Klein-Gordon case of Appendix A, the two second class
constraints ${\cal P}^{\tau}_{rad} - E_{rad} \approx 0$, $
X^{\tau}_{rad} - \tau \approx 0$, eliminating the last two
collective variables, select a symplectic sub-manifold of the
surface of constant energy ${\cal P}^{\tau}_{rad} = E_{rad}$ of the
radiation field.

\bigskip

C) The multipoles $ {\bf H}_{rad}(\vec k)$, ${\bf K}_{rad}(\vec k)$,
$\lambda_{rad}(\vec k)$, $\rho_{rad}(\vec k)$.

\bigskip

\textit{Therefore the two subsystems (particles and radiation
field),  although  not coupled in the equations of motion due to
${\cal E}_{(int)} = {\cal E}_{matter} + {\cal E}_{rad}$, are
nevertheless effectively interacting through the rest-frame
constraints and their gauge fixings as clear from Eq.(\ref{4.17}).}

\bigskip

The canonical variables ${\hat {\vec q}}_+$, $ {\hat {\vec
\kappa}}_+ = {\hat {\vec \kappa}}_{12}$, ${\hat {\vec \rho}}_q$,
${\hat {\vec \pi}}_q$, describing the canonical (Coulomb-dressed)
Newton-Wigner 3-center of mass and the relative motion of the
2-particle subsystem (see Subsection IIA), could be found by using
Eqs.(\ref{4.7}) and (\ref{4.8}). Then we could re-express
$\mathcal{E} _{(int)}$ of Eq.(\ref{4.13}) and ${\vec
{\mathcal{J}}}_{(int)}$ of Eq.(\ref{4.14}) in terms of them. In
particular these variables would allow to write
$\mathcal{E}_{(int)}$ of Eq.(\ref{4.13}) in the form

\begin{eqnarray}
\mathcal{E}_{(int)} &\approx& c\, \sqrt{{\hat {\mathcal{M}}}^2_o\,
c^2 + {\hat {\vec \pi}}^2_{(12)3}} +   \Big(function\, of\, {\hat
{\vec \rho}}_q, {\hat {\vec \pi}}_q,
{\hat {\vec \pi}}_{(12)3}....\Big) + {\cal P}^{\tau}_{rad},  \nonumber \\
&&{}  \nonumber \\
{\hat {\mathcal{M}}}_o\, c &=& \sqrt{m_1^2\, c^2 + {\hat {\vec
\pi}}_q^2} + \sqrt{m_2^2\, c^2 + {\hat {\vec \pi}}_q^2}.
  \label{4.19}
\end{eqnarray}

In this way the 2-particle subsystem is visualized as an effective
pseudo-particle (an atom after quantization) of mass ${\hat
{\mathcal{M}}}_o$ plus interactions. However a drawback of these
variables is that the Coulomb interaction depends upon ${\hat {\vec
\rho}} _{12}$ and not upon ${\hat {\vec \rho}}_q$, variables which
do not coincide for ${\hat {\vec \kappa}} \not= 0$.

\subsection{Coming back with the Inverse Canonical Transformation}

If we use the inverse canonical transformation (I-3.10) of I,
Eqs.(\ref{4.7})-(\ref{4.8}) and the consequences
(\ref{4.11})-(\ref{4.15}) of ${\vec {\mathcal{K}}}_{(int)} \approx
0$, we can get the expression of ${\vec {\cal P}}_{em}$, ${\vec
{\cal J}}_{em}$ and ${\cal P}^{\tau}_{em}$ of Eqs.(\ref{4.1}) in
terms of the radiation field and of the Coulomb-dressed particles
from the comparison of the internal Poincare' generators before and
after the canonical transformation.

\bigskip

Eqs. (\ref{4.1}) and (\ref{4.2}) imply

\bea
  {\vec {\cal P}}_{em} &=& {1\over c}\, \int d^3\sigma\, \Big({\vec
  \pi}_{\perp} \times \vec B\Big)(\tau
  ,\vec \sigma ) \approx \nonumber \\
  &\approx& {\vec {\cal P}}_{rad} - \sum_{i=1}^2\, {\frac{{
Q_i}}{c}}\, {\frac{{\partial\, {\hat T}_i(\tau )}}{{\partial\, {\hat
{\vec \eta}}_i}}} - {\frac{1 }{2}}\, {\frac{{Q_1\, Q_2}}{c}}\,
\Big({\frac{{\partial}}{{\partial\, {\hat {\vec \eta}}_1 }}} -
{\frac{{\partial}}{{\partial\, {\hat {\vec \eta}}_2}}}\Big)\,
{\hat {\cal K}}_{12}(\tau ) =\nonumber \\
 &=& {1\over c}\, \int d^3\sigma\, \Big({\vec
  \pi}_{\perp rad} \times {\vec B}_{rad}\Big)(\tau
  ,\vec \sigma ) + O({1\over c})\approx  \nonumber \\
 &\approx& - {\hat {\vec \pi}}_{(12)3} + O({1\over c}),\nonumber \\
&&{}  \label{4.20}
\end{eqnarray}

 \bigskip

\bea
 {\vec {\cal J}}_{em} &=& {\frac{1}{c}}\, \int d^3\sigma\,\, \vec \sigma
\times \Big({\vec \pi}_{\perp} \times \vec B\Big)(\tau ,\vec \sigma
) \approx  \nonumber \\
 &\approx& {\vec S}_{rad} - \Big({\hat {\vec \eta}} - {\frac{1}{2}}\,
{\hat {\vec \rho}}_{(12)3}\Big)
\times {\hat {\vec \pi}}_{(12)3} -  \nonumber \\
&&- \sum_{i=1}^2\, {\frac{{Q_i}}{c}}\, \Big({\hat {\vec \eta}}_i
\times { \frac{{\partial\, {\hat T}_i(\tau )}}{{\partial\, {\hat
{\vec \eta}}_i}}} + {\hat {\vec \kappa}} _i \times
{\frac{{\partial\, {\hat T}_i(\tau )}}{{\partial\, {\hat
{\vec \kappa}}_i}}}\Big) -\nonumber \\
&&- {\frac{1}{2}}\, {\frac{{Q_1\, Q_2}}{c}}\, \sum_{i=1}^2\,
\Big({\hat { \vec \eta}}_i \times {\frac{{\partial\, {\hat {\cal
K}}_{12}(\tau )}}{{\partial\, {\hat {\vec \eta}}_i}}} + {\hat {\vec
\kappa}}_i \times {\frac{{\partial\, {\hat {\cal K}} _{12}(\tau
)}}{{\partial\, {\hat {\vec \kappa}}_i}}}\Big) =\nonumber \\
 &=& {\vec S}_{rad} + O({1\over c}),
  \label{4.21}
\end{eqnarray}

\bigskip

\bea
 {\cal P}^{\tau}_{em} &=& {1\over {2\, c}}\, \int d^3\sigma\,
 \Big({\vec \pi}_{\perp}^2 + {\vec B}^2\Big)(\tau ,\vec \sigma )
 =\nonumber \\
 &=& {\cal P}^{\tau}_{rad} + \sum_{i=1}^2\, \Big(\sqrt{m_i^2\, c^2
 + \Big({\frac{{m_i}}{m}}\, {\hat {\vec \pi}}
_{(12)3} + (-)^{i+1}\, {\hat {\vec \pi}}_{12}\Big)^2} -
 \sqrt{m_i^2\, c^2 + {\vec \kappa} _i^2}\Big) +\nonumber \\
 &+& {\frac{{Q_1\, Q_2}}{{4\pi\, c\, |{\hat
{\vec \rho }}_{12}|}}} - {\frac{{Q_1\, Q_2}}{{4\pi\, c\, |{\vec
\rho}_{12}|}}}  + {1\over c}\, V_{Darwin}({\hat {\vec \rho}}_{12},
{\hat {\vec \pi}}_{12}, { \hat {\vec \pi}}_{(12)3}) +\nonumber \\
 &+& \sum_{i=1}^2\, {{Q_i}\over c}\, {\frac{{{\vec \kappa}_i \cdot
{\vec A} _{\perp}(\tau , {\vec \eta}_i(\tau ))}}{\sqrt{m_i^2\, c^2 +
{\vec \kappa}_i^2 }}} =\nonumber \\
 &=& {1\over {2\, c}}\, \int d^3\sigma\,
 \Big({\vec \pi}_{\perp rad}^2 + {\vec B}_{rad}^2\Big)(\tau ,\vec \sigma )
 + O({1\over {c^2}}).
  \label{4.22}
\end{eqnarray}

\bigskip

This is in accord with Eq.(I-2.49) whose semi-relativistic limit
$a_{em\, \lambda}(\tau ,\vec k) = a_{\lambda }(\vec{k})\,
e^{-i\,\omega (\vec{k})\,\tau } + \omega (\vec{k})\, {
\vec{\epsilon}}_{\lambda }(\vec{k}) \cdot \sum_{i=1}^{2}\,
\frac{Q_{i}}{m_{i}\, c} \, {\frac{{e^{-i\,\vec{k} \cdot
{\vec{\eta}}_{i}(\tau )}}}{{{\vec{k}}^{4}}}\, \vec{k} \times
\Big({\vec{\kappa}}_{i}(\tau ) \times \vec{k}\Big)}\, + O( c^{-3})$
says the Fourier coefficients of the transverse electro-magnetic
field differ from those of a transverse radiation field for particle
terms of order $O(c^{-1})$. This is the error when we replace the
electro-magnetic field with a radiation field, as is often done in
atomic physics.

\vfill\eject

\section{The Multipolar Expansion of the Particle Energy-Momentum Tensor}

As said in Section I of paper I, in the rest-frame instant form  the
isolated system of charged particles plus the transverse
electro-magnetic field can be described as a decoupled non-covariant
center of mass carrying the invariant mass and the spin of the
isolated system, i.e. some type of pole-dipole structure. The
invariant mass and the spin are evaluated by means of the
energy-momentum tensor $T^{AB}(\tau ,\vec \sigma ) =
T^{AB}_{matter}(\tau ,\vec \sigma ) + T^{AB}_{em}(\tau ,\vec \sigma
)$ determined by using the action (I-2.1) of the parametrized
Minkowski theory.\medskip

Let us now consider the {\it open} subsystem composed only by the
particles, whose non-conserved energy-momentum tensor is
$T^{AB}_{matter}(\tau ,\vec \sigma )$. The study of its multipolar
expansion (see Ref.\cite{11}) allows one to replace the extended
subsystem with its multipoles when some analyticity conditions are
satisfied and then to define a \textit{ pole-dipole} approximation
of the subsystem. As shown in Ref.\cite{11}, this can be done also
for single strongly bound groups of particles, to be used to
describe atoms after quantization, inside the particle subsystem by
identifying the effective energy-momentum of each group.\bigskip

Till now this is the only description of a relativistic (either free
or interacting with the environment) composite system like an atom
as a collective point-like system endowed with multipolar
properties. The lowest approximation of the composite system is the
\textit{pole-dipole approximation}, in which the system is simulated
with a point-like particle (monopole) of mass $M_c$ moving along the
world-line of an effective 4-center of motion  and carrying the spin
dipole ${\vec {\mathcal{J}}}_c$ (to be used for the magnetic dipole
moment). In the rest-frame instant form the 4-center of motion has
the world-line $w_c^{\mu}(\tau ) = Y^{\mu}(\tau ) +
\epsilon^{\mu}_r(\vec h)\, \zeta^r_c(\tau )$ and 4-momentum
$P^{\mu}_c = h^{\mu}\, M_c\, c + \epsilon^{\mu}_r(\vec h)\,
\mathcal{P}_c^r$.\medskip

Let us remark that this pole-dipole approximation  is not the
"dipole approximation" of the semi-relativistic atomic physics,
which will be studied in the next Section.

\bigskip

As shown in Ref.\cite{11}, given a non-isolated cluster of particles
the main problem is the determination of an \textit{effective
4-center of motion} described by 3-coordinates $ \zeta^r_c(\tau )$
and with world-line $w^{\mu}_c(\tau ) = Y^{\mu}(\tau ) +
\epsilon^{\mu}_r(\vec h)\, \zeta^r_c(\tau )$ in the rest frame of
the isolated system \footnote{ See Ref. \cite{11} for a discussion
of the main choices present in the literature.}. The unit 4-velocity
of this center of motion is $u^{\mu}_c(\tau ) = {\dot w}
_c^{\mu}(\tau ) / \sqrt{1 - {\dot {\vec \zeta}}^2_c(\tau )}$ with
${\dot w} _c^{\mu}(\tau ) = h^{\mu} + \epsilon^{\mu}_r(\vec h)\,
{\dot \zeta} ^r_c(\tau )$ (from paper I we have $h^{\mu} =
u^{\mu}(P) = P^{\mu}/Mc$). By using $\delta\, z^{\mu}(\tau ,\vec
\sigma ) = \epsilon^{\mu}_r(\vec h)\, (\sigma^r - \zeta^r(\tau ))$
we can define the Dixon multipoles of the cluster with respect to
the world-line $w^{\mu}_c(\tau )$

\begin{equation}
q_c^{r_1..r_nAB}(\tau ) = \int d^3\sigma \, [\sigma^{r_1} -
\zeta_c^{r_1}(\tau )] .. [\sigma^{r_n} - \zeta_c^{r_n}(\tau )]\,
T^{AB}_{matter}(\tau ,\vec \sigma ).
  \label{5.1}
\end{equation}
\medskip

The mass and momentum monopoles, and the mass, momentum and spin
dipoles are, respectively \medskip

\begin{eqnarray*}
q^{\tau\tau}_c &=& M_c,\qquad q_c^{r\tau} = \mathcal{P}_c^r,  \nonumber \\
q_c^{r\tau\tau} &=& - \mathcal{K}^r_c - M_c\, \zeta^r_c(\tau ) =
M_c\, (R^r_c(\tau ) - \zeta^r_c(\tau )),\qquad q_c^{ru\tau} =
p^{ru}_c(\tau ) -
\zeta^r_c(\tau ) \, \mathcal{P}^u_c,  \nonumber \\
&&{}  \nonumber \\
p^{ru}_c &=& \int d^3\sigma\, \sigma^r\, T^{u\tau}_{matter}(\tau
,\vec \sigma ) = \sum_{i = 1}^2\, \eta^r_i(\tau )\, \kappa_i^u(\tau
) -  \nonumber \\
&-& \sum_{i=1}^2\, Q_i\, \int d^3\sigma\, c(\vec \sigma - {\vec
\eta}_i(\tau ))\, \Big(\partial^r\, A^u_{\perp} + \partial_u\,
A^r_{\perp}\Big)(\tau
,\vec \sigma ),  \nonumber \\
p_c^{ru} + p^{ur}_c &=& \sum_{i=1}^2\, \Big(\eta^r_i(\tau )\,
\kappa^u_i(\tau ) + \eta^u_i(\tau )\, \kappa_i^r(\tau )\Big) -  \nonumber \\
&-& 2\, \sum_{i=1}^2\, Q_i\, \int d^3\sigma\, c(\vec \sigma - {\vec
\eta} _i(\tau ))\, \Big(\partial^r\, A^u_{\perp} + \partial_u\,
A^r_{\perp}\Big)
(\tau ,\vec \sigma )  \nonumber \\
p_c^{ru} - p_c^{ur} &{\buildrel {def}\over =}& \epsilon^{ruv}\,
\mathcal{J}^v_c,
\end{eqnarray*}

\begin{eqnarray}
S^{\mu\nu}_c &=& [\epsilon^{\mu}_r(\vec h)\, h^{\nu} -
\epsilon^{\nu}_r(\vec h)\, h^{\mu}]\, q_c^{r\tau\tau} +
\epsilon^{\mu}_r(\vec h)\, \epsilon^{\nu}_u(\vec h)\, (q^{ru\tau}_c
- q_c^{ur\tau}) =  \nonumber \\
&=& [\epsilon^{\mu}_r(\vec h)\, h^{\nu} - \epsilon^{\nu}_r(\vec h)\,
h^{\mu}]\, M_c\, (R^r_c - \zeta^r_c) +  \nonumber \\
&+& \epsilon^{\mu}_r(\vec h)\, \epsilon^{\nu}_u(\vec h)\, \Big[
\epsilon^{ruv}\, \mathcal{J}^v_c - (\zeta^r_c\, \mathcal{P}_c^u -
\zeta^u_c\, \mathcal{P}_c^r)
\Big],  \nonumber \\
&&{}  \nonumber \\
&&\Rightarrow \, m^{\mu}_{c(P)} = - S^{\mu\nu}_c\, h_{\nu} = -
\epsilon^{\mu}_r(\vec h)\, q_c^{r\tau\tau},
  \label{5.2}
\end{eqnarray}

\noindent where ${\vec {\cal J}}_c$ and ${\vec {\cal K}}_c$ are the
angular momentum and the boost associated with the chosen definition
of effective center of motion.

\bigskip

As shown in Ref.\cite{11}, it is convenient to choose the {\it
center of energy} ${ \vec R}_c = - {\vec {\mathcal{K}}}_c/M_c\, c$
as center of motion. ${\vec R}_c$ and ${\vec {\mathcal{P}}}_c$
(momentum monopole) are the {\it non-canonical} variables describing
the \textit{monopole}, i.e. the collective pseudo-particle of
non-conserved mass $M_c$ (mass monopole).\medskip

On the world-line of this collective pseudo-particle there is also a
\textit{spin dipole} (the anti-symmetric part of the momentum
dipole)  and the symmetric part of the \textit{momentum dipole}
\footnote{This symmetric part of the momentum dipole depends on the
electromagnetic potential and is connected with the electric dipole:
it is directed along the associated relative momentum if one
introduces a suitable definition of a relative variable.}. See
Ref.\cite{11}, Eqs. (5.23) - (5.26).\medskip

Since we are in a Hamiltonian formulation, the constitutive relation
between monopole 3-velocity ${\frac{{d\, {\vec \zeta}_c(\tau
)}}{{d\, \tau}}}$ and monopole (non conserved) 3-momentum ${\vec
{\mathcal{P}}}_c$ must not be added by hand, but can be determined
for each choice of collective center of motion \cite{11}.\medskip

As shown in Ref.\cite{11}, the total 4-momentum $P^{\mu}_c =
\epsilon^{\mu}_A(\vec h)\, q^{A\tau}_c = h^{\mu}\, M_o +
\epsilon^{\mu}_r(\vec h)\, \mathcal{P}_c^r$ and the spin dipole
tensor $ S^{\mu\nu}_c$ obey the Papapetrou-Dixon-Souriau equations.

\subsection{The Multipolar Expansion after
the Canonical Transformation.}

Let us consider the \textit{open} subsystem formed by the particles.
\medskip

From Eqs.(\ref{4.13}) and (\ref{4.14}) we get

\begin{eqnarray}
\mathcal{E}_{matter}(\tau ) &=& c\, \sqrt{m_1^2\, c^2 + ({\hat {\vec
\pi}} _{12} + {\frac{{m_1}}{m}}\, {\hat {\vec \pi}}_{(12)3})^2} +
c\, \sqrt{ m_2^2\, c^2 + ({\hat {\vec \pi}}_{12} -
{\frac{{m_2}}{m}}\, {\hat {\vec \pi}}
_{(12)3})^2} +  \nonumber \\
&+& {\frac{{Q_1\, Q_2}}{{4\pi\, |{\hat {\vec \rho}}_{12}|}}} +
V_{DARWIN}({ \hat {\vec \rho}}_{12}; {\hat {\vec \pi}}_{12} +
{\frac{{m_1}}{m}}\, {\hat { \vec \pi}}_{(12)3}; - {\hat {\vec
\pi}}_{12} + {\frac{{m_2}}{m}}\, {\hat {
\vec \pi}}_{(12)3})\, {\buildrel \circ \over {=}}\, constant,  \nonumber \\
&&{}  \nonumber \\
{\vec {\mathcal{P}}}_{matter}(\tau ) &=& \sum_{i=1}^2\, {\hat {\vec
\kappa}} _i(\tau ) \approx {\hat {\vec \pi}}_{(12)3}\, {\buildrel
\circ \over {=}}\, constant,  \nonumber \\
&&{}  \nonumber \\
{\vec {\mathcal{J}}}_{matter}(\tau ) &=& \sum_i\, {\hat {\vec
\eta}}_i(\tau ) \times {\hat {\vec \kappa}}_i(\tau ) =
 {\hat {\vec \rho}}_{12} \times {\hat {\vec \pi}}_{12} + \Big({\hat
{\vec \eta}} + {\frac{1}{2}}\, {\hat {\vec \rho}}_{(12)3}\Big)
\times {\hat {\vec \pi}}_{(12)3}\, {\buildrel \circ \over {=}}\,
constant,\nonumber \\
 &&{}
  \label{5.3}
\end{eqnarray}

\noindent while from Eq.(\ref{4.2}) we get

\begin{eqnarray}
{\vec {\mathcal{K}}}_{matter} &=& - \sum_{i=1}^2\, {\hat {\vec
\eta}}_i\, \sqrt{m_i^2\, c^2 + {\hat {\vec \kappa}}_i^2} -  \nonumber \\
&-& {\frac{1}{2}}\, {\frac{{Q_1\, Q_2}}{c}}\, \Big[{\hat {\vec
\eta}}_1\, { \frac{{{\hat {\vec \kappa}}_1 \cdot
\Big({\frac{1}{2}}\, {\frac{{\partial\, { \hat
{\mathcal{K}}}_{12}({\hat {\vec \kappa}}_1, {\hat {\vec \kappa}}_2,
{ \hat {\vec \rho}}_{12})}}{{\partial\, {\hat {\vec \rho}}_{12}}}} -
2\, {\vec A}_{\perp S2}({\hat {\vec \kappa}}_2, {\hat {\vec
\rho}}_{12})\Big)}}{\sqrt{m_1^2\, c^2 + {\hat {\vec \kappa}}_1^2}}} +  \nonumber \\
&+& {\hat {\vec \eta}}_2\, {\frac{{{\hat {\vec \kappa}}_2 \cdot
\Big({\frac{1 }{2}}\, {\frac{{\partial\, {\hat
{\mathcal{K}}}_{12}({\hat {\vec \kappa}}_1, {\hat {\vec \kappa}}_2,
{\hat {\vec \rho}}_{12})}}{{\partial\, {\hat {\vec \rho}}_{12}}}} -
2\, {\vec A}_{\perp S1}({\hat {\vec \kappa}}_1, {\hat {\vec
\rho}}_{12})\Big)}}{\sqrt{m_2^2\, c^2 + {\hat {\vec \kappa}}_2^2}}}
\Big] -\nonumber \\
&-& {\frac{1}{2}}\, {\frac{{Q_1\, Q_2}}{c}}\, \Big(\sqrt{m_1^2\, c^2
+ { \hat {\vec \kappa}}_1^2}\, {\frac{{\partial}}{{\partial\, {\hat
{\vec \kappa} }_1}}} + \sqrt{m_2^2\, c^2 + {\hat {\vec
\kappa}}_2^2}\, {\frac{{\partial}}{{
\partial\, {\hat {\vec \kappa}}_2}}} \Big)\, {\hat {\mathcal{K}}}_{12}({
\hat {\vec \kappa}}_1, {\hat {\vec \kappa}}_2, {\hat {\vec
\rho}}_{12}) -\nonumber \\
&-& {\frac{{Q_1\, Q_2}}{{4\pi\, c}}}\, \int d^3\sigma\,
\Big({\frac{{{\hat { \vec \pi}}_{\perp S1}(\vec \sigma - {\hat {\vec
\eta}}_1, {\hat {\vec \kappa} }_1)}}{{|\vec \sigma - {\hat {\vec
\eta}}_2|}}} + {\frac{{{\hat {\vec \pi}} _{\perp S2}(\vec \sigma -
{\hat {\vec \eta}}_2, {\hat {\vec \kappa}}_2)}}{{
|\vec \sigma - {\hat {\vec \eta}}_1|}}} \Big) -  \nonumber \\
&-& {\frac{{Q_1\, Q_2}}{c}}\, \int d^3\sigma\, \vec \sigma\,\,
\Big({\hat { \vec \pi}}_{\perp S1}(\vec \sigma - {\hat {\vec
\eta}}_1, {\hat {\vec \kappa} }_1) \cdot {\hat {\vec \pi}}_{\perp
S2}(\vec \sigma - {\hat {\vec \eta}}_2, {
\hat {\vec \kappa}}_2) +  \nonumber \\
&+& {\hat {\vec B}}_{S1}(\vec \sigma - {\hat {\vec \eta}}_1, {\hat
{\vec \kappa}}_1) \cdot {\hat {\vec B}}_{ S2}(\vec \sigma - {\hat
{\vec \eta}}_2, {\hat {\vec \kappa}}_2) \Big) =  \nonumber \\
&&{}  \nonumber \\
&{\buildrel {def}\over {=}}& - {\frac{1}{c}}\,
\mathcal{E}_{matter}\, {\vec R}_c(\tau ).
  \label{5.4}
\end{eqnarray}

\bigskip

While the natural choice for the effective center of motion is
center of energy ${\hat {\vec \zeta}}_c = {\vec R}_c = - c\, {{{\vec
{\cal K}}_{matter}}\over {{\cal E}_{matter}}}$, Eq.(\ref{5.3})
suggests the other possibility ${\hat {\vec \zeta}}_{matter}(\tau )
= {\hat {\vec \eta}}_{12} = {\hat { \vec \eta}} + {\frac{1}{2}}\,
{\hat {\vec \rho}}_{(12)3}$ with ${\hat {\vec \eta}}$ given in
Eq.(\ref{4.11}): it would imply the relation ${\vec
{\mathcal{J}}}_{matter}(\tau )= {\hat {\vec \rho}}_{12} \times {\hat
{\vec \pi}}_{12} + {\hat {\vec \zeta}}_{matter}(\tau ) \times {\hat
{\vec \pi}}_{(12)3}$ and , differently from the choice ${\vec
\zeta}_c = {\vec R}_c$, ${\hat {\vec \zeta}}_{matter}$ is a
canonical variable like ${\vec \eta}_{12}$ (however ${\hat {\vec
\pi}}_{(12)3}$ is not the conjugate variable).

\vfill\eject

\section{The Dipole Approximation of Atomic Physics}

Let us now look for the relativistic generalization of the {\it
electric dipole approximation} of semi-relativistic atomic physics.
To this end we consider a 2-particle system and we replace the
particle 3-positions ${\vec \eta}_i$ and 3-momenta ${\vec
\kappa}_i$, $i=1,2$, with the naive center of mass and relative
variables of Eqs. (\ref{2.3}): ${\vec \eta}_{12}$, ${\vec
\kappa}_{12}$, ${\vec \rho}_{12}$, ${\vec \pi}_{12}$.\medskip

The standard {\it electric dipole moment} is directed along the
relative variable ${\vec \rho}_{12}$. Usually only neutral systems
with opposite charges of the particles are considered in
applications of the dipole approximation. As a consequence it is
convenient to introduce the following notation for the
Grassmann-valued charges

\bea
 &&Q_1 = Q + {\cal Q},\qquad Q_2 = - Q + {\cal Q},\qquad Q^2_1 =
 Q^2_2 = 0,\quad Q_1\, Q_2 = Q_2\, Q_1 \not= 0,\nonumber \\
 &&{}\nonumber \\
 &&{}\nonumber \\
 &&Q = {1\over 2}\, (Q_1 - Q_2),\qquad {\cal Q} = {1\over 2}\,
 (Q_1 + Q_2),\qquad Q^2 = - Q_1\, Q_2,\quad {\cal Q}^2 = Q_1\,
 Q_2\,\, if\, {\cal Q} \not= 0.\nonumber \\
 &&{}
 \label{6.1}
 \eea

The restriction to opposite charges is done by introducing the
constraint ${\cal Q} \approx 0$, which implies $e = Q \approx Q_1
\approx - Q_2$. This allows us to discard the terms in $Q_1\, Q_2$
coming from ${\cal Q}^2$ from those coming from $Q^2$: only these
terms will produce effects of order $e^2$ in a neutral system with
Grassmann regularization (it eliminates the $e^2$ terms coming from
$Q^2_1$ and $Q^2_2$).

Strictly speaking it is only after the quantization of the
Grassmann-valued charges $Q_i$, sending each of them in a two-level
system with charges $( + e, 0)$ or $(- e, 0)$, that we are really
allowed to use $e = Q_1 = - Q_2$.
\bigskip

The natural definition of a Grassmann-valued electric dipole moment
is

\beq
 \vec d(\tau ) = Q\, {\vec \rho}_{12}(\tau ) =
 {{Q_1 - Q_2}\over 2}\, {\vec \rho}_{12}(\tau )\,
 \rightarrow_{e = Q_1 \approx - Q_2}\,\, e\, {\vec \rho}_{12}(\tau ).
 \label{6.2}
 \eeq
\medskip

The alternative definition

\beq
 \vec D(\tau ) = \sum_{i=1}^2\, Q_i\, {\vec
\eta}_i(\tau ) = \vec d(\tau ) + 2\, {\cal Q}\, \Big[{\vec
\eta}_{12}(\tau ) + {{m_2 - m_1}\over {2m}}\, {\vec \rho}_{12}(\tau
)\Big] \rightarrow_{{\cal Q} \approx 0}\, \vec d(\tau ),
 \label{6.3}
 \eeq

\noindent is equivalent to Eq.(\ref{6.2}) for neutral systems.

\medskip
The electric dipole has not to be confused with the Dixon spin
dipole, which is oriented along the spin $\vec S = {\vec \rho}_{12}
\times {\vec \pi}_{12} $ and determines the direction of the {\it
magnetic dipole moment}.\bigskip

The length $|{\vec \rho}_{12}(\tau )|$ is of the \textit{size of the
atom}, i.e. a few Bohr radii. For a monochromatic electromagnetic
wave, i.e. for ${\vec A} _{\perp}(\tau , {\vec \eta}_{12}(\tau ))
\approx \vec a\, e^{i\, \vec k \cdot {\vec \eta}_{12}(\tau )}$ with
$|\vec k| = {\frac{{2\pi}}{{\lambda_{em} }}}$, we have
${\frac{1}{{|{\vec A}_{\perp}(\tau , {\vec \eta}_{12}(\tau ))|} }}\,
|\Big({\vec \rho}_{12}(\tau ) \cdot {\frac{{\partial}}{{\partial\, {
\vec \eta}_{12}}}}\Big)\, {\vec A}_{\perp}(\tau , {\vec
\eta}_{12}(\tau ))| \approx {\vec \rho}_{12}(\tau ) \cdot \vec k
\approx 2\pi\, {\frac{{size\,\, of\,\, atom}}{{\lambda_{em}}}}$.
This ratio is $<<\, 1$ in the \textit{long wavelength approximation}
$\lambda_{em}\, >>\, size\,\, of\,\, atom$ (i.e. for
radio-frequency, infrared, visible or ultraviolet radiation).

\bigskip

To get the dipole approximation \cite{7} we use Eq.(\ref{2.3}) and
we make the following expansion ($m_3 \equiv m_1$)

\begin{eqnarray}
{\vec A}_{\perp}(\tau , {\vec \eta}_i(\tau )) &=& {\vec
A}_{\perp}\Big(\tau , { \vec \eta}_{12}(\tau ) + (-)^{i+1}\,
{\frac{{m_{i+1}}}{m}}\, {\vec \rho}
_{12}(\tau )\Big) =  \nonumber \\
&=& {\vec A}_{\perp}(\tau , {\vec \eta}_{12}(\tau )) + (-)^{i+1}\,
{\frac{{ m_{i+1}}}{m}}\, \Big({\vec \rho}_{12}(\tau ) \cdot
{\frac{{\partial}}{{
\partial\, {\vec \eta}_{12}}}}\Big)\, {\vec A}_{\perp}(\tau , {\vec \eta}
_{12}(\tau )) +   \nonumber \\
 &+& O([{\vec \rho}_{12} \cdot {\vec \partial}_{{\vec \eta}_{12}}]^2 {\vec A}_{\perp}).
  \label{6.4}
\end{eqnarray}

\bigskip

For the internal Poincare' energy generator of Eqs.(\ref{4.1}) we
have ($\mu = m_1\, m_2 /m$)

\begin{eqnarray*}
&&\mathcal{E}_{(int)} = M\, c^2 =\nonumber \\
 &&{}\nonumber \\
 &=& c\, \Big(\sqrt{m_1^2\, c^2 + \Big({\frac{{m_1}}{m}}\, {\vec
\kappa} _{12}(\tau ) + {\vec \pi}_{12}(\tau )\Big)^2} +
\sqrt{m_2^2\, c^2 + \Big({ \frac{{m_2}}{m}}\, {\vec
\kappa}_{12}(\tau ) - {\vec \pi}_{12}(\tau )\Big)^2}
\Big) -  \nonumber \\
 &&{}\nonumber \\
 &-& \Big({\frac{{m_1\, Q_1}}{\sqrt{m_1^2\, c^2 +
\Big({\frac{{m_1}}{m}}\, { \vec \kappa}_{12}(\tau ) + {\vec
\pi}_{12}(\tau )\Big)^2}}} + {\frac{{m_2\, Q_2}}{{\ \sqrt{m_2^2\,
c^2 + \Big({\frac{{m_2}}{m}}\, {\vec \kappa} _{12}(\tau ) - {\vec
\pi}_{12}(\tau )\Big)^2} }}}\Big)\nonumber \\
 &&{\frac{{{\vec \kappa}_{12}(\tau ) }}{m}}\, \cdot {\vec
A}_{\perp}(\tau ,{\vec \eta}_{12}(\tau )) -  \nonumber \\
 &&{}\nonumber \\
 &-& \Big({\frac{{\ Q_1}}{\sqrt{m_1^2\, c^2 +
\Big({\frac{{m_1}}{m}}\, {\vec \kappa}_{12}(\tau ) + {\vec
\pi}_{12}(\tau )\Big)^2}}} - {\frac{{\ Q_2}}{{\ \sqrt{m_2^2\, c^2 +
\Big({\frac{{m_2}}{m}}\, {\vec \kappa}_{12}(\tau ) - { \vec
\pi}_{12}(\tau )\Big)^2} }}}\Big)  \nonumber \\
 &&{\vec \pi}_{12}(\tau )\, \cdot {\vec A}_{\perp}(\tau ,{\vec \eta}
_{12}(\tau )) -  \nonumber \\
 \end{eqnarray*}

\bea
 &-& {\frac{{\mu}}{m}}\, \Big({\frac{{\ Q_1}}{\sqrt{m_1^2\, c^2 +
\Big({\frac{ {m_1}}{m}}\, {\vec \kappa}_{12}(\tau ) + {\vec
\pi}_{12}(\tau )\Big)^2}}} - { \frac{{\ Q_2}}{{\ \sqrt{m_2^2\, c^2 +
\Big({\frac{{m_2}}{m}}\, {\vec \kappa}
_{12}(\tau ) - {\vec \pi}_{12}(\tau )\Big)^2} }}}\Big)  \nonumber \\
&& {\vec \kappa}_{12} \cdot \Big({\vec \rho}_{12}(\tau ) \cdot
{\frac{{\partial}}{{\partial\, {\vec \eta}_{12}}}}\Big)\,{\vec
A}_{\perp}(\tau , {\vec \eta}_{12}(\tau )) -  \nonumber \\
 &&{}\nonumber \\
&-& \Big({\frac{{m_2\, Q_1}}{\sqrt{m_1^2\, c^2 +
\Big({\frac{{m_1}}{m}}\, { \vec \kappa}_{12}(\tau ) + {\vec
\pi}_{12}(\tau )\Big)^2}}} + {\frac{{m_1\, Q_2}}{{\ \sqrt{m_2^2\,
c^2 + \Big({\frac{{m_2}}{m}}\, {\vec \kappa}
_{12}(\tau ) - {\vec \pi}_{12}(\tau )\Big)^2} }}}\Big)  \nonumber \\
&& {\frac{{{\vec \pi}_{12}(\tau )}}{m}} \cdot \Big({\vec
\rho}_{12}(\tau ) \cdot {\frac{{\partial}}{{\partial\, {\vec
\eta}_{12}}}}\Big)\,{\vec A}
_{\perp}(\tau , {\vec \eta}_{12}(\tau )) +  \nonumber \\
 &&{}\nonumber \\
 &+& {\frac{{Q_1\, Q_2}}{{4\pi\, |{\vec \rho}_{12}(\tau )|}}} +
{\frac{1}{2}} \, \int d^{3}\sigma \, [{\vec{\pi }}_{\perp }^{2} +
{\vec{B}}^{2}](\tau , \vec{\sigma}) +
  O([{\vec \rho}_{12} \cdot {\vec \partial}_{{\vec \eta}_{12}}]^2 {\vec A}_{\perp}).
 \label{6.5}
  \eea

\medskip

Its semi-relativistic limit with the restriction $e = Q_1 \approx
 - Q_2$ is

\bea
 {\cal E}_{(int)} &\rightarrow_{c \rightarrow \infty, e = Q_1 \approx
 Q_2}&  mc^2 +{\frac{{{\vec \kappa}
^2_{12}(\tau )}}{{2m}}} + {\frac{{{\vec \pi}_{12}^2(\tau
)}}{{2\mu}}} - {{e^2}\over {4\pi\, |{\vec \rho}_{12}(\tau )|}}
 -\nonumber \\
 &-& {\frac{e}{c}}\, {\frac{ {{\vec \pi}_{12}(\tau )}}{{\mu}}} \cdot
{\vec A}_{\perp}(\tau ,{\vec \eta}
_{12}(\tau )) -  \nonumber \\
 &-& {\frac{e}{c}}\, {\frac{{{\vec \kappa}_{12}(\tau )}}{m}}\, \cdot
\Big({ \vec \rho}_{12}(\tau ) \cdot {\frac{{\partial}}{{\partial\,
{\vec \eta}_{12}}
}}\Big)\,{\vec A}_{\perp}(\tau , {\vec \eta}_{12}(\tau )) -  \nonumber \\
&-& {\frac{e}{c}}\, {\frac{{m_2 - m_1}}{m}}\, {\frac{{{\vec
\pi}_{12}(\tau )} }{{\mu}}}\, \cdot \Big({\vec \rho}_{12}(\tau )
\cdot {\frac{{\partial}}{{
\partial\, {\vec \eta}_{12}}}}\Big)\,{\vec A}_{\perp}(\tau , {\vec \eta}
_{12}(\tau )) + \nonumber \\
 &+& O(c^{-2}) + O([{\vec \rho}_{12} \cdot {\vec \partial}_{{\vec
 \eta}_{12}}]^2 {\vec A}_{\perp}).  \nonumber \\
&&{}
 \label{6.6}
 \eea

In this way we recover a semi-classical version of Eqs.(L3), (L4)
and (14.34) of Ref. \cite{7}, without the $e^2$ terms corresponding
to $Q_i^2 = 0$.\bigskip

If we could evaluate the internal 3-center of mass ${\vec q}_+$ (see
Subsection IIA) as a function of ${\vec \eta}_{12}$, ${\vec
\kappa}_{12} = {\vec {\kappa}}_+$, ${\vec \rho}_{12}$, ${\vec
\pi}_{12}$, then the results of Section IV would allow us to write
the internal Poincare' generators (\ref{4.1}) in the following form
(see Subsection IIA for the relative variables ${\vec \rho}_q$,
${\vec \pi}_q$)

\begin{eqnarray*}
 \mathcal{E}_{(int)} &=& M\, c^2 \approx c\, \sqrt{\mathcal{M}_o^2\,
c^2 + { \vec \kappa}_{+}^2} + \Big(function\, of\, {\vec q}_+,
{\vec \kappa}_+, {\vec \rho}_q, {\vec \pi}_q\Big) +\nonumber \\
 &+& {\cal P}^{\tau}_{em} + O({\vec \rho}_{12}{}^2),
\end{eqnarray*}

\begin{eqnarray*}
 \mathcal{\vec{P}}_{(int)} &=& {\vec \kappa}_+ + {\vec {\cal P}}_{em}
\approx 0,  \nonumber \\
 &&{}  \nonumber \\
 \mathcal{J}_{(int)}^{r} &=& {\vec \eta}_{12} \times {\vec \kappa}_{12} + { \vec \rho}_{12}
\times {\vec \pi}_{12} + {\vec {\cal J}}_{em} =\nonumber \\
 &=& {\vec q}_+ \times {\vec \kappa}_+ + {\vec \rho}_q \times {\vec
 \pi}_q + {\vec {\cal J}}_{em},
\end{eqnarray*}

\begin{eqnarray}
 {\vec {\cal K}}_{(int)} &=& {\vec {\cal K}}[{\vec q}_+, {\vec \kappa}_+, {\vec \rho}_q,
 {\vec \pi}_q, {\vec A}_{\perp}, {\vec \pi}_{\perp}] \approx 0.
  \label{6.7}
\end{eqnarray}

\noindent ${\vec {\cal K}}_{(int)} \approx 0$ determines ${\vec q}_+
= {\hat {\vec \eta}} + ..$ as it was done in Section IV for ${\hat
{\vec \eta}}$. In this way we would get $\mathcal{E}_{(int)}$ as a
function only of the relative variables ${\vec \rho}_q$, ${\vec
\pi}_q$ in a form useful for the electric dipole approximation.

\bigskip

While the external decoupled (canonical non-covariant) 4-center of
mass ${\tilde x}^{\mu}$, $P^{\mu}$, has an effective mass $M =
\mathcal{E}_{(int)} / c^2$ and a spin ${\vec {\bar S}} = {\vec
{\mathcal{J}}} _{(int)}$, the particle subsystem (the atom) has the
effective mass ${\cal M}_o$ of Eq.(\ref{6.7}), a position ${\vec
q}_+$, a 3-momentum ${\vec \kappa}_+$ and a spin given by the matter
part of ${\vec {\cal J}}_{(int)}$. These quantities replace $M_c$,
${\vec \zeta}_c$, ${\vec {\cal P}}_c$ and ${\vec {\cal J}}_c$ of the
pole-dipole approximation to the multipolar expansion of the
previous Section and give a {\it canonical pole-dipole} description
of the atom. We have the following replacements\medskip

${\vec \zeta}_c(\tau ) \mapsto\, {\vec \eta}_{12}(\tau )\,\, or\,\,
{\vec q} _+(\tau )$,\hfill\break

${\vec {\mathcal{P}}}_c(\tau )\, \mapsto\,\, {\vec \kappa}_{12}(\tau
)$, \hfill\break

$M_c(\tau )\, \mapsto\,\, \mathcal{M}_o(relative\, variables) +
Coulomb\, potential\, + interaction\, with\, the\,
electro-magnetic\, field$ ,\hfill\break

${\vec {\mathcal{J}}}_c(\tau )\, \mapsto\,\, matter\, part\, of\,
{\vec { \mathcal{J}}}_{(int)}$.\hfill\break

\medskip

The results of Section IV allow us to eliminate the overall internal
center of mass and introduce a dependence on the variables ${\hat
{\vec \pi}}_{(12)3}\, {\buildrel \circ \over {=}} \, constant\, of\,
motion$  and ${\hat {\vec \rho}} _{(12)3}$, describing the relative
motion of the atom with respect to the collective variables of the
electro-magnetic field configuration.

\bigskip

Usually atoms are described not as the quantization of extended open
subsystems with an effective 4-center of motion as in the previous
Section, but as point-like systems ({\it monopole} approximation:
$M_c\, \mapsto\, m$, ${\vec \zeta} _c(\tau )\, \mapsto\, {\vec
\eta}_{12}$, ${\vec { \mathcal{P}}}_c(\tau )\, \mapsto\, {\vec
\kappa}_{12}$) with some additional structure (higher multipoles)
describing the energy levels and the interaction with an
electro-magnetic field. After quantization this point-like
description of positive-energy atoms leads to the effective
Schroedinger equation of Ref.\cite{14} used to describe the external
propagation  (its de Broglie wave) in atom interferometry.
Conceptually this effective Schroedinger equation should be derived
by studying the positive-energy sector of solution of some wave
equation with a fixed mass and a spin (spin dipole) \footnote{For
instance the Dirac equation is used for the coupling to external
gravitational fields and the study of gravito-inertial effects
\cite{14}.}, which couples to the magnetic field. As shown in
Ref.\cite{16}, the description of charged positive-energy spinning
particles in the rest-frame instant form can be made by using
Grassmann variables for the spin.\medskip

In this approximation with a fixed mass one looses all the internal
structure of the atom, described by its energy levels. To remedy it
one adds a finite-level structure to the point particle at the
quantum level: this allows to consider more realistic approximations
for the coupling to electric fields in the electric dipole
approximation. The simplest model is the two-level atom
approximation \cite{7,17}.\medskip

As it will be shown in paper III, we can define a system in the
rest-frame instant form, which after quantization leads to a
two-level atom whose electric dipole interacts with the electric
field after the transition to the electric dipole representation.
There will be extra Grassmann degrees of freedom for the description
of the two levels.\medskip

Therefore, we must now find the relativistic generalization of the
electric dipole representation in the limit of the dipole
approximation.

\vfill\eject

\section{The Relativistic Electric Dipole Representation}

In atomic physics the electric dipole approximation suggested the
introduction of the {\it electric dipole representation}, where the
interaction term ${\vec A}_{\perp} \cdot {\vec \kappa}_i$ is
replaced with the interaction of the electric field with   the
electric dipole $\vec d$ of Eq.(\ref{6.2}), $\vec d \cdot {\vec
\pi}_{\perp}$. In this way there is no explicit dependence on the
transverse electro-magnetic potential: only the transverse electric
and magnetic fields appear.\bigskip

This representation is a particular case of an equivalent
formulation of electro-dynamics, as explained in Chapter IV of
Ref\cite{9}. Equivalent formulations are obtained\medskip

a) by a change of the gauge of the electro-magnetic field (here we
are using the radiation gauge, a special case of Coulomb
gauge);\medskip

b) by a unitary transformation $e^{i\, \hat S}$ corresponding to a
classical canonical transformation whose generating function
$S$\medskip

$b_1$) is determined by a change of the particle Lagrangian by means
of a total time derivative ${{dS}\over {dt}}$, with the
electro-magnetic field considered as an external field with a given
time dependence; the electric dipole representation is obtained with
the {\it G$\o$ppert-Mayer unitary transformation} (see p.635 of the
Appendix of Ref.\cite{8} and pp. 266-275 of Chapter IV of
Ref.\cite{9}), which is useful when $\sum_i\, Q_i = 0$ and whose
classical generating function is $S_o = {1\over c}\, \vec d(\tau )
\cdot {\vec A}_{\perp}(\tau ,{\vec \eta}_{12}(\tau ))$ with $\vec
d(\tau )$ of Eq.(\ref{6.2}) (use is done of the dipole
approximation) \footnote{Strictly speaking the Appendix of
Ref.\cite{8} (see Eqs.(4) and subsequent one) makes the separation
of the matter part from the radiation field in the Coulomb gauge and
in the unitary transformation uses only the radiation electric
field. Therefore our canonical transformation of I is the analogous
separation in the radiation gauge (we get Darwin as an extra
bonus!). The final Hamiltonian of Eq.(\ref{4.13}) has the
interaction reintroduced by the vanishing of the internal boosts
(for the $1/c$ expansion see Eq.(\ref{4.5})). As a consequence a
canonical transformation with generating function $S^{^{\prime}}_o =
{1\over c}\, {\hat {\vec d}}(\tau ) \cdot {\vec A}_{\perp rad}(\tau
,{\vec \eta}_{12}(\tau ))$ should now work like it happened in the
Appendix of Ref.\cite{8}.} \footnote{A generalization of the
G$\o$ppert-Mayer transformation is given in Ref.\cite{9}. Its
generating function is $S_Z = { \frac{1}{c}}\, \vec d(\tau ) \cdot
\int^1_o d\lambda\, {\vec A}_{\perp}(\tau , \lambda\, {\vec
\eta}_1(\tau ) + (1 - \lambda )\, {\vec \eta}_2(\tau )) = {
\frac{1}{c}}\, \vec d(\tau ) \cdot \int^1_o d\lambda\, {\vec
A}_{\perp}(\tau ,{\vec \eta}_{12}(\tau ) + {\frac{{\lambda\, m_2 -
(1 - \lambda )\, m_1}}{m}} \, {\vec \rho}_{12}(\tau )) =
{\frac{1}{c}}\, \vec d(\tau ) \cdot \int^1_o d\lambda\, \Big[{\vec
A}_{\perp}(\tau , {\vec \eta}_{12}(\tau )) - ({\frac{{ m_1}}{m}} -
\lambda )\, \Big({\vec \rho}_{12}(\tau ) \cdot {\frac{{\partial}
}{{\partial\, {\vec \eta}_{12}}}}\Big) {\vec A}_{\perp}(\tau ,{\vec
\eta} _{12}(\tau )) + ..\Big]$, with the integral taken along the
straight-line joining the two charges. The first term is the
G$\o$ppert-Mayer generating function.};\medskip

$b_2$) is not determined by a change of Lagrangian but still with
the electro-magnetic field considered as an external prescribed
field \footnote{See the {\it Hennenberger unitary transformation},
given at pp.275-279 of Ref.\cite{9}, changing the positions instead
of the momenta: the generating function is $\tilde S =
{\frac{1}{c}}\, \sum_i\, {\frac{{Q_i}}{{m_i}}}\, { \vec
\kappa}_i(\tau ) \cdot \vec Z(\tau ,\vec 0)$ with $\vec Z(\tau ,\vec
\eta ) = - \int^{\tau}\, d\tau_1\, {\vec A}_{external}(\tau_1, \vec
\eta )$. When the electromagnetic field is dynamical and not
external, it is considered a quantized radiation field and the
Hennenberg unitary transformation leads to the Pauli-Fierz-Kramers
transformation.};\medskip

$b_3$) is connected to a change of Lagrangian with a dynamical
electro-magnetic field but with the dipole approximation replaced by
a description of the localized system of charges by polarization and
magnetization densities \footnote{See the {\it Power-Zienau-Woolley
transformation} of pp. 280-297 of Ref.\cite{9} generalizing the
G$\o$ppert-Meyer one.}.

\bigskip

In Appendix L of Ref.\cite{7}, the semi-relativistic electric dipole
representation is obtained by starting from the particle Hamiltonian
of a neutral 2-particle system in the dipole approximation (the
electro-magnetic field is considered external), by evaluating the
corresponding Lagrangian, by eliminating a suitable total time
derivative ${{dS}\over {dt}}$ and by reverting to the Hamiltonian.

\subsection{The Semi-Relativistic Electric Dipole Representation
with Grassmann-Valued Charges}

In Subsection 1 of Appendix B the results of Ref.\cite{7} on the
electric dipole representation are revisited  in the case of
Grassmann-valued electric charges ($Q_i^2 = 0$, $Q_1\, Q_2 \not= 0$)
in the radiation gauge for the electro-magnetic field, which is
treated as an external field, by starting from  the Hamiltonian
(${{d\, F}\over {dt}} = c\, {{d\, F}\over {d\tau}} = \{ F, H\, c\}$)

\beq
 H\, c = \sum_{i=1}^2\, {{({\vec \kappa}_i(\tau ) - {{Q_i}\over c}\, {\vec
A}_{\perp}(\tau ,{\vec \eta}_i(\tau )))^2}\over {2 m_i}} =
  \sum_{i=1}^2\, {{{\vec \kappa}^2_i(\tau )}\over {2m_i}} -
 \sum_{i=1}^2\, {{Q_i}\over c}\, {{{\vec \kappa}_i(\tau )}\over {m_i}}\,
 \cdot {\vec A}_{\perp}(\tau ,{\vec \eta}_i(\tau )),
 \label{7.1}
 \eeq

\medskip

As shown in Appendix B, the final Hamiltonian (\ref{b6}) becomes the
R$\o$ntgen Hamiltonian (\ref{7.2}) of the electric dipole
representation, given in Eq.(14.37) of Ref.\cite{7}, for $2\, {\cal
Q} = Q_1 + Q_2 \approx 0$, $e = Q_1$

\bea
 H_1\, c &=& {\frac{{{\vec {\bar \kappa}}_{12}^2(\tau )}}{{2m}}} + {\frac{{{\vec
{\bar \pi}}_{12}^2(\tau )}}{{2\mu}}} + V({\vec \rho}_{12}(\tau )) +  \nonumber \\
&+& {\frac{e}{{mc}}}\, {\vec {\bar \kappa}}_{12}(\tau ) \cdot
\Big[{\vec \rho} _{12}(\tau ) \times \vec B(\tau ,{\vec
\eta}_{12}(\tau ))\Big] + \nonumber \\
&+& {\frac{e}{{2\,\mu\, c}}}\, {\frac{{m_2 - m_1}}{m}}\, {\vec {\bar
\pi}}_{12}(\tau ) \cdot \Big[{\vec \rho}_{12}(\tau ) \times \vec
B(\tau ,{\vec \eta}_{12}(\tau ))\Big] -  \nonumber \\
&-& {\frac{e}{c}}\, {\vec \rho}_{12}(\tau ) \cdot {\vec
\pi}_{\perp}(\tau ,{ \vec \eta}_{12}(\tau )) - {\frac{e}{{2c}}}\,
{\frac{{m_2 - m_1}}{m}}\, \Big({ \vec \rho}_{12}(\tau ) \cdot
{\frac{{\partial}}{{\partial\, {\vec \eta}_{12}} }}\Big)\, {\vec
\rho}_{12}(\tau ) \cdot {\vec \pi}_{\perp}(\tau ,{\vec \eta}
_{12}(\tau )) -  \nonumber \\
 &-& {{e^2}\over c}\, {{3 \mu}\over {4 m^2}}\, \Big[{\vec \rho}_{12}(\tau ) \times \vec
B(\tau ,{\vec \eta}_{12}(\tau ))\Big]^2.\nonumber \\
 &&{}
 \label{7.2}
 \eea

\medskip

\noindent Note that the last term has not the coefficient
${{e^2}\over {8 \mu\, c}}$ of Eq.(15.37) or (L.14) of Ref.\cite{7},
because in our calculation the terms $Q_i^2 = 0$ are missing before
imposing the condition $Q_1 + Q_2 \approx 0$.

\medskip

The resulting generating function $S$ of Eq.(\ref{b3}) is an
extension to the next order of $S_o$, the generating function of the
classical G$\o$ppert-Mayer transformation.

\subsection{A Generating Function from a Relativistic Lagrangian with External
Electro-Magnetic Field.}

Let us now try to define a relativistic electric dipole
representation in the framework of the dipole approximation with the
same method but starting from a particle Hamiltonian given by
${1\over c}\, \Big({\cal E}_{(int)} - {\cal E}_{em}\Big)$ in accord
with Eq.(\ref{4.1}). In Subsection 2 of Appendix B we determine a
relativistic Lagrangian for the particles in the dipole
approximation treating the electro-magnetic field as an external
one, following the same method as in semi-relativistic atomic
physics \cite{7}. This identifies the generator of a canonical
transformation, which will be used to find an electric dipole
representation.
\bigskip

If we use  the dipole approximation for neutral systems, ${\cal Q}
\approx 0$,  the emerging total time derivative identifies the
following generating function $\tilde S$, given in Eq.(\ref{b14})
and coinciding with $S$ of Eq.(\ref{b3}) at the lowest order,
(Eqs.(\ref{2.3}), (\ref{6.2}) and (\ref{6.4}) are used)

\bea
 {\tilde S}{|}_{{\cal Q} \approx 0} &=& {{Q_1 - Q_2}\over {2\, c}}\, {\vec \rho}_{12}(\tau )
 \cdot \Big[ {\vec A}_{\perp}(\tau ,{\vec \eta}_{12}(\tau )) + {{m_2 - m_1}\over {2\, m}}\,
  {{\partial}\over {\partial\, {\vec \eta}_{12}}}\,
 \Big({\vec \rho}_{12}(\tau ) \cdot {\vec A}_{\perp}(\tau
 ,{\vec \eta}_{12}(\tau ))\Big)\Big] +\nonumber \\
 &+& O([{\vec \rho}_{12} \cdot {\vec \partial}_{{\vec
 \eta}_{12}}]^2 {\vec A}_{\perp}) = S{|}_{{\cal Q} \approx 0} +
 O([{\vec \rho}_{12} \cdot {\vec \partial}_{{\vec
 \eta}_{12}}]^2 {\vec A}_{\perp}),\nonumber \\
 &&{}\nonumber \\
 &&{}\nonumber \\
 S{|}_{{\cal Q} \approx 0} &=&  {1\over c}\, \vec d(\tau ) \cdot \Big[
 {\vec A}_{\perp}(\tau ,{\vec \eta}_{12}(\tau )) + {{m_2 - m_1}\over {2\, m}}\,
  {{\partial}\over {\partial\, {\vec \eta}_{12}}}\,
 \Big({\vec \rho}_{12}(\tau ) \cdot {\vec A}_{\perp}(\tau
 ,{\vec \eta}_{12}(\tau ))\Big)\Big] =\nonumber \\
 &=& {\tilde S}_1 + O([{\vec \rho}_{12} \cdot {\vec \partial}_{{\vec
 \eta}_{12}}]^2 {\vec A}_{\perp}),\nonumber \\
 &&{}\nonumber \\
 {\tilde S}_1 &=&{1\over {2\, c}}\, \vec d(\tau ) \cdot
 \Big[{\vec A}_{\perp}(\tau ,{\vec \eta}_1(\tau ))
 + {\vec A}_{\perp}(\tau ,{\vec \eta}_2(\tau ))\Big] +
 O([{\vec \rho}_{12} \cdot {\vec \partial}_{{\vec
 \eta}_{12}}]^2 {\vec A}_{\perp}).\nonumber \\
 &&{}
 \label{7.3}
 \eea

\noindent  We also find that ${\tilde S}{|}_{{\cal Q} \approx 0}$
and $S{|}_{{\cal Q} \approx 0}$ differ by terms $O([{\vec \rho}_{12}
\cdot {\vec \partial}_{{\vec \eta}_{12}}]^2 {\vec A}_{\perp})$ from
a generating function ${\tilde S}_1 = {1\over {2c}}\, \vec d(\tau )
\cdot \sum_{i=1}^2\, { \vec A}_{\perp}(\tau , {\vec \eta}_i(\tau
))$, which could have been defined independently by using  the
dipole approximation.

\medskip

The inverse Legendre transformation from the resulting Lagrangian
(\ref{b15}) is the Hamiltonian of Eq.(\ref{b20}) with the ${\vec
{\cal A}}_i(\tau )$ given in Eq.(\ref{b16}). If we add to it ${\cal
E}_{em}$, we get the Hamiltonian $M_{e.d.r.}c$ in the dipole
approximation for the electric dipole representation replacing
${\cal P}^{\tau}_{(int)} = M\, c$ of Eq.(\ref{4.1}) of the standard
representation (${\vec \kappa}_i^{'}(\tau )$ are the new momenta of
Eqs.(\ref{b17}))

\begin{eqnarray*}
M_{e.d.r}\, c &=& {1\over c}\, {\cal E}_{e.d.r.} = \sum_{i=1}^2\,
\Big(\sqrt{m_i^{2}\,c^{2} + {\vec{\kappa}}_i{}^{{'}\, 2}(\tau )} -
\frac{ {\vec{ \kappa}}_i^{'}(\tau ) \cdot \mathcal{\vec{A}}_i(\tau
)}{\sqrt{m_i^{2}\,c^{2} + {\vec{ \kappa}}_i{}^{{'}\, 2}(\tau
)}}\Big) + \frac{Q_{1}Q_{2}}{4\pi\, c \,|{\vec{\rho}}_{12}(\tau )|} -\nonumber \\
 &-&{{Q_1 - Q_2}\over {2 c}}\, \Big[
{\vec{\rho}}_{12}(\tau ) \cdot {\vec \pi}_{\perp }(\tau
,{\vec{\eta}}_{12}(\tau )) - \frac{m_2 - m_1}{2m}\, {\vec
\rho}_{12}(\tau ) \cdot \Big({\vec \rho}_{12}(\tau ) \cdot
{{\partial}\over {\partial\, {\vec \eta}_{12}}}\,
 {\vec \pi}_{\perp }(\tau ,{\vec{\eta}}_{12}(\tau ))\Big)\Big]-  \nonumber \\
 &-&{{Q_1 + Q_2}\over {2 c}}\,
\Big[\frac{m_2 - m_1}{m}\, {\vec{\rho}}_{12}(\tau ) \cdot {\vec
\pi}_{\perp }(\tau , {\vec{\eta}}_{12}(\tau )) -\nonumber \\
 &-&\frac{m^{2} - 2 m_1\, m_2}{2m^{2}}\, {\vec{\rho}}_{12}(\tau ) \cdot
\Big({\vec{ \rho}}_{12}(\tau ) \cdot {{\partial}\over {\partial\,
{\vec \eta}_{12}}}\, {\vec \pi}_{\perp }(\tau
,{\vec{\eta}}_{12}(\tau
))\Big)\Big] + {1\over c}\, {\cal E}_{em}\nonumber \\
 &&{}\nonumber \\
 &\rightarrow_{{\cal Q} \approx 0}& \sum_{i=1}^2\, \Big(
 \sqrt{m_i^{2}\,c^{2} +
{\vec{\kappa}}_i{}^{{'}\, 2}(\tau )} - \frac{ {\vec{
\kappa}}^{'}_i(\tau ) \cdot \mathcal{\vec{A}}_i(\tau
)}{\sqrt{m_i^{2}\,c^{2} + {\vec{ \kappa}}_i{}^{{'}\, 2}(\tau
)}}\Big) +
\frac{Q_{1}Q_{2}}{4\pi\, c \,|{\vec{\rho}}_{12}(\tau )|} -\nonumber \\
 &-&{{Q_1 - Q_2}\over {2 c}}\, \Big[
{\vec{\rho}}_{12}(\tau ) \cdot {\vec \pi}_{\perp }(\tau
,{\vec{\eta}}_{12}(\tau )) - \frac{m_2 - m_1}{2m}\, {\vec
\rho}_{12}(\tau ) \cdot \Big({\vec \rho}_{12}(\tau ) \cdot
{{\partial}\over {\partial\, {\vec \eta}_{12}}}\,
 {\vec \pi}_{\perp }(\tau ,{\vec{\eta}}_{12}(\tau ))\Big)\Big] +  \nonumber \\
 &+& {1\over c}\,{\cal E}_{em},
 \end{eqnarray*}

 \bea
 \mathcal{\vec{A}}_i(\tau )
 &\rightarrow_{{\cal Q} \approx 0}& - {Q\over {2c}}\,
 {\vec{\rho}}_{12}(\tau ) \times \vec{B}(\tau ,{\vec \eta}_{12} (\tau
 )).
 \label{7.4}
\eea

This Hamiltonian becomes the R$\o$ntgen Hamiltonian $H_1$ of
Eq.(\ref{7.2}) in the semi-relativistic limit. However it has not
the form expected in the electric dipole representation.

\subsection{The Relativistic Electric Dipole Representations Induced
by the Generating Functions $S$ and ${\tilde S}_1$.}

To try to find a relativistic electric dipole representation, let us
study the classical {\it point} canonical transformation generated
by $T_S = e^{ \{ ., S\}} $ with the generating function $S$ of
Eqs.(\ref{b3})-(\ref{7.3}) (coinciding with $\tilde S$ in the dipole
approximation) \footnote{Since $S = Q_1\, {\cal S}_1 + Q_2\, {\cal
S}_2$ depends only on the coordinates, we get $ \tilde F = T_{S}\, F
= F + Q_1\, \{ F, {\cal S}_1\} + Q_2\, \{ F, {\cal S}_2\} + Q_1\,
Q_2\, [\{ \{ F, {\cal S}_1\}, {\cal S}_2 \} + \{ \{ F, {\cal S}_2\},
{\cal S}_1\}] = F + Q_1\, \{ F, {\cal S}_1\} + Q_2\, \{ F, {\cal
S}_2\}$, since $\{ \{ F, {\cal S}_a\}, {\cal S}_b\} = 0$.} by
considering also the electro-magnetic field dynamical
(Eq.(\ref{7.4}) was obtained by considering it as an external
field). In the canonical basis ${\vec \eta}_{12}(\tau )$, ${\vec
\kappa}_{12}(\tau )$, ${\vec \rho}_{12}(\tau )$, ${\vec
\pi}_{12}(\tau )$, ${\vec A}_{\perp}(\tau ,\vec \sigma)$, ${\vec
\pi}_{\perp}(\tau ,\vec \sigma )$, the variables ${\vec A}_{\perp}$,
${\vec \eta} _{12}$, ${\vec \rho}_{12}$  remain unchanged. Instead
for the momenta and the internal energy of Eq.(\ref{4.1}) we get

\begin{eqnarray*}
 {\vec \kappa}_{12}(\tau ) &{\buildrel T_S\over \rightarrow}& {\vec
 \kappa}_{12}^{'}(\tau ) = {\vec \kappa}_{12}(\tau ) -
  {{Q_1 - Q_2}\over {2c}}\, {{\partial}\over {\partial\, {\vec \eta}_{12}}}\,
 \Big[{\vec \rho}_{12}(\tau ) \cdot {\vec A}_{\perp}(\tau ,{\vec \eta}_{12}(\tau ))
 +\nonumber \\
 &+& {{m_2 - m_1}\over {2m}}\, \Big({\vec \rho}_{12}(\tau ) \cdot {{\partial}\over
 {\partial\, {\vec \eta}_{12}}}\Big)\, \Big({\vec \rho}_{12}(\tau ) \cdot
 {\vec A}_{\perp}(\tau ,{\vec \eta}_{12}(\tau ))\Big)\Big],\nonumber \\
 &&{}\nonumber \\
 {\vec \pi}_{12}(\tau ) &{\buildrel T_S\over \rightarrow}& {\vec
 \pi}_{12}^{'}(\tau ) = {\vec \pi}_{12}(\tau ) -
  {{Q_1 - Q_2}\over {2c}}\, \Big[{\vec A}_{\perp}(\tau ,{\vec \eta}_{12}(\tau ))
 +\nonumber \\
 &+& {{m_2 - m_1}\over {2m}}\, \Big(\Big({\vec \rho}_{12}(\tau ) \cdot {{\partial}\over
 {\partial\, {\vec \eta}_{12}}}\Big)\, {\vec A}_{\perp}(\tau ,{\vec \eta}_{12}(\tau ))
 + {{\partial}\over {\partial\, {\vec \eta}_{12}}}\, \Big({\vec \rho}_{12}(\tau ) \cdot
 {\vec A}_{\perp}(\tau ,{\vec \eta}_{12}(\tau ))\Big)\Big)\Big],\nonumber \\
 &&{}\nonumber \\
 &&{\vec \kappa}_i^{'}(\tau ) = {{m_i}\over m}\, {\vec
 \kappa}_{12}^{'}(\tau ) + (-)^{i+1}\, {\vec \pi}_{12}^{'}(\tau ),
 \end{eqnarray*}

\bea
 \pi^r_{\perp}(\tau ,\vec \sigma ) &{\buildrel T_S\over \rightarrow}&
 \pi^{{'}\, r}_{\perp}(\tau ,\vec \sigma ) = \pi^r_{\perp}(\tau ,\vec \sigma )
 + {{Q_1 - Q_2}\over {2}}\, \rho^s_{12}(\tau )\,
 P^{sr}_{\perp}(\vec \sigma )\, \Big[\delta^3(\vec \sigma - {\vec \eta}_{12}(\tau ))
 +\nonumber \\
 &+& {{m_2 - m_1}\over {2m}}\, \Big({\vec \rho}_{12}(\tau ) \cdot {{\partial}\over
 {\partial\, {\vec \eta}_{12}}}\Big)\, \delta^3(\vec \sigma - {\vec \eta}_{12}(\tau ))\Big].
 \label{7.5}
 \eea
\bigskip

If we apply $T_S = e^{\{ ., S\}}$ to ${\cal E}_{(int)}$ of
Eq.(\ref{4.1}) we get its expression in the new canonical basis
${\vec \eta}_{12}(\tau )$, ${\vec \kappa}^{'}_{12}(\tau )$, ${\vec
\rho}_{12}(\tau )$, ${\vec \pi}^{'}_{12}(\tau )$, ${\vec
A}_{\perp}(\tau ,\vec \sigma )$, ${\vec \pi}^{'}_{\perp}(\tau ,\vec
\sigma )$

\begin{eqnarray*}
 {1\over c}\, {\cal E}^{'}_{(int)} &=& M^{'}\, c = \sum_{i=1}^2\,
 \sqrt{m_i^2\, c^2 + {\vec \kappa}_i{}^{{'}\, 2}(\tau )} + {{Q_1\,
 Q_2}\over {4\pi\, c\, |{\vec \rho}_{12}(\tau )|}} +\nonumber \\
 &+& \sum_{i=1}^2\, (-)^{i+1}\, {{Q_i}\over c}\, {{{\vec \kappa}^{'}_i(\tau )
 \cdot \Big[{\vec F}_i(\tau ) - (-)^{i+1}\, {\vec A}_{\perp}(\tau
 ,{\vec \eta}_i(\tau ))\Big]}\over {\sqrt{m_i^2\, c^2 +
 {\vec \kappa}_i{}^{{'}\, 2}(\tau )}}} -\nonumber \\
 &-& {{Q_1 + Q_2}\over {2 c}}\, \sum_{i=1}^2\, (-)^{i+1}\, {{{\vec
 \kappa}_i^{'}(\tau ) \cdot {\vec F}_i(\tau )}\over
 {\sqrt{m_i^2\, c^2 + {\vec \kappa}_i{}^{{'}\, 2}(\tau )}}}
 -\nonumber \\
 &-& {{Q_1\, Q_2}\over {2 c^2}}\, \sum_{i=1}^2\, {{{1\over 2}\, {\vec F}_i(\tau ) -
 {\vec A}_{\perp}(\tau ,{\vec \eta}_i(\tau ))}\over {\sqrt{m_i^2\, c^2
 + {\vec \kappa}_i{}^{{'}\, 2}(\tau )}}}\, \cdot \Big[{\vec F}_i(\tau ) +
 {\vec \kappa}_i^{'}(\tau )\, {{{\vec \kappa}_i^{'}(\tau ) \cdot {\vec F}_i(\tau )}
 \over {m_i^2\, c^2 + {\vec \kappa}_i{}^{{'}\, 2}(\tau )}}\Big] -\nonumber \\
 &-&{{Q_1 - Q_2}\over {2 c}}\, \Big[{\vec \rho}_{12}(\tau ) \cdot
 {\vec \pi}_{\perp}^{'}(\tau ,{\vec \eta}_{12}(\tau )) + {{m_2 - m_1}\over {2m}}\,
 \Big({\vec \rho}_{12}(\tau ) \cdot {{\partial}\over
 {\partial\, {\vec \eta}_{12}}}\Big)\, \Big({\vec \rho}_{12}(\tau ) \cdot
 {\vec \pi}_{\perp}^{'}(\tau ,{\vec \eta}_{12}(\tau ))\Big)\Big] -\nonumber \\
 &-& {{Q_1\, Q_2}\over {2\, c}}\, I(\tau ) + {1\over {2c}}\, \int
 d ^3\sigma\, \Big[{\vec \pi}_{\perp}{}^{{'}\, 2} + {\vec B}^2\Big](\tau ,\vec \sigma )
 \end{eqnarray*}

\bea
 {\vec F}_i(\tau) &=& {{m_i}\over m}\, {{\partial}\over {\partial\, {\vec \eta}_{12}}}\,
 \Big[{\vec \rho}_{12}(\tau ) \cdot {\vec A}_{\perp}(\tau ,{\vec \eta}_{12}(\tau ))
 +\nonumber \\
 &+& {{m_2 - m_1}\over {2m}}\, \Big({\vec \rho}_{12}(\tau ) \cdot {{\partial}\over
 {\partial\, {\vec \eta}_{12}}}\Big)\, \Big({\vec \rho}_{12}(\tau ) \cdot
 {\vec A}_{\perp}(\tau ,{\vec \eta}_{12}(\tau ))\Big)\Big] +\nonumber \\
 &+& (-)^{i+1}\, \Big[{\vec A}_{\perp}(\tau ,{\vec \eta}_{12}(\tau ))
 +\nonumber \\
 &+& {{m_2 - m_1}\over {2m}}\, \Big(\Big({\vec \rho}_{12}(\tau ) \cdot {{\partial}\over
 {\partial\, {\vec \eta}_{12}}}\Big)\, {\vec A}_{\perp}(\tau ,{\vec \eta}_{12}(\tau ))
 + {{\partial}\over {\partial\, {\vec \eta}_{12}}}\, \Big({\vec \rho}_{12}(\tau ) \cdot
 {\vec A}_{\perp}(\tau ,{\vec \eta}_{12}(\tau
 ))\Big)\Big)\Big],\nonumber \\
 &&{}\nonumber \\
 I(\tau ) &=& lim_{\vec \sigma \rightarrow {\vec \eta}_{12}(\tau
 )}\, \rho_{12}^r(\tau )\, \rho_{12}^s(\tau )\, P^{rs}_{\perp}(\vec
 \sigma )\, \Big[\delta^3(\vec \sigma - {\vec \eta}_{12}(\tau ))
 +\nonumber \\
 &+& {{m_2 - m_1}\over m}\, \Big({\vec \rho}_{12}(\tau ) \cdot {{\partial}\over
 {\partial\, {\vec \eta}_{12}}}\Big)\, \delta^3(\vec \sigma - {\vec \eta}_{12}(\tau ))
 +\nonumber \\
 &+& \Big({{m_2 - m_1}\over {2m}}\Big)^2\, \Big({\vec \rho}_{12}(\tau ) \cdot {{\partial}\over
 {\partial\, {\vec \eta}_{12}}}\Big)^2\, \delta^3(\vec \sigma - {\vec \eta}_{12}(\tau ))\Big].
 \label{7.6}
 \eea

\medskip

It can be checked that in the dipole approximation and in the
semi-relativistic limit one recovers the R$\o$ntgen Hamiltonian
$H_1$ (\ref{7.2}) of the semi-relativistic electric dipole
representation: ${\cal E}^{'}_{(int)} = M^{'}\, c^2 = m\, c^2 + H_1
+ {\cal E}^{'}_{em} + O(c^{-2})$.\medskip

However, at the order $O({{Q_1\, Q_2}\over {c}})$ a singular term
$I(\tau )$ appears. It comes from the ${\vec \pi}^2_{\perp}(\tau
,\vec \sigma )$ term of the original electro-magnetic energy
density. Notwithstanding the Grassmann regularization, we get this
diverging term as a consequence of the dipole approximation: both
${\vec \eta}_i(\tau )$ collapse in the collective variable ${\vec
\eta}_{12}(\tau)$ describing the motion of the electric dipole.
Moreover the transverse vector potential is still present in
Eq.(\ref{7.6}) like in Eq.(\ref{7.4}).
\bigskip

The same problem arises if we consider the point canonical
transformation $T_{{\tilde S}_1} = e^{\{ ., {\tilde S}_1\}}$
generated by the function ${\tilde S}_1 = {1\over {2c}}\, \vec
d(\tau ) \cdot \sum_i\, { \vec A}_{\perp}(\tau , {\vec \eta}_i(\tau
))$ defined in Eq.(\ref{7.3}) \footnote{In this case we have
${\tilde S}_1 = (Q_1 - Q_2)\, {\bar S}_1$ with ${\bar S}_1 = {1\over
{4c}}\, {\vec \rho}_{12}(\tau ) \cdot \sum_i\, { \vec
A}_{\perp}(\tau , {\vec \eta}_i(\tau ))$ depending only on the
coordinates. Therefore we get $\tilde F = T_{{\tilde S}_1}\, F = F +
(Q_1 - Q_2)\, \{ F, {\bar S}_1\} - Q_1\, Q_2\, \{ \{ F, {\bar
S}_1\}, {\bar S}_1\}$.}. Also in this case we get expressions of
type of Eq.(\ref{7.6}), which we omit, with a similar singular term.

\medskip

The use of $\tilde S$, instead of $S$ or ${\tilde S}_1$ does not
change the situation. Therefore it is not clear how to define a
relativistic electric dipole representation, free from singularities
and connected with the semi-relativistic one in the dipole
approximation, when the electro-magnetic field is dynamical as it
happens in the rest-frame instant form.

\subsection{A Relativistic Lagrangian with Dynamical Electro-Magnetic Field
and the Induced Relativistic Electric Dipole Representation}

However a relativistic representation in which the singular term is
replaced by a {\it contact interaction} at order $O(c^{-1})$ can be
defined by defining a Lagrangian for the isolated system "charged
particles plus the electro-magnetic field" on the instantaneous
Wigner 3-spaces. This is done in Subsection 3 of Appendix B. By
modifying this Lagrangian with a total time derivative, we can
identify a new generating function $S_2 = {1\over c}\,
\sum_{i=1}^2\, Q_i\, {\vec \eta}_i(\tau ) \cdot {\vec
A}_{\perp}(\tau ,{\vec \eta}_i(\tau ))$, whose dipole approximation
is given in Eq.(\ref{b28}), leading to this new representation.

\medskip

Again the variables ${\vec A}_{\perp}(\tau ,\vec \sigma )$, ${\vec
\eta} _{12}(\tau )$, ${\vec \rho}_{12}(\tau )$  remain unchanged
under the {\it point} canonical transformation generated by $T_{S_2}
= e^{\{ ., S_2\}}$ with generating function  $S_2$. Instead for the
momenta  we get \footnote{ We use $F^{sr} = - \epsilon^{sru}\, B^u$,
${\bar \kappa}^r_i\,\eta^s_i\, {\frac{{\partial\,
A^s_{\perp}}}{{\partial \eta^r_i}}} = {\bar \kappa}^r_i\, \eta^s_i\,
\Big(F^{sr} + {\frac{{\partial\, A^r_{\perp}}}{{\partial\,
\eta^s_i}}}\Big) = {\vec {\bar \kappa}} \cdot {\vec \eta}_i \times
\vec B + { \bar \kappa}^r_i\, \eta^s_i\, {\frac{{\partial
A^r_{\perp}}}{{\partial\, \eta^s_i}}} = - {\vec \eta}_i \times {\vec
{\bar \kappa}}_i \cdot \vec B + { \bar \kappa}^r_i\, \eta^s_i\,
{\frac{{\partial A^r_{\perp}}}{{\partial\, \eta^s_i}}}$.}

\bigskip

\bea
 \kappa^r_i(\tau ) &{\buildrel T_{S_2}\over \rightarrow}& {\bar \kappa}^r_i(\tau ) =
\kappa^r_i(\tau ) - {\frac{{Q_i}}{c}}\, \Big[A^r_{\perp}(\tau ,
{\vec \eta}_i(\tau )) + {\vec \eta}_i(\tau ) \cdot
{\frac{{\partial\, {\vec A}_{\perp}(\tau ,{\vec \eta}
_i(\tau ))}}{{\partial\, \eta^r_i}}}\Big],  \nonumber \\
 &&{}  \nonumber \\
 \pi^r_{\perp}(\tau ,\vec \sigma ) &{\buildrel T_{S_2}\over \rightarrow}& {\bar
\pi}^r_{\perp}(\tau ,\vec \sigma ) = \pi^r_{\perp}(\tau ,\vec \sigma
) +  \sum_{i=1}^2\, Q_i\, P^{rs}_{\perp}(\vec \sigma )\,
\delta^3(\vec \sigma - {
\vec \eta}_i(\tau ))\, \eta^s_i(\tau ),  \nonumber \\
&&{}  \nonumber \\
&& \kappa^r_i(\tau ) - {\frac{{Q_i}}{c}}\, A^r_{\perp}(\tau ,{\vec
\eta} _i(\tau ))\, {\buildrel T_{S_2}\over \rightarrow}\, {\bar
\kappa}^r_i(\tau ) + {\frac{{Q_i}}{c}}\, { \vec \eta}_i(\tau ) \cdot
{\frac{{\partial\, {\vec A}_{\perp}(\tau ,{\vec
\eta}_i(\tau ))}}{{\partial\, \eta^r_i}}},  \nonumber \\
&&{}  \nonumber \\
&&\int d^3\sigma\, {\vec \pi}^2_{\perp}(\tau ,\vec \sigma )\,
{\buildrel T_{S_2}\over \rightarrow}\, \int d^3\sigma\, {\vec {\bar
\pi}}^2_{\perp}(\tau ,\vec \sigma ) -  \nonumber \\
&-& 2\, \sum_{i=1}^2\, Q_i\, {\vec \eta}_i(\tau ) \cdot {\vec { \bar
\pi}}_{\perp}(\tau ,{\vec \eta}_i(\tau )) + 2\, Q_1\, Q_2\,
\eta^r_1(\tau )\, \eta^s_2(\tau )\, P^{rs}_{\perp}({\vec
\eta}_1(\tau ))\, \delta^3({\vec \eta}_1(\tau )
- {\vec \eta}_2(\tau )).\nonumber \\
 &&{}
 \label{7.8}
 \eea

The internal energy of Eq. (\ref{4.1}) is replaced by the following
expression

\begin{eqnarray*}
{{\mathcal{E}_{(int)}}\over c} &=&M\, c \, {\buildrel T_{S_2}\over
\rightarrow}\nonumber \\
 &&{}  \nonumber \\
 \rightarrow && {{{\bar {\mathcal{E}}}_{(int)}}\over c} = \bar M\, c =
\sum_{i=1}^2\, \sqrt{ m_{i}^{2}\, c^2 + \sum_r\, \Big( {\bar
\kappa}^r_i(\tau ) + {\frac{{Q_i}}{c}}\, {\vec \eta}_i(\tau ) \cdot
{\frac{{\partial\, {\vec A}_{\perp}(\tau ,{\vec \eta}_i(\tau ))}}
{{\partial\, \eta^r_i}}} \,      \Big)^2} +  \nonumber \\
 &+& \frac{Q_1\, Q_2}{4\pi\, c\, \mid
{\vec \rho}_{12}(\tau )\mid } + {\frac{1}{2c}}\, \int d^{3}\sigma \,
[{\vec{{ \bar \pi} }}_{\perp
}^{2} + {\vec{B}}^{2}](\tau ,\vec{\sigma}) - \nonumber \\
&-&  \sum_{i=1}^2\, Q_i\, {\vec \eta}_i(\tau ) \cdot
{\vec {\bar \pi}}_{\perp}(\tau ,{\vec \eta}_i(\tau )) +  \nonumber \\
&+& {\frac{{Q_1\, Q_2}}{{c}}}\, \eta^r_1(\tau )\, \eta^s_2(\tau )\,
P^{rs}_{\perp}({\vec \eta}_1(\tau ))\, \delta^3({\vec \eta}_1(\tau )
- {\vec \eta}_2(\tau )) =
\end{eqnarray*}

\bea
 &=& \sum_{i=1}^2\, \sqrt{m_i^2\, c^2 + {\vec {\bar \kappa}}_i^2(\tau
 )} +\nonumber \\
 &+& \sum_{i=1}^2\, {{Q_i}\over c}\, {{{\vec \eta}_i(\tau ) \cdot
 \Big( {\vec {\bar \kappa}}_i(\tau ) \cdot {\frac{{\partial\, }}{{\partial\,
 {\vec \eta}_i}}}\Big) \, {\vec A}_{\perp}(\tau ,{\vec \eta} _i(\tau ))
 }\over {\sqrt{m_i^2\, c^2 + {\vec {\bar \kappa}}_i^2(\tau )}}}  +\nonumber \\
  &+& \frac{Q_1\, Q_2}{4\pi\, c\, \mid
{\vec \rho}_{12}(\tau )\mid } + {\frac{1}{2c}}\, \int d^{3}\sigma \,
[{\vec{{ \bar \pi} }}_{\perp
}^{2} + {\vec{B}}^{2}](\tau ,\vec{\sigma}) - \nonumber \\
&-&  \sum_{i=1}^2\, Q_i\, {\vec \eta}_i(\tau ) \cdot
{\vec {\bar \pi}}_{\perp}(\tau ,{\vec \eta}_i(\tau )) +  \nonumber \\
&+& {\frac{{Q_1\, Q_2}}{{c}}}\, \eta^r_1(\tau )\, \eta^s_2(\tau )\,
P^{rs}_{\perp}({\vec \eta}_1(\tau ))\, \delta^3({\vec \eta}_1(\tau )
- {\vec \eta}_2(\tau )).
 \label{7.9}
 \eea
 \medskip

The Lorentz boost ${\vec {\cal K}}_{(int)}$  of Eqs.(\ref{4.1}) is
replaced by the following expression

\bea
  {\cal K}^r_{(int)} &{\buildrel T_{S_2}\over
\rightarrow}& {\bar {\cal K}}^r_{(int)} =\nonumber \\
 &&{}\nonumber \\
 &=& - \sum_{i=1}^2\, \eta_i^r(\tau )\, \Big[
 \sqrt{m_i^2\, c^2 + {\vec {\bar \kappa}}_i^2(\tau
 )} -\nonumber \\
 &-& \sum_{i=1}^2\, {{Q_i}\over c}\,
{{{\vec \eta}_i(\tau ) \cdot \Big({\vec {\bar \kappa}}_i(\tau )
\cdot {{\partial}\over {\partial\, {\vec \eta}_i}}\Big)\, {\vec
A}_{\perp}(\tau ,{\vec \eta}_i(\tau ))}\over {\sqrt{m_i^2\, c^2 +
{\vec {\bar \kappa}}_i^2(\tau )}}}\Big] +\nonumber \\
 &+& {\frac{Q_1\, Q_2}{4\pi\, c}}\, \Big[{{\eta_1^r(\tau ) +
 \eta_2^r(\tau )}\over {|{\vec \eta}_1(\tau ) - {\vec \eta}_2(\tau )|}}
 -\nonumber \\
 &-& \int {{d^3\sigma}\over {4\pi}}\, \Big({{\sigma^r - \eta_2^r(\tau )}
 \over {|\vec \sigma - {\vec \eta}_1(\tau )|\, |\vec \sigma -
 {\vec \eta}_2(\tau )|^3}} + {{\sigma^r - \eta_1^r(\tau )}
 \over {|\vec \sigma - {\vec \eta}_2(\tau )|\, |\vec \sigma -
 {\vec \eta}_1(\tau )|^3}}\Big)\Big] +  \nonumber \\
 &+&\sum_{i=1}^2\, {{Q_i}\over {4\pi\, c}}\, \int d^{3}\sigma\,
 {{{\bar \pi}_{\perp }^{r}(\tau ,\vec{\sigma})}\over {|\vec{\sigma} -
 {\vec{\eta}}_{i}(\tau )|}} -\nonumber \\
 &-& {{Q_1\, Q_2}\over {4\pi\, c}}\, \sum_s\, \Big[\eta^s_1(\tau )\,
 P^{sr}_{\perp}({\vec \eta}_1(\tau ))\, + \eta_2^s(\tau )\,
 P^{sr}_{\perp}({\vec \eta}_2(\tau ))\Big]\, {1\over
 {|{\vec \eta}_1(\tau ) - {\vec \eta}_2(\tau )|}} -\nonumber \\
 &-& {\frac{1}{2c}}\, \int d^{3}\sigma\, \sigma ^{r}\,
({{\vec{\bar \pi}}}_{\perp }^{2} + {{\vec{B} }} ^{2})(\tau
,\vec{\sigma}) -\nonumber \\
 &-& {{Q_1\, Q_2}\over {2\, c}}\, \sum_{mnu}\, \eta^m_1(\tau )\, \eta^n_2(\tau )\,
 \Big[P^{mu}_{\perp}({\vec \eta}_1(\tau ))\, \Big(\eta^r_1(\tau )\,
 P^{un}_{\perp}({\vec \eta}_1(\tau ))\, \delta^3({\vec \eta}_1(\tau )
 - {\vec \eta}_2(\tau ))\Big) +\nonumber \\
 &+& P^{mu}_{\perp}({\vec \eta}_2(\tau ))\, \Big(\eta^r_2(\tau )\,
 P^{un}_{\perp}({\vec \eta}_2(\tau ))\, \delta^3({\vec \eta}_1(\tau )
 - {\vec \eta}_2(\tau ))\Big) \Big].
 \label{7.10}
 \eea
\medskip

The new form of ${\vec {\cal P}}_{(int)}$ and ${\vec {\cal
J}}_{(int)}$ of Eq.(\ref{4.1}) after the canonical transformation is

\bea
 {\bar {\cal P}}^r_{(int)} &=& \sum_{i=1}^2\, \Big({\bar \kappa}_i^r(\tau ) +
 {{Q_i}\over c}\, A^r_{\perp}(\tau ,{\vec \eta}_i(\tau ))\Big) + {1\over c}\,
 \int d^3\sigma\, \Big({\vec {\bar \pi}}_{\perp} \times \vec B\Big)^r(\tau
 ,\vec \sigma ),\nonumber \\
 &&{}\nonumber \\
 {\vec {\bar {\cal J}}}_{(int)} &=& \sum_{i=1}^2\, {\vec
 \eta}_i(\tau ) \times {\vec {\bar \kappa}}_i(\tau ) + {1\over c}\,
 \int d^3\sigma\, \vec \sigma \times \Big({\vec {\bar \pi}}_{\perp}
 \times \vec B\Big)(\tau ,\vec \sigma ).
 \label{7.11}
 \eea

\bigskip

In Eqs.(\ref{7.9})-(\ref{7.11}) the transverse potential ${\vec
A}_{\perp}(\tau ,\vec \sigma )$ has to be re-expressed in terms of
the magnetic field $\vec B(\tau ,\vec \sigma )$.\medskip

In Abelian electro-magnetism, where the 2-form $F_{rs}\, d\sigma^r\,
d\sigma^s$ is closed, the vector potential ${\vec A}_P(\tau ,\vec
\sigma )$ of the radial (or Poincare') gauge, satisfying $\vec
\sigma \cdot {\vec A}_P(\tau ,\vec \sigma ) = 0$, leads to the exact
1-form ${\vec A}_P(\tau ,\vec \sigma ) \cdot d\vec \sigma$, so that
the Poincare' lemma implies

\beq
 {\vec A}_P(\tau ,{\vec \eta}_i(\tau )) = - \int_0^1 d\lambda\,
\lambda {\vec \eta}_i(\tau ) \times \vec B(\tau ,\lambda\, {\vec
\eta}_i(\tau )).
 \label{7.12}
 \eeq

\medskip

The transverse potential ${\vec A}_{\perp}(\tau ,\vec \sigma )$ of
the radiation gauge can be connected to ${\vec A}_P(\tau ,\vec
\sigma )$ by means of a gauge transformation such that ${\vec
A}_{\perp}(\tau ,\vec \sigma ) = {\vec A}_P(\tau ,\vec \sigma ) +
\vec \partial\, \lambda (\tau ,\vec \sigma )$ with the following
consequence of transversality

\bea
 \triangle\, \lambda (\tau ,\vec \sigma ) &=& - \vec \partial \cdot
 {\vec A}_P(\tau ,\vec \sigma ) =\nonumber \\
 &=& - \int_o^1 d\lambda\, \lambda\, \vec \sigma \cdot \Big[\vec
 \partial \times \vec B(\tau ,\lambda\, \vec \sigma )\Big],
 \label{7.13}
 \eea

\noindent whose solution is (we disregard solutions of the
homogeneous equation)

\beq
 \lambda (\tau ,\vec \sigma ) = \int d^3\sigma_1\, {{\int_o^1 d\lambda\,
 \lambda\, {\vec \sigma}_1 \cdot \Big[{\vec \partial}_1 \times \vec B(\tau
 , \lambda\, {\vec \sigma}_1)\Big]}\over {4\pi\, |\vec \sigma - {\vec
 \sigma}_1|}}.
 \label{7.14}
 \eeq

\medskip

Therefore we get

\bea
 {\vec A}_{\perp}(\tau ,\vec \sigma ) &=& - \int_o^1 d\lambda\,
 \lambda\, \vec \sigma \times \vec B(\tau ,\lambda\, \vec \sigma )
 +\nonumber \\
 &+& \vec \partial\,\, \int d^3\sigma_1\, {{\int_o^1 d\lambda\,
 \lambda\, {\vec \sigma}_1 \cdot \Big[{\vec \partial}_1 \times \vec B(\tau
 , \lambda\, {\vec \sigma}_1)\Big]}\over {4\pi\, |\vec \sigma - {\vec
 \sigma}_1|}}.
 \label{7.15}
 \eea

\bigskip

The term depending on ${\vec A}_{\perp}(\tau ,\vec \sigma )$ in the
energy (\ref{7.9}) and in the boost (\ref{7.10}) becomes (we use the
spin notation ${\vec {\bar S}}_i = {\vec \eta}_i \times {\vec {\bar
\kappa }}_i$)

\bea
 &&{\vec \eta}_i(\tau ) \cdot \Big({\vec {\bar \kappa}}_i(\tau ) \cdot
 {{\partial}\over {\partial\, {\vec \eta}_i}}\Big)\, {\vec
 A}_{\perp}(\tau ,{\vec \eta}_i(\tau )) =\nonumber \\
 &&{}\nonumber \\
 &=& - {\vec S}_i(\tau ) \cdot \int_o^1 d\lambda\, \lambda\, \vec
 B(\tau ,{\vec \eta}_i(\tau )) +\nonumber \\
 &+& \Big({\vec \eta}_i(\tau ) \cdot
 {{\partial}\over {\partial\, {\vec \eta}_i}}\Big)\,
 \Big({\vec {\bar \kappa}}_i(\tau ) \cdot
 {{\partial}\over {\partial\, {\vec \eta}_i}}\Big)\,
\int d^3\sigma_1\, {{\int_o^1 d\lambda\,
 \lambda\, {\vec \sigma}_1 \cdot \Big[{\vec \partial}_1 \times \vec B(\tau
 , \lambda\, {\vec \sigma}_1)\Big]}\over {4\pi\, |{\vec \eta}_i(\tau ) - {\vec
 \sigma}_1|}}.
 \label{7.16}
 \eea

\noindent With this equation we can also evaluate $ {\vec {\bar
\kappa}}_i(\tau ) \cdot \Big({\vec \eta}_i(\tau ) \cdot
{{\partial}\over {\partial\, {\vec \eta}_i}}\Big)\, {\vec
A}_{\perp}(\tau ,{\vec \eta}_i(\tau )) = {\vec \eta}_i(\tau ) \cdot
\Big({\vec {\bar \kappa}}_i(\tau ) \cdot {{\partial}\over
{\partial\, {\vec \eta}_i}}\Big)\, {\vec A}_{\perp}(\tau ,{\vec
\eta}_i(\tau )) + {\vec S}_i(\tau ) \cdot \vec B(\tau ,{\vec
\eta}_i(\tau ))$.

\bigskip

As a consequence we get a {\it generalized electric dipole
representation}: besides interactions of the particles with the
transverse electric field at their position and  interactions with a
delocalized magnetic field ( one of them is with the particle
angular momentum), there is a contact $O(c^{-1})$ term replacing the
divergent term arising from the previous canonical transformations.

\bigskip

If we make the dipole approximation and the semi-relativistic limit
of ${\bar {\cal E}}_{(int)}$ we get

\begin{eqnarray*}
 {\bar {\cal E}}_{(int)} &\rightarrow_{c \rightarrow \infty}& mc^2 +{\frac{{{\vec
\kappa}^2_{12}(\tau )}}{{2m}}} + {\frac{{{\vec \pi}_{12}^2(\tau
)}}{{2\mu}}}
-  \nonumber \\
&-& {\frac{1}{c}}\, \Big[{\frac{{Q_1 + Q_2}}{m}}\, {\vec
\eta}_{12}(\tau ) \times {\vec {\bar \kappa}}_{12}(\tau ) +
\Big({\frac{{Q_1}}{{m_1}}} - {\frac{{Q_2}}{{m_2}}}\Big)  \nonumber \\
&& \Big({\vec \eta}_{12}(\tau ) \times {\vec {\bar \pi}}_{12}(\tau )
+ { \frac{{\mu}}{m}}\, {\vec \rho}_{12}(\tau ) \times {\vec {\bar
\kappa}}_{12}(\tau )\Big) +  \nonumber \\
&+& {\frac{1}{m}}\, \Big({\frac{{m_2}}{{m_1}}}\, Q_1 +
{\frac{{m_1}}{{m_2}}} \, Q_2\Big)\, {\vec \rho}_{12}(\tau ) \times
{\vec \pi}_{12}(\tau )\Big]
\cdot \vec B(\tau ,{\vec \eta}_{12}(\tau )) -  \nonumber \\
&-& {\frac{1}{c}}\, \Big[{\frac{{\mu}}{m}}\, ({\frac{{Q_1}}{{m_1}}}
- {\frac{ {Q_2}}{{m_2}}})\, {\vec \eta}_{12}(\tau ) \times {\vec
{\bar \kappa}}_{12}(\tau) +  \nonumber \\
&+& {\frac{1}{m}}\, ({\frac{{m_2}}{{m_1}}}\, Q_1 +
{\frac{{m_1}}{{m_2}}}\, Q_2)\, {\vec \eta}_{12}(\tau ) \times {\vec
{\bar \pi}}_{12}(\tau )\Big] \cdot \Big({\vec \rho}_{12}(\tau )
\cdot {\frac{{\partial}}{{\partial\, {
\vec \eta}_{12}}}}\Big) \vec B(\tau ,{\vec \eta}_{12}(\tau )) -  \nonumber \\
&-& {\frac{1}{c}}\, \Big[{\frac{{\mu}}{{m^2}}}\,
\Big({\frac{{m_2}}{{m_1}}} \, Q_1 + {\frac{{m_1}}{{m_2}}}\,
Q_2\Big)\, {\vec \rho}_{12}(\tau ) \times {
\vec \kappa}_{12}(\tau ) +  \nonumber \\
&+& {\frac{1}{{m^2}}}\, \Big({\frac{{m_2^2}}{{m_1}}}\, Q_1 -
{\frac{{m_1^2}}{ {m_2}}}\, Q_2\Big)\, {\vec \rho}_{12}(\tau ) \times
{\vec \pi}_{12}(\tau ) \Big] \cdot \Big({\vec \rho}_{12}(\tau )
\cdot {\frac{{\partial}}{{\partial\, {\vec \eta}_{12}}}}\Big)\, \vec
B(\tau ,{\vec \eta}_{12}(\tau )) +\nonumber \\
&+& {\frac{1}{c}}\, \Big[{\frac{{Q_1 + Q_2}}{m}}\, {\vec {\bar
\kappa}} _{12}(\tau ) + ({\frac{{Q_1}}{{m_1}}} -
{\frac{{Q_2}}{{m_2}}})\, {\vec {\bar \pi}}_{12}(\tau )\Big]\, \cdot
\Big({\vec \eta}_{12}(\tau ) \cdot {\frac{{\partial} }{{\partial\,
{\vec \eta}_{12}}}}\Big)\, {\vec A}_{\perp}(\tau ,{\vec \eta}
_{12}(\tau )) +  \end{eqnarray*}

\bea &+& {\frac{1}{{m\, c}}}\, \Big[{\frac{{m_2\, Q_1 - m_1\,
Q_2}}{m}}\, {\vec { \bar \kappa}}_{12}(\tau ) +
({\frac{{m_2}}{{m_1}}}\, Q_1 +
{\frac{{m_1}}{{m_2}}}\, Q_2)\, {\vec {\bar \pi}}_{12}(\tau )\Big]  \nonumber \\
&&\Big({\vec \rho}_{12}(\tau ) \cdot {\frac{{\partial}}{{\partial\,
{\vec \eta}_{12}}}}\Big)\, {\vec A}_{\perp}(\tau ,{\vec
\eta}_{12}(\tau )) -\nonumber \\
&-&  \Big[(Q_1 + Q_2)\, {\vec \eta}_{12}(\tau ) + {\frac{{ m_2\, Q_1
- m_1\, Q_2}}{m}}\, {\vec \rho}_{12}(\tau )\Big]
\cdot {\vec {\bar\pi}}_{\perp}(\tau , {\vec \eta}_{12}(\tau )) -  \nonumber \\
&-&  {\frac{{m_2\, Q_1 - m_1\, Q_2}}{m}}\, {\vec \eta} _{12}(\tau )
\cdot \Big({\vec \rho}_{12}(\tau ) \cdot
{\frac{{\partial}}{{\partial\, {\vec \eta}_{12}}}}\Big) {\vec {\bar
\pi}}_{\perp}(\tau ,{\vec \eta}_{12}(\tau )) -  \nonumber \\
&-&  {\frac{{m_2^2\, Q_1 + m^2_1\, Q_2}}{{m^2}}}\, {\vec \rho}
_{12}(\tau ) \cdot \Big({\vec \rho}_{12}(\tau ) \cdot
{\frac{{\partial}}{{\partial\, {\vec \eta}_{12}}}}\Big)\, {\vec
{\tilde \pi}}_{\perp}(\tau , {\vec \eta}_{12}(\tau )) +  \nonumber \\
&+& {\frac{{Q_1\, Q_2}}{{4\pi\, |{\vec \rho}_{12}(\tau )|}}} +
{\frac{1}{2}} \, \int d^{3}\sigma \, [{\vec{{\bar \pi} }}_{\perp
}^{2} + {\vec{B}} ^{2}](\tau ,\vec{\sigma}) +\nonumber \\
 &+& Q_1\, Q_2\, \sum_{rs}\, \eta^r_1(\tau )\, \eta_2^s(\tau )\,
 P^{rs}_{\perp}({\vec \eta}_1(\tau ))\, \delta^3({\vec \eta}_1(\tau
 ) - {\vec \eta}_2(\tau )) +\nonumber \\
 &+& O\Big(({\vec \rho}_{12}\cdot {{\partial}\over {\partial\, {\vec
\eta}_{12}}})^2\,
{\vec A}_{\perp}\Big) + O(c^{-2}).\nonumber \\
 &&{}
 \label{7.17}
 \eea

For neutral systems, $e = Q \approx Q_1 \approx - Q_2$, ${\cal Q}
\approx 0$, we get

\begin{eqnarray*}
 {\bar {\cal E}}_{(int)} &\rightarrow_{{\cal Q} \approx 0}& mc^2 +{\frac{{{\vec
\kappa}^2_{12}(\tau )}}{{
2m}}} + {\frac{{{\vec \pi}_{12}^2(\tau )}}{{2\mu}}} -  \nonumber \\
&-& {\frac{e}{{c}}}\, \Big({\frac{1}{{\mu}}}\, {\vec \eta}_{12}(\tau
) \times {\vec {\bar \pi}}_{12}(\tau ) + {\frac{1}{m}}\, {\vec
\rho}_{12}(\tau ) \times {\vec {\bar \kappa}}_{12}(\tau ) +  \nonumber \\
&+& {\frac{{m_2 - m_1}}{{\mu\, m}}}\, {\vec \rho}_{12}(\tau ) \times
{\vec \pi}_{12}(\tau )\Big) \cdot \vec B(\tau ,{\vec \eta}_{12}(\tau
)) -\nonumber \\
&-& {\frac{e}{{m\, c}}}\, \Big[ {\vec \eta}_{12}(\tau ) \times {\vec
{\bar \kappa}}_{12}(\tau) + {\frac{{m_2 - m_1}}{{\mu}}}\,
{\vec \eta}_{12}(\tau ) \times  \nonumber \\
&\times& {\vec {\bar \pi}}_{12}(\tau )\Big] \cdot \Big({\vec
\rho}_{12}(\tau ) \cdot {\frac{{\partial}}{{\partial\, {\vec
\eta}_{12}}}}\Big) \vec B(\tau , {\vec \eta}_{12}(\tau )) -
 \end{eqnarray*}

\bea
 &-& {\frac{e}{{mc}}}\, \Big[ {\frac{{m_2 - m_1}}{m}}\, {\vec
\rho}_{12}(\tau ) \times {\vec \kappa}_{12}(\tau ) +  \nonumber \\
&+& {\frac{{m_2^3 - m_1^3}}{{m_1\, m_2\, m}}}\, {\vec
\rho}_{12}(\tau ) \times {\vec \pi}_{12}(\tau )\Big] \cdot
\Big({\vec \rho}_{12}(\tau ) \cdot { \frac{{\partial}}{{\partial\,
{\vec \eta}_{12}}}}\Big)\, \vec B(\tau ,{\vec
\eta}_{12}(\tau )) +  \nonumber \\
&+& {\frac{e}{{\mu\, c}}}\, {\vec {\bar \pi}}_{12}(\tau )\,
\Big({\vec \eta} _{12}(\tau ) \cdot {\frac{{\partial}}{{\partial\,
{\vec \eta}_{12}}}}\Big)\,
{\vec A}_{\perp}(\tau ,{\vec \eta}_{12}(\tau )) +  \nonumber \\
&+& {\frac{e}{{m\, c}}}\,\Big[{\vec {\bar \kappa}}_{12}(\tau ) +
{\frac{{m_2 - m_1}}{{\mu}}}\, {\vec {\bar \pi}}_{12}(\tau )\Big]  \nonumber \\
&&\Big({\vec \rho}_{12}(\tau ) \cdot {\frac{{\partial}}{{\partial\,
{\vec \eta}_{12}}}}\Big)\, {\vec A}_{\perp}(\tau ,{\vec
\eta}_{12}(\tau )) -\nonumber \\
&-& e\, \Big[{\vec \rho}_{12}(\tau ) \cdot {\vec {\bar \pi}}
_{\perp}(\tau , {\vec \eta}_{12}(\tau )) + {\vec \eta}_{12}(\tau )
\cdot \Big({\vec \rho}_{12}(\tau ) \cdot
{\frac{{\partial}}{{\partial\, {\vec \eta} _{12}}}}\Big) {\vec {\bar
\pi}}_{\perp}(\tau ,{\vec \eta}_{12}(\tau )) \Big] -  \nonumber \\
&-& e\, {\frac{{m_2 - m_1}}{m}}\, {\vec \rho}_{12}(\tau ) \cdot
\Big({\vec \rho}_{12}(\tau ) \cdot {\frac{{\partial}}{{\partial\,
{\vec \eta} _{12}}}} \Big)\, {\vec
{\bar \pi}}_{\perp}(\tau ,{\vec \eta}_{12}(\tau )) -\nonumber \\
&-& {\frac{{e^2}}{{4\pi\, |{\vec \rho}_{12}(\tau )|}}} +
{\frac{1}{2}} \, \int d^{3}\sigma \, [{\vec{{\bar \pi} }}_{\perp
}^{2} + {\vec{B}} ^{2}](\tau ,\vec{\sigma}) +\nonumber \\
 &+& Q_1\, Q_2\, \sum_{rs}\, \eta^r_1(\tau )\, \eta_2^s(\tau )\,
 P^{rs}_{\perp}({\vec \eta}_1(\tau ))\, \delta^3({\vec \eta}_1(\tau
 ) - {\vec \eta}_2(\tau )) +\nonumber \\
 &+& O\Big(({\vec \rho}_{12}\cdot {{\partial}\over {\partial\, {\vec
\eta}_{12}}})^2\,
{\vec A}_{\perp}\Big) + O(c^{-2}).\nonumber \\
 &&{}
 \label{7.18}
 \eea

With respect to the R$\o$ntgen Hamiltonian (\ref{7.2}) there are
numerical factors ${1\over 2}$ of difference and extra terms
containing the transverse vector potential and  effects due to the
motion of the collective variable ${\vec \eta}_{12}(\tau )$.

\medskip

Eq.(\ref{7.9}) seems to be the simplest relativistic extension of
the semi-relativistic electric dipole representation in the
framework of the rest-frame instant form.

\vfill\eject

\section{Conclusions}

In this second paper we concluded the classical treatment of the
isolated system of charged positive-energy scalar particles with
Grassmann-valued electric charges with mutual Coulomb interaction
plus the transverse electromagnetic field in the radiation gauge in
the rest-frame instant form of dynamics. In the first paper we
showed how every isolated system (particles, fields, strings,
fluids)  is described in the inertial frame centered on its
Fokker-Planck external center of inertia: the isolated systems is
replaced by a decoupled canonical non-covariant Newton-Wigner center
of mass (described by the frozen Jacobi data $\vec z$, $\vec h$)
carrying its internal mass and spin in such a way that there is an
associated external realization of the Poincare' group. The dynamics
inside the isolated system is defined by construction in the
instantaneous Wigner 3-spaces and is described only by means of
relative canonical variables. This happens  because the internal
realization of the Poincare' group built from the internal energy
momentum tensor of the system is unfaithful: there are the rest
frame conditions ${\vec {\cal P}}_{(int)} \approx 0$ and its gauge
fixing ${\vec {\cal K}}_{(int)} \approx 0$ (vanishing of the
internal Lorentz boosts), whose role is the elimination of the
internal 3-center of mass and of its three momentum.\bigskip

In this paper we have studied how to find collective variables in
the instantaneous Wigner 3-spaces such that it is possible to solve
the equations ${\vec {\cal K}}_{(int)} \approx 0$ explicitly. While
for particle systems there are no problems, the construction of
collective variables for the configurations of massive and massless
fields is nontrivial and only holds for a subset of the field
configurations with finite Poincare' generators. We adapted to the
rest-frame instant form the methods developed by Longhi and
collaborators in Refs.\cite{5,6} and used them to find a solution of
the equations ${\vec {\cal K}}_{(int)} \approx 0$ for the transverse
radiation field. This made possible to solve this equation also for
the  isolated system of charged particles with mutual Coulomb
interaction plus the transverse electro-magnetic field after the
canonical transformation of paper I. \bigskip

The final result  in an arbitrary inertial frame is that inside each
instantaneous Wigner 3-space, with origin in the external covariant
Fokker-Pryce center of inertia, the internal 3-center of mass
becomes a function of the relative canonical variables, so that the
dynamics is described only in terms of them.\medskip

If we consider an isolated 2-particle system (but the same
conclusions hold for an arbitrary isolated system) we have the
following situation on the instantaneous Wigner 3-space:\medskip

A) the particles are described by the two Wigner 3-vectors ${\vec
\eta}_i(\tau ) = {\vec \eta}_{12}(\tau ) + (-)^{i+1}\, {{m_i}\over
m}\, {\vec \rho}_{12}(\tau ) \approx {\vec f}({\vec \rho}_{12}(\tau
), {\vec \pi}_{12}(\tau )) + (-)^{i+1}\, {{m_i}\over m}\, {\vec
\rho}_{12}(\tau )$ with ${\vec \rho}_{12}(\tau ) = {\vec
\eta}_1(\tau ) - {\vec \eta}_2(\tau )$;\bigskip

B) as shown by Eq.(I.2.19) the external non-covariant canonical
center of mass, ${\tilde x}^{\mu}(\tau ) = Y^{\mu}(\tau ) +
\epsilon^{\mu}_r(\vec h)\, {\tilde \sigma}^r$, is identified by the
3-vector ${\tilde {\vec \sigma}} = {{- {\vec {\bar S}} \times \vec
h}\over {Mc\, (1 + \sqrt{1 + {\vec h}^2})}}$. We could then try to
define two new  particle 3-vectors ${\tilde {\vec \eta}}_i(\tau ) =
{\tilde {\vec \sigma}} + (-)^{i+1}\, {{m_i}\over m}\, {\vec
\rho}_{12}(\tau )$ and describe the dynamics with them and their
conjugate momenta rather than with $\vec z$, $\vec h$, ${\vec
\rho}_{12}$, ${\vec \pi}_{12}$: this would be equivalent to the
non-relativistic treatment with ${\vec x}_{(n)\, i}$ and ${\vec
p}_{(n)\, i}$. However the obstruction to do this is the
non-covariance of the external canonical center of mass, which
implies that ${\tilde {\vec \sigma}}$ is not a Wigner spin-1
3-vector. The presence of interactions in the Lorentz boosts and the
associated No-Interaction theorem, which imply the existence of
three relativistic collective variables replacing the unique
non-relativistic center of mass and spanning the non-covariance
M$\o$ller world-tube around the external Fokker-Pryce center of
inertia, lead to a description in which the decoupled canonical
center of mass breaks manifest Lorentz covariance and is a global
quantity which cannot be locally determined.

\bigskip

Then we clarified how to get the multipolar expansion of the
energy-momentum of the particle subsystem and its non-canonical
pole-dipole approximation, not to be confused with the canonical
pole-dipole structure carried by the external Newton-Wigner center
of mass. Thinking to strongly bound clusters of particles as
classical atoms, it is found how to define their effective  4-center
of motion.\bigskip

Finally we have studied the relativistic electric dipole
approximation. We have shown that it is non trivial to find a
relativistic generalization of the electric dipole representation
used in atomic physics. When the electro-magnetic field is
dynamical, as requested by the rest-frame instant form, the point
canonical transformations suggested by the semi-relativistic
approach tend to generate singular terms at the order $O(c^{-1})$.
They are connected to the use of the dipole approximation. From a
study of the Lagrangian on the instantaneous Wigner 3-spaces it is
possible to define a relativistic representation in which the
singular terms are replaced by contact interactions. In this new
relativistic representation, whose semi-relativistic limit is
different from the standard electric dipole representation, there
are interactions only with the transverse electric and magnetic
fields, like in the semi-relativistic case.
\bigskip

In the third paper we will quantize the rest-frame instant form of
our formulation of relativistic particle mechanics. This will allow
us to consider relativistic bound states of clusters of particles as
atoms. The coupling to a quantized transverse electro-magnetic
fields will allow us to define relativistic atomic physics and
relativistic generalizations of the two-level atom. Future
applications will be relevant to study relativistic entanglement and
atom interferometry.

\vfill\eject

\appendix

\section{Definition of the Collective Variables for the Massive Klein-Gordon
Field.}

The first formulation of collective and relative variables for a
field was done in Ref.\cite{5} for the free massive Klein-Gordon
(KG) field. It was then adapted to the rest-frame instant form in
Ref.\cite{13}. Here we give the main results to introduce the
methodology to be used to find such variables for the transverse
radiation field starting from Ref.\cite{6}, where the results of
Ref.\cite{5} were extended to a scalar massless field.

\subsection{The Massive Scalar Field}

In Ref.\cite{13} the scalar massive field $\tilde \phi (x)$ was
reformulated in the rest-frame instant form as the  field $\phi
(\tau ,\vec \sigma ) = \tilde \phi (z_W(\tau ,\vec \sigma ))$ on the
Wigner hyper-planes $z^{\mu}_W(\tau ,\vec \sigma ) = Y^{\mu}(\tau )
+ \epsilon^{\mu}_r(\vec h)\, \sigma^r$. Its conjugate momentum is
$\pi (\tau ,\vec \sigma )$, $\{ \phi (\tau ,\vec \sigma ), \pi (\tau
,{\vec \sigma}_1) \} = \delta^3 (\vec \sigma - {\vec \sigma}_1)$. A
first canonical transformation replaces these canonical variables
with their Fourier coefficients $a(\tau ,\vec k)$, $a^*(\tau ,\vec
k)$, whose non-vanishing Poisson brackets are $\{ a(\tau ,\vec k),
a^*(\tau ,{\vec k}_1) \} = - i\, (2\pi )^3\, 2\, \omega (k)\,
\delta^3(\vec k - {\vec k}_1)$, $\omega (k) = \sqrt{m^2\, c^2 +
{\vec k}^2}$, $k = \sqrt{{\vec k}^2}$ \footnote{The quantity $k_A\,
\sigma^A = \omega (k)\, \tau - \vec k \cdot \vec \sigma$ is built
with $\sigma^A = (\tau ;\vec \sigma )$ and $k^A = (k^{\tau} = \omega
(k); \vec k)$. It is a Lorentz scalar because on the instantaneous
Wigner 3-spaces the 3-vectors $\vec \sigma$ and $\vec k$ are Wigner
spin-1 3-vectors and $k^{\tau} = \omega (k)$ is a Lorentz scalar.
For the free KG field the Fourier coefficients and modulus-phase
variables are $\tau$-independent: in this Appendix we leave a formal
$\tau$-dependence, because the formalism can be used also when
interactions are present as in Ref.\cite{13}.}.

Then a second canonical transformation replaces the Fourier
coefficients with the modulus-phase canonical variables $I(\tau
,\vec k) = a^*(\tau ,\vec k)\, a(\tau ,\vec k)$, $\varphi (\tau
,\vec k) = {1\over {2\, i}}\, ln\, {{a (\tau ,\vec k)}\over
{a^*(\tau ,\vec k)}}$, whose non-vanishing Poisson brackets are $\{
I(\tau ,\vec k), \varphi (\tau ,{\vec k}_1) \} = (2\pi )^3\, 2\,
\omega (k)\, \delta^3(\vec k - {\vec k}_1)$. The final expression of
the original fields is ($d\tilde k = {{d^3k}\over { (2\pi )^3\, 2\,
\omega (k)}}$)

\bea
 \phi (\tau ,\vec \sigma ) &=& \int d\tilde k\, \sqrt{I(\tau ,\vec k)}\,
 \Big(e^{i\, \varphi (\tau ,\vec k) - i\, [\omega (k)\, \tau - \vec k
 \cdot \vec \sigma]} + e^{- i\, \varphi (\tau ,\vec k) + i\, [\omega (k)\,
 \tau - \vec k \cdot \vec \sigma]}\Big),\nonumber \\
 \pi (\tau ,\vec \sigma ) &=& - i\, \int d\tilde k\, \omega (k)\,
 \sqrt{I(\tau ,\vec k)}\, \Big(e^{i\, \varphi (\tau ,\vec k) - i\,
 [\omega (k)\, \tau - \vec k \cdot \vec \sigma]} - e^{- i\, \varphi
 (\tau ,\vec k) + i\, [\omega (k)\, \tau - \vec k \cdot \vec \sigma]}\Big).
 \label{a1}
 \eea
\medskip

The internal Poincare' generators are (we must have
$a_{\lambda}(\tau ,\hat k),\quad \vec \partial a_{\lambda}(\tau
,\hat k) \in L_2(R^3, d^3k)$ for the existence of the following ten
integrals )

\bea
 {\cal P}_{\phi}^{\tau} &=& M_{\phi}\, c = \int d\tilde k\, \omega
 (k)\, I(\tau ,\vec k),\qquad {\vec {\cal P}}_{\phi} = \int d\tilde
 k\, \vec k\, I(\tau ,\vec k) \approx 0,\nonumber \\
 {\vec {\cal J}}_{\phi} &=& {\vec {\bar S}}_{\phi} = \int d\tilde
 k\, I(\tau ,\vec k)\, \vec k \times {{\partial}\over {\partial\,
 \vec k}}\, \varphi (\tau ,\vec k),\nonumber \\
 {\vec {\cal K}}_{\phi} &=& - {\vec {\cal P}}_{\phi}\, \tau - \int
 d\tilde k\, \omega (k)\, I(\tau ,\vec k)\, {{\partial}\over
 {\partial\, \vec k}}\, \varphi (\tau ,\vec k) \approx 0.
 \label{a2}
 \eea
\medskip

In particular we have

\beq
 \{ \varphi (\tau ,\vec k), {\cal P}^A_{\phi} \} = - k^A,
 \label{a3}
 \eeq

\noindent where ${\cal P}^A_{\phi} = ({\cal P}^{\tau}_{\phi}; {\vec
{\cal P}}_{\phi})$ and $k^A = (\omega (k); \vec k)$.

\medskip
The  collective canonical variables are assumed to have the form

\bea
 X_{\phi}^\tau &=& \int d\tilde k\,  \omega (k)\, F^{\tau}(k)\,
\varphi (\tau ,\vec k) ,\qquad {\cal P}^{\tau}_{\phi} = M_{\phi}\,
c, \nonumber \\
 {\vec X}_{\phi} &=& \int d\tilde k\, \vec k\, F(k) \varphi
(\tau ,\vec k),\qquad {\vec {\cal P}}_{\phi} \approx 0,
 \label{a4}
 \eea

\noindent where the two Lorentz-scalar functions $F^{\tau}(k)$,
$F(k)$ have to be determined by the canonical conditions $\{
X^{\tau}_{\phi}, {\cal P}^{\tau}_{\phi} \} = - 1$, $\{ X^r_{\phi},
{\cal P}^s_{\phi} \} = \delta^{rs}$, $\{ X^{\tau}_{\phi}, X^r_{\phi}
\} = \{ X^r_{\phi}, X^s_{\phi} \} = \{ X^{\tau}_{\phi}, {\cal
P}^r_{\phi} \} = \{ X^r_{\phi}, {\cal P}^{\tau}_{\phi} \} = 0$. The
solution of these conditions is \cite{13}

\bea
 F^{\tau}(k) &=& {{16 \pi^2\, e^{-  4\pi\,{{k^2}\over {m^2\, c^2}}
}}\over {mc\, k^2\, \sqrt{m^2\, c^2 + k^2} }},  \qquad  F(k) = -
48\, \pi^2\, {{\sqrt{m^2\, c^2 + k^2}}\over {mc\, k^4}} \, e^{-
4\pi\,{{k^2}\over {m^2\, c^2}} },\nonumber \\
 &&{}\nonumber \\
 &&\Downarrow\nonumber \\
 &&{}\nonumber \\
 &&\int d\tilde k\, \omega^2(k)\, F^{\tau}(k) = 1,\qquad \int
 d\tilde k\, \omega (k)\, k^r\, F^{\tau}(k) = 0,\nonumber \\
 &&\int d\tilde k\, k^r\, k^s\, F(k) = - \delta^{rs},\qquad \int
 d\tilde k\, \omega (k)\, k^r\, F(k) = 0.
 \label{a5}
 \eea
\medskip

This result implements the original construction of Ref.\cite{5}
adapted to the rest-frame instant form. The basic idea is to
consider a real Lorentz-scalar function ${\cal F}(\vec k, {\cal
P}^B_{\phi})$ normalized in the following way

\beq
  \int d\tilde k\, k^A\, {\cal F}(\vec k, {\cal P}^B_{\phi}),\qquad k^A
  = \Big(\omega (k); \vec k\Big) = {\cal P}^A_{\phi}.
 \label{a6}
 \eeq

Then, by using Eq.(\ref{a3}) and (\ref{a6}), one can check that the
collective variable

\beq
  X^A_{\phi} = \int d\tilde k\, {{\partial\, {\cal F}(\vec k,
 {\cal P}^B_{\phi})}\over {\partial\, {\cal P}_{\phi\, A}}}\,
 \varphi (\tau ,\vec k),
 \label{a7}
 \eeq

\noindent satisfies $\{ X^A_{\phi}, X^B_{\phi} \} = 0$, $\{
X^A_{\phi}, {\cal P}^B_{\phi} \} = - \eta^{AB}$.

Eqs.(\ref{a5}) imply that the function ${\cal F}(\vec k, {\cal
P}^B_{\phi})$ has the form

\beq
  {\cal F}(\vec k, {\cal P}^B_{\phi}) = \omega (k)\,
 F^{\tau}(k)\, {\cal P}^{\tau}_{\phi} - F(k)\, \vec k \cdot {\vec
 {\cal P}}_{\phi}.
 \label{a8}
 \eeq
\medskip

Since the functions $F^{\tau}(k)$ and $F(k)$ are singular at $\vec k
= 0$, the collective variables exist for field configurations whose
phase variables satisfy $\varphi (\tau ,\vec k)\, \rightarrow_{k
\rightarrow 0}\, k^{\eta}$ with $\eta > 0$.\medskip

Let us remark that for field configurations $\phi (\tau ,\vec \sigma
)$ such that the Fourier transform $\hat \phi (\tau ,\vec k)$ has
compact support in a sphere centered at $\vec k = 0$ of volume V, we
get $X^{\tau}_{\phi} = - {\frac{ 1}{V}}\, \int
{\frac{{d^3k}}{{\omega (k)}}}\, \varphi (\tau ,\vec k)$, ${\vec X}
_{\phi} = {\frac{1}{V}}\, \int d^3k\, {\frac{{3\vec k}}{{{\vec
k}^2}}}\, \varphi (\tau ,\vec k)$.\bigskip

The canonical relative variables, having vanishing Poisson brackets
with the collective ones, are \cite{5,13}

\bea
 {\bf H}(\tau ,\vec k) &=& \int d\tilde q\, {\cal G}(\vec k, \vec
 q)\, \Big[I(\tau ,\vec q) - \omega (q)\, F^{\tau}(q)\, \int d{\tilde q}_1\,
 \omega (q_1)\, I(\tau ,{\vec q}_1) +\nonumber \\
 &+& F(q)\, \vec q \cdot \int d{\tilde q}_1\, {\vec q}_1\, I(\tau
 ,{\vec q}_1)\Big],\nonumber \\
 {\bf K}(\tau ,\vec k) &=& {\cal D}_{\vec k}\, \varphi (\tau, \vec
 k),\nonumber \\
 &&{}\nonumber \\
 && \{ {\bf H}(\tau ,\vec k), {\bf K}(\tau ,{\vec k}_1) \} = (2\pi )^3\,
 2\, \omega (k)\, \delta^3(\vec k - {\vec k}_1),\nonumber \\
 &&\{ {\bf H}(\tau ,\vec k), {\bf H}(\tau ,{\vec k}_1) \} =
 \{ {\bf K}(\tau ,\vec k), {\bf K}(\tau ,{\vec k}_1) \} =
 0.\nonumber \\
 &&{}
 \label{a9}
 \eea
\medskip

Here (see Ref.\cite{5}) ${\cal G}(\vec q, \vec k)$ is the Green
function of the operator ${\cal D}_{\vec q} = 3 - m^2c^2\,
\triangle_{LB} = 3 - m^2c^2\, {\vec \partial}^2_{\vec q} - 2\, \vec
q \cdot {\vec \partial}_{\vec q} - (\vec q \cdot {\vec
\partial}_{\vec q})^2\quad$ \footnote{$\triangle_{LB}$ is the
Laplace-Beltrami operator of the mass-shell sub-manifold $H^1_3$ of
the KG field. The operator ${\cal D}_{\vec q}$ is scalar, formally
self-adjoint with respect to the Lorentz invariant measure for the
massive case and has $k^A$ as null eigenvectors, ${\cal D}_{\vec
q}\, k^A = 0$.}: ${\cal D}_{\vec q}\, {\cal G}(\vec q, \vec k) =
(2\pi )^3\, 2\, \omega (k)\, \delta^3(\vec k - \vec q)$.\medskip

Another needed distribution is $\triangle (\vec k,\vec q) = (2\pi
)^3\, 2\, \omega (k)\, \delta^3(\vec k - \vec q) - F^{\tau}(k)\,
\omega (k)\, \omega (q) + F(k)\, \vec k \cdot \vec q$ enjoying the
semi-group property $\int d\tilde q\, \triangle (\vec k, \vec q)\,
\triangle (\vec q, {\vec k}_1) = \triangle (\vec k, {\vec k}_1)$.
\bigskip

The final expression of the original fields in terms of the
collective and relative canonical variables is

\bea
 \phi (\tau ,\vec \sigma ) &=& 2\, \int d\tilde k\, A(\tau ,\vec
 k)\, cos\, \Big(\vec k \cdot \vec \sigma + B(\tau ,\vec k)\Big),
 \nonumber \\
 \pi (\tau ,\vec \sigma ) &=& - 2\, \int d\tilde k\, \omega (k)\,
 A(\tau ,\vec k)\, sin\, \Big(\vec k \cdot \vec \sigma +
 B(\tau ,\vec k)\Big),\nonumber \\
 &&{}\nonumber \\
 A(\tau ,\vec k) &=& \sqrt{I(\tau ,\vec k)} = \sqrt{F^{\tau}(k)\,
 \omega (k)\, {\cal P}^{\tau}_{\phi} - F(k)\, \vec k \cdot {\vec
 {\cal P}}_{\phi} + {\cal D}_{\vec k}\, {\bf H}(\tau ,\vec k)},\nonumber \\
 B(\tau ,\vec k) &=& \varphi (\tau ,\vec k) - \omega (k)\, \tau
 =\nonumber \\
 &=& - \vec k \cdot {\vec X}_{\phi} - \omega (k)\, (\tau - X^{\tau}_{\phi})
 + \int d{\tilde k}_1\, d{\tilde k}_2\, {\bf K}(\tau ,{\vec k}_1)\,
 {\cal G}({\vec k}_1, {\vec k}_2)\, \triangle ({\vec k}_1, \vec k).
 \nonumber \\
 &&{}
 \label{a10}
 \eea
\bigskip

The classical field configurations which can be described in this
way must belong to a function space such that:\medskip

a) the internal Poincare' generators (\ref{a2}) are finite (this is
required in every approach to be able to build the Poincare'
representation);\medskip

b) the collective and relative variables are well defined: as a
consequence of the existence of the Poincare' generators and of the
study of the zero modes of the operator ${\vec D}_{\vec q}$ done in
Ref.\cite{5}, we must have: $b_1$) $|I(\tau ,\vec k)| \rightarrow \,
k^{- 3 - \sigma}$, $\sigma > 0$, $|\varphi (\tau ,\vec k)|
\rightarrow\, k$ for $k \rightarrow \infty$; $b_2$) $|I(\tau ,\vec
k)| \rightarrow\, k^{- 3 + \epsilon}$, $\epsilon > 0$, $|\varphi
(\tau ,\vec k)| \rightarrow\, k^{\eta}$, $\eta > 0$, for $k
\rightarrow 0$.\bigskip

The field configurations satisfying these conditions may be
described in terms of the following multipolar structure in the
rest-frame instant form:\medskip

A) An effective particle $X^A_{\phi} = (X^{\tau}_{\phi}; {\vec
X}_{\phi})$ whose conjugate momentum ${\cal P}^A_{\phi} = ({\cal
P}^{\tau}_{\phi} = M_{\phi}\, c; {\vec {\cal P}}_{\phi} \approx 0)$
describes the mass and momentum components of the Dixon {\it
monopole} \cite{13} in the multi-polar analysis of the
energy-momentum tensor of the KG field. While ${\vec X}_{\phi}$ is a
{\it 3-center of phase} for the field configuration,
$X^{\tau}_{\phi}$ is an internal time variable conjugate to ${\cal
P}^{\tau}_{\phi} = M_{\phi}c$. For particle systems an analogue of
$X^{\tau}_{\phi}$ does not exist, because the invariant mass is a
function of the particle canonical variables and not an independent
function like $M_{\phi}$. As shown in Ref.\cite{13}, the rest-frame
Hamilton equations imply that ${\cal P}^{\tau}_{\phi} = M_{\phi}c$
is a constant of motion and that $X^{\tau}_{\phi}$ satisfies the
equation ${{d X^{\tau}_{\phi}}\over {d\tau}} \cir - 1$.\medskip

B) A set of canonical relative variables ${\bf H}(\tau ,\vec k)$,
${\bf K}(\tau ,\vec k)$, which are constant of motion for a free
field due to the complete integrability (Liouville theorem) of the
free KG field. In terms of them we can build canonical multipoles
with respect to the Fokker-Pryce center of inertia, origin of the
instantaneous Wigner 3-spaces, and then all Dixon multipoles
\cite{13}.\medskip

C) The canonical variables ${\vec X}_{\phi}$, ${\vec {\cal
P}}_{\phi}$ are not independent: we have ${\vec {\cal P}}_{\phi}
\approx 0$ from the rest-frame conditions and ${\vec X}_{\phi}
\approx - {1\over {M_{\phi}\, c}}\, \int d\tilde k\, \omega (k)\,
{\bf H}(\tau ,\vec k)\, {{\partial}\over {\partial\, \vec k}}\, {\bf
K}(\tau ,\vec k)$ from the elimination of the internal 3-center of
mass implied by ${\vec {\cal K}}_{\phi} \approx 0$.\medskip

D) The variables ${\cal P}^{\tau}_{\phi} = M_{\phi}c$ and
$X^{\tau}_{\phi}$ have the following interpretation. If we consider
a {\it constant energy surface} ${\cal E}_E$, i.e. $M_{\phi}\, c^2
\approx E$, in the KG phase space, then ${\cal E}_E$ is not a
symplectic submanifold due to the presence of the extra variable
$X^{\tau}_{\phi}$. However, if we add the $\tau$-dependent gauge
fixing $X^{\tau}_{\phi} - a\, \tau \approx 0$ ($X^{\tau}_{\phi} \cir
- 2\, \tau$ from Hamilton equations) to the constraint $M_{\phi}\,
c^2 - E \approx 0$ and we go to Dirac brackets, then ${\cal E}_E$
becomes a phase space spanned by the canonical variables ${\bf
H}(\tau ,\vec k)$, ${\bf K}(\tau ,\vec k)$ for every value of $E$.
The Dirac Hamiltonian becomes a numerical constant $H_D = E/c$ and,
as shown from Eqs.(\ref{a6}), the fields become $\tau$-independent
(they only depend upon constants of motion for the free KG theory).

\subsection{The Massless Scalar Field}

In Ref.\cite{6} it was shown that the construction based on
Eqs.(\ref{a6}) and (\ref{a7}) also holds for the massless scalar
field. Therefore we have to find the limit of the function ${\cal
F}(\vec k, {\cal P}^B_{\phi})$ for $m \rightarrow 0$. Now we have
$\omega (k) = k = |\vec k|$.\medskip

By using the definition of the delta function in the form $\delta
(x) = lim_{\epsilon \rightarrow 0}\, \Big({1\over {\sqrt{\pi\,
\epsilon}}}\, e^{- x^2/\epsilon}\Big)$ with $\epsilon = m^2\, c^2/
4\pi \rightarrow 0$, we get that the functions $F^{\tau}(k)$ and
$F(k)$ of Eqs.(\ref{a5}) become the following distribution in the
massless limit

\beq
 F^{\tau}(k) = {{8\, \pi^2}\over {k^3}}\, \delta (k),\qquad
 F(k) = - {{24\, \pi^2}\over {k^3}}\, \delta (k).
 \label{a11}
 \eeq

It can be checked that the conditions contained in Eqs.(\ref{a5})
are still satisfied since $\delta (k)$ is an even function of $k$
and we have

\bea
 \int d\tilde k\, \omega^2(k)\, F^{\tau}(k) &=& 2\,
 \int_0^{\infty}\, dk\, \delta (k) = \int_{-\infty}^{\infty}\, dk\,
 \delta (k) = 1,\nonumber \\
 \int d\tilde k\, k^r\, k^s\, F(k) &=& - 2\, \delta^{rs}\,
 \int_0^{\infty}\, dk\, \delta (k) = - \delta^{rs}\, \int_{-\infty}
 ^{\infty}\, dk\, \delta (k) = - \delta^{rs}.
 \label{a12}
 \eea
\medskip

The collective variables (\ref{a7}) are now well defined for the
field configurations whose  phase variables behave as $\varphi (\tau
,\vec k)\, \rightarrow_{k \rightarrow 0}\, k$ and have the form

\bea
 X^{\tau}_{\phi} &=& {1\over {2\pi}}\, \int\, {{d^3k}\over {k^3}}\,
 \delta (k)\, \varphi (\tau ,\vec k), \nonumber \\
 {\vec X}_{\phi} &=& {3\over {2\pi}}\, \int\, {{d^3k}\over {k^3}}\,
 \vec k\, \delta (k)\, \varphi (\tau ,\vec k).
 \label{a13}
 \eea
\bigskip

As shown in Ref.\cite{6} the relative canonical variables, having
zero Poisson brackets with the collective ones,  are again given by
Eqs.(\ref{a9}), but now $\omega (k) = k$ and ${\cal G}(\vec k, \vec
q)$ is the Green function of the operator ${\cal D}_{\vec q} = 3 -
2\, \vec q \cdot {\vec \partial}_{\vec q} - (\vec q \cdot {\vec
\partial}_{\vec q})^2\quad$ \footnote{Like in the massive case this operator
is scalar, formally self-adjoint with respect the Lorentz invariant
measure for the massless case and has $k^A = (|\vec k|; \vec k)$ as
null eigenvectors.}.\medskip

Also Eqs.(\ref{a10}) for the reconstruction of the massless KG
fields hold.\medskip

Now, the field configurations having well defined collective and
relative canonical variables are restricted by the requirement of
the existence of the Poincare' generators. This requires \cite{6}:
i) $|I(\tau ,\vec k)| \rightarrow\, k^{- 3 - \delta}$, $\delta > 0$,
and $|\varphi (\tau ,\vec k)| \rightarrow\, k$ for $k \rightarrow
\infty$; ii) $|I(\tau ,\vec k)| \rightarrow\, k^{- 2 + \epsilon}$,
$\epsilon > 0$, and $|\varphi (\tau ,\vec k)| \rightarrow\,
k^{\alpha}$, $\alpha > - \delta$ for $k \rightarrow 0$. However an
extra infinite set of integral restrictions comes from the study of
the zero modes $w_{lm}(\vec k)$ of the operator ${\vec D}_{\vec q}$
done in Ref.\cite{6}. If we define the quantities $P_{lm} = \int
d\tilde k\, w_{lm}(\vec k)\, I(\tau ,\vec k)$, then the restrictions
are $P_{lm} \equiv \int d\tilde k\, w_{lm}(\vec k)\, {\cal F}(\vec
k, {\cal P}^B_{\phi})$ for $l \geq 2$ (the $P_{lm}$ with $l > 2$ are
called super-translations, being connected with the generators of
the algebra of BMS group studied in this framework in
Ref.\cite{15}).

\vfill\eject

\section{Relativistic Lagrangians for the Electric Dipole Approximation}

\subsection{The Standard Electric Dipole Representation with
Grassmann-valued Electric Charges}

Let us revisit the standard electric dipole representation starting
from the Hamiltonian (\ref{7.1}) with Grassmann-valued charges. we
go to the collective and relative variables (\ref{2.3}) and we make
the dipole approximation (\ref{6.2}) (i.e. we neglect terms of order
$O\Big(({\vec \rho}_{12} \cdot {{\partial}\over {\partial\, {\vec
\eta}_{12}}})^2\, {\vec A}_{\perp}\Big)$), we get from the Hamilton
equations ($\mu = m_1\, m_2/m$, $\dot a = {{d\, a}\over {d\tau}}$,
$\tau = ct$, $\int d\tau\, L(\tau ) =  \int dt\, \tilde L(t)$)

\bea
 {\vec \kappa}_{12}(\tau ) &=& mc\, {\dot {\vec \eta}}_{12}(\tau ) +
 {Q\over c}\, \Big({\vec \rho}_{12}(\tau ) \cdot {{\partial}\over {\partial\,
{\vec \eta}_{12}}}\Big)\, {\vec A}_{\perp}(\tau ,{\vec
\eta}_{12}(\tau )) +\nonumber \\
 &+& {{{\cal Q}}\over c}\, {{m_2 - m_1}\over m}\, \Big({\vec \rho}_{12}(\tau ) \cdot {{\partial}\over {\partial\,
{\vec \eta}_{12}}}\Big)\, {\vec A}_{\perp}(\tau ,{\vec
\eta}_{12}(\tau )),\nonumber \\
 {\vec \pi}_{12}(\tau ) &=& \mu c\, {\dot {\vec \rho}}_{12}(\tau ) +
 {Q\over c}\, \Big[{\vec A}_{\perp}(\tau ,{\vec \eta}_{12}(\tau )) +
 {{m_2 - m_1}\over {m}}\,\Big({\vec \rho}_{12}(\tau ) \cdot {{\partial}\over {\partial\,
{\vec \eta}_{12}}}\Big)\, {\vec A}_{\perp}(\tau ,{\vec
\eta}_{12}(\tau )) \Big] +\nonumber \\
 &+& {{{\cal Q}}\over c}\, \Big[{{m_2 - m_1}\over m}\, {\vec A}_{\perp}(\tau ,{\vec \eta}_{12}(\tau )) +
 {{m_1^2 + m_2^2}\over {m^2}}\, \Big({\vec \rho}_{12}(\tau ) \cdot {{\partial}\over {\partial\,
{\vec \eta}_{12}}}\Big)\, {\vec A}_{\perp}(\tau ,{\vec
\eta}_{12}(\tau ))\Big],
 \label{b1}
 \eea

\noindent where the notation of Eq.(\ref{6.1}) has been used.
\medskip

Then by means of the inverse Legendre transformation we get the
Lagrangian

\bea
 L &=& {{m c}\over 2}\, {\dot {\vec \eta}}^2_{12}(\tau ) + {{\mu c}\over 2}\,
{\dot {\vec \rho}}_{12}^2(\tau ) +\nonumber \\
 &+& \Big({Q\over c} + {{{\cal Q}}\over c}\, {{m_2 - m_1}\over
 m}\Big)\, {\dot {\vec \eta}}_{12}(\tau ) \cdot \Big({\vec \rho}_{12}(\tau )
\cdot {{\partial}\over {\partial\, {\vec \eta}_{12}}}\Big)\, {\vec
A}_{\perp}(\tau ,{\vec \eta}_{12}(\tau )) +\nonumber \\
 &+& {{2\, {\cal Q}}\over c}\, {\dot {\vec
\eta}}_{12}(\tau ) \cdot {\vec A}_{\perp}(\tau ,{\vec
\eta}_{12}(\tau )) + \Big({Q\over c} + {{{\cal Q}}\over c}\, {{m_2 -
m_1}\over m}\Big)\, {\dot {\vec \rho}}_{12}(\tau ) \cdot {\vec
A}_{\perp}(\tau ,{\vec \eta}_{12}(\tau )) +\nonumber \\
 &+&\Big({Q\over c}\, {{m_2 - m_1}\over {m}} + {{{\cal Q}}\over c}\,
 {{m_1^2 + m_2^2}\over {m^2}},\Big)\,{\dot {\vec \rho}}_{12}(\tau )
 \cdot \Big({\vec \rho}_{12}(\tau )
\cdot {{\partial}\over {\partial\, {\vec \eta}_{12}}}\Big)\, {\vec
A}_{\perp}(\tau ,{\vec \eta}_{12}(\tau )).
 \label{b2}
 \eea

\medskip

Let us write $L = L_1 + {{d S}\over {d\tau}}$ with the following
function $S$

\bea
 S &=& {{m_2\, Q_1 - m_1\, Q_2}\over {2\, m\, c}}\, {\vec
\rho}_{12}(\tau ) \cdot {\vec A}_{\perp}(\tau ,{\vec \eta}_{12}(\tau
)) +\nonumber \\
 &+& {{m_2^2\, Q_1 + m_1^2\, Q_2}\over {2m^2}}\, {\vec
 \rho}_{12}(\tau ) \cdot \Big({\vec \rho}_{12}(\tau )
\cdot {{\partial}\over {\partial\, {\vec \eta}_{12}}}\Big)\, {\vec
A}_{\perp}(\tau ,{\vec \eta}_{12}(\tau )) =\nonumber \\
 &=& {Q\over c}\, \Big[{\vec \rho}_{12}(\tau ) \cdot {\vec A}_{\perp}(\tau ,{\vec \eta}_{12}(\tau ))
 + {{m_2 - m_1}\over {2m}}\, {\vec
 \rho}_{12}(\tau ) \cdot \Big({\vec \rho}_{12}(\tau )
\cdot {{\partial}\over {\partial\, {\vec \eta}_{12}}}\Big)\, {\vec
A}_{\perp}(\tau ,{\vec \eta}_{12}(\tau ))\Big] +\nonumber \\
 &+& {{{\cal Q}}\over c}\, \Big[{{m_2 - m_1}\over m}\, {\vec \rho}_{12}(\tau )
 \cdot {\vec A}_{\perp}(\tau ,{\vec \eta}_{12}(\tau )) +
 {{m_1^2 + m_2^2}\over {2m^2}}\, {\vec
 \rho}_{12}(\tau ) \cdot \Big({\vec \rho}_{12}(\tau )
\cdot {{\partial}\over {\partial\, {\vec \eta}_{12}}}\Big)\, {\vec
A}_{\perp}(\tau ,{\vec \eta}_{12}(\tau ))\Big] \nonumber \\
&&{}\nonumber \\
 &\rightarrow_{{\cal Q} \approx 0}&\,\, {Q\over {c}}\, {\vec \rho}_{12}(\tau )
\cdot \Big[{\vec A}_{\perp}(\tau ,{\vec \eta}_{12}(\tau )) + {{m_2 -
m_1}\over {2m}}\, \Big({\vec \rho}_{12}(\tau ) \cdot
{{\partial}\over {\partial\, {\vec \eta}_{12}}}\Big)\, {\vec
A}_{\perp}(\tau ,{\vec \eta}_{12}(\tau ))\Big],
 \label{b3}
 \eea

\noindent where we used the notation of Eq.(\ref{6.1}). For neutral
systems, ${\cal Q} \approx 0$, $S = S_o + O\Big(({\vec \rho}_{12}
\cdot {{\partial}\over {\partial\, {\vec \eta}_{12}}})\, {\vec
A}_{\perp}\Big)$ is the extension to the next order of the classical
counterpart $S_o$ of the generator of the G$\o$ppert-Mayer unitary
transformation.

 \medskip

If we use ${{d\, {\vec A}_{\perp}(\tau ,{\vec \eta}_{12}(\tau
))}\over {d\tau}} = - {\vec \pi}_{\perp}(\tau ,{\vec \eta}_{12}(\tau
)) + \Big({\dot {\vec \eta}}_{12}(\tau ) \cdot {{\partial}\over
{\partial\, {\vec \eta}_{12}}}\Big)\, {\vec A}_{\perp}(\tau ,{\vec
\eta}_{12}(\tau ))$, $L_1$ takes the following form

\bea
 L_1 &=& {{m c}\over 2}\, {\dot {\vec \eta}}_{12}^2(\tau ) + {{\mu c}\over
2}\, {\dot {\vec \rho}}_{12}^2(\tau ) + {{2\, {\cal Q}}\over c}\,
{\dot {\vec \eta}}_{12}(\tau ) \cdot {\vec A}_{\perp}(\tau ,{\vec
\eta}_{12}(\tau )) +\nonumber \\
 &+& \Big({Q\over c} + {{{\cal Q}}\over c}\, {{m_2 - m_1}\over m}\Big)\,
 \Big[{\vec \rho}_{12}(\tau ) \cdot {\vec
\pi}_{\perp}(\tau ,{\vec \eta}_{12}(\tau )) -  {\dot {\vec
\eta}}_{12}(\tau ) \cdot \Big({\vec \rho}_{12}(\tau ) \times \vec
B(\tau ,{\vec \eta}_{12}(\tau ))\Big)\Big] +\nonumber \\
 &+& \Big({Q\over c}\, {{m_2 - m_1}\over {2m}} + {{{\cal Q}}\over c}\,
 {{m_1^2 + m_2^2}\over {2m^2}}\Big)\, \Big[{\vec \rho}_{12}(\tau )
 \cdot \Big({\vec \rho}_{12}(\tau ) \cdot
{{\partial}\over {\partial\, {\vec \eta}_{12}}}\Big)\, {\vec
\pi}_{\perp}(\tau ,{\vec \eta}_{12}(\tau )) -\nonumber \\
&-& {\dot {\vec \rho}}_{12}(\tau ) \cdot \Big({\vec \rho}_{12}(\tau
) \times \vec B(\tau ,{\vec \eta}_{12}(\tau ))\Big)\Big].
 \label{b4}
 \eea
 \medskip

We have used $\vec B(\tau ,{\vec \eta}_{12}(\tau )) = -
{{\partial}\over {\partial\, {\vec \eta}_{12}}} \times {\vec
A}_{\perp}(\tau ,{\vec \eta}_{12}(\tau ))$, ${\vec \rho}_{12}(\tau )
\times \vec B(\tau ,{\vec \eta}_{12}(\tau )) = \Big({\vec
\rho}_{12}(\tau ) \cdot {{\partial}\over {\partial\, {\vec
\eta}_{12}}}\Big)\, {\vec A}_{\perp}(\tau ,{\vec \eta}_{12}(\tau ))
- {{\partial}\over {\partial\, {\vec \eta}_{12}}}\, \Big({\vec
\rho}_{12}(\tau ) \cdot {\vec A}_{\perp}(\tau ,{\vec \eta}_{12}(\tau
))\Big)$. The new momenta are

\bea
 &&{\vec {\bar \kappa}}_{12}(\tau ) = m\, c\, {\dot {\vec \eta}}_{12}(\tau
) + {{2\, {\cal Q}}\over c}\, {\vec A}_{\perp}(\tau ,{\vec
\eta}_{12}(\tau )) - \Big({Q\over c} + {{{\cal Q}}\over c}\, {{m_2 -
m_1}\over m}\Big)\, {\vec \rho}_{12}(\tau ) \times \vec
B(\tau ,{\vec \eta}_{12}(\tau )),\nonumber \\
 &&{\vec {\bar \pi}}_{12}(\tau ) = \mu\, c\, {\dot {\vec \rho}}_{12}(\tau )
- \Big({Q\over c}\, {{m_2 - m_1}\over {2m}} + {{{\cal Q}}\over c}\,
{{m_1^2 + m_2^2}\over {2m^2}}\Big)\, {\vec \rho}_{12}(\tau ) \times
\vec B(\tau ,{\vec \eta}_{12}(\tau )).
 \label{b5}
 \eea
\medskip

Then by Legendre transformation we get the Hamiltonian of the
electric dipole representation in the dipole approximation

\bea
 H_1 &=& {{{\vec {\bar \kappa}}^2_{12}(\tau )}\over {2 m\, c}} + {{{\vec
{\bar \pi}}_{12}^2(\tau )}\over {2 \mu\, c}} - \Big({Q\over c} +
{{{\cal Q}}\over c}\, {{m_2 - m_1}\over m}\Big)\,
 {\vec \rho}_{12}(\tau ) \cdot {\vec \pi}_{\perp}(\tau ,{\vec
\eta}_{12}(\tau )) -\nonumber \\
 &-& \Big({Q\over c}\, {{m_2 - m_1}\over {2m}} + {{{\cal Q}}\over c}\,
{{m_1^2 + m_2^2}\over {2m^2}}\Big)\, {\vec \rho}_{12}(\tau ) \cdot
\Big({\vec \rho}_{12}(\tau ) \cdot {{\partial}\over {\partial\,
{\vec \eta}_{12}}}\Big)\, {\vec \pi}_{\perp}(\tau ,{\vec
\eta}_{12}(\tau )) +\nonumber \\
 &+& {1\over {m c} }\, \Big({Q\over c} + {{{\cal Q}}\over c}\, {{m_2 -
m_1}\over m}\Big)\, {\vec {\bar \kappa}}_{12}(\tau ) \cdot
\Big({\vec \rho}_{12}(\tau ) \times \vec B(\tau ,{\vec
\eta}_{12}(\tau ))\Big) +\nonumber \\
 &+& {1\over {\mu\, c}}\, \Big({Q\over c}\, {{m_2 - m_1}\over {2m}} + {{{\cal Q}}\over c}\,
{{m_1^2 + m_2^2}\over {2m^2}}\Big)\,
 {\vec {\bar \pi}}_{12}(\tau )\, \cdot \Big({\vec \rho}_{12}(\tau )
\times \vec B(\tau ,{\vec \eta}_{12}(\tau ))\Big) -\nonumber \\
 &-& Q_1\, Q_2\, {{3\, \mu}\over {4 m^2\, c^2}}\, \Big({\vec
\rho}_{12}(\tau ) \times \vec B(\tau ,{\vec \eta}_{12}(\tau ))\Big)^2
 -\nonumber \\
 &-& {{2\, {\cal Q}}\over {m c^2}}\,
\Big[ {\vec {\bar \kappa}}_{12}(\tau ) \cdot {\vec A}_{\perp}(\tau
,{\vec \eta}_{12}(\tau )) + \Big(Q + {\cal Q}\, {{m_2 - m_1}\over
m}\Big)\, {\vec A}_{\perp}(\tau ,{\vec \eta}_{12}(\tau )) \cdot
\Big({\vec \rho}_{12}(\tau ) \times \vec B(\tau ,{\vec
\eta}_{12}(\tau ))\Big) \Big] +\nonumber \\
 &+& {{4\, {\cal Q}^2}\over {2\, m\, c^3}}\, {\vec A}^2_{\perp}(\tau
 ,{\vec \eta}_{12}(\tau )).
 \label{b6}
 \eea
 \medskip

For $2\, {\cal Q} = Q_1 + Q_2 \approx 0$, $e = Q_1$, we get the
R$\o$ntgen Hamiltonian (\ref{7.2}) of the electric dipole
representation, given in Eq.(14.37) of Ref.\cite{7}.

\subsection{A Lagrangian for the Internal Motion of the Particles
with an External Electro-Magnetic Field}

Let us now follow the method of Appendix L of Ref.\cite{7} of
determining the Lagrangian from the Hamiltonian by considering the
electro-magnetic field as an external field.\medskip

We begin with the internal energy ${1\over c}\, {\cal E}_{(int)} =
{\cal P}^{\tau}_{(int)} = Mc$ of Eqs.(\ref{4.1}) and we do the
inverse Legendre transformation on the particle variables to find an
effective Lagrangian for the particles. After the dipole
approximation and by using $e = Q_1 \approx - Q_2$, a term in the
resulting Lagrangian will be a total time derivative identifying the
generating function $\tilde S$. Then we will do the Legendre
transformation of the Lagrangian without the total time derivative.
The semi-relativistic limit of the resulting Hamiltonian turns out
to be the R$\o$ntgen Hamiltonian (\ref{7.2}) of Ref.\cite{7}.

\medskip

By ignoring the field energy ${\cal E}_{em}$ in Eq.(\ref{4.1}), the
inverse Legendre transformation produces the Lagrangian

\begin{eqnarray}
L(\tau ) &=&\vec{\kappa}_{1} \cdot \frac{d{\vec{\eta}}_{1}(\tau
)}{d\tau} + \vec{\kappa}_{2} \cdot \frac{d{\vec{\eta}}_{2}(\tau
)}{d\tau} - M\, c =\nonumber \\
 &=&- m_{1}\, c\, \sqrt{1 -  \left(
 \frac{d{\vec{\eta}}_{1}(\tau )}{d\tau}\right) ^{2}}
- m_{2}\, c\, \sqrt{1 -  \left( \frac{d{\vec{\eta}}_{2}(\tau
)}{d\tau}\right) ^{2}}-\, \frac{Q_{1}\,Q_{2}}{4\pi\, c
\,|\vec{\eta}_{1}(\tau ) -
\vec{\eta}_{2}(\tau )|} +\nonumber \\
&+&{\frac{{Q_{1}}}{c}\, {\vec{A}}_{\perp }(\tau
,{\vec{\eta}}_{1}(\tau ))} \cdot \frac{d{\vec{\eta}}_{1}}{d\tau} +
{\frac{{Q_{2}}}{c}\, {\vec{A}}_{\perp }(\tau ,{\
\vec{\eta}}_{2}(\tau ))} \cdot \frac{d{\vec{\eta}}_{2}}{d\tau}.
\label{b7}
\end{eqnarray}

Switching to the center of mass and relative variables of
Eqs.(\ref{2.3}) (here we put $\vec \eta = {\vec \eta}_{12}$ and
$\vec \rho = {\vec \rho}_{12}$) we obtain

\begin{eqnarray}
L(\tau ) &=&- m_{1}\, c\, \sqrt{1 -  \left( \frac{d{\vec{\eta}(\tau
)}}{d\tau} + \frac{m_{2}}{m}\, \frac{d{\vec{\rho}(\tau
)}}{d\tau}\right)^{2}} - m_{2}\, c\, \sqrt{1 - \left( \frac{d{
\vec{\eta}(\tau )}}{d\tau} -
\frac{m_{1}}{m}\,\frac{d{\vec{\rho}(\tau )}}{d\tau}\right)
^{2}} -\nonumber \\
&-& \,\frac{Q_{1}\,Q_{2}}{4\pi\, c \,|{\vec{\rho}}(\tau )|} + \nonumber \\
 &+&\frac{{Q_{1}}}{c}\, \Big(\vec{A}_{\perp }(\tau ,\vec{\eta}(\tau ))
 + \frac{m_{2}}{m}\, {{\partial}\over {\partial\, \vec \eta}}\,
 \vec{\rho}(\tau ) \cdot \vec{A}_{\perp }(\tau ,\vec{\eta}(\tau ))\Big)
\cdot \left( \frac{d{\vec{\eta}}(\tau )}{d\tau} + \frac{m_{2}}{m}\,
\frac{d{\vec{\rho}}(\tau )}{d\tau}\right) +\nonumber \\
 &+&\frac{{Q_{2}}}{c}\, (\vec{A}_{\perp }(\tau ,\vec{\eta}(\tau )) - \frac{m_{1}}{m}\,
{{\partial}\over {\partial \vec \eta}}\, \vec{\rho}(\tau ) \cdot
\vec{A}_{\perp }(\tau ,\vec{\eta}(\tau ))) \cdot \left(
\frac{d{\vec{\eta}}(\tau )}{d\tau} - \frac{m_{1}}{m}\,
\frac{d{\vec{\rho}}(\tau )}{d\tau}\right),
 \label{b8}
\end{eqnarray}

\noindent in which we have expanded the vector potentials about
${\vec{\eta}}$ by using the dipole approximation. By using the
notation  $Q = {1\over 2}\, (Q_{1} - Q_{2})$, ${\cal Q} = {1\over
2}\, (Q_{1} + Q_{2})$, of Eq.(\ref{6.1}) after some algebra  we find
\footnote{Note that we divided the $\frac{d{\vec{\rho}}(\tau
)}{d\tau} \cdot {{\partial}\over {\partial\, \vec \eta}}\,
\Big(\vec{\rho}(\tau ) \cdot \vec{A}_{\perp }(\tau ,\vec \eta (\tau
))\Big)$ term into separate and equal parts following the treatment
of Ref.\cite{7}: this will produce the correct ${1\over 2}$
factors.}

\begin{eqnarray}
L(\tau ) &=& - m_{1}\, c\, \sqrt{1 -  \left(
\frac{d{\vec{\eta}}(\tau )}{d\tau} + \frac{m_{2}}{m}\,
\frac{d{\vec{\rho}}(\tau )}{d\tau}\right) ^{2}} - m_{2}\, c\,
\sqrt{1 -  \left( \frac{d{\vec{\eta}}(\tau )}{d\tau} -
\frac{m_{1}}{m}\, \frac{d{\vec{\rho}}(\tau )}{d\tau}\right) ^{2}}
+\nonumber \\
 &+&\frac{Q_{1}\, Q_{2}}{4\pi\, c \,|{\vec{\rho}}(\tau )|} + {Q\over c}\, \Big(\vec{\rho}(\tau )
 \cdot \vec{\pi}_{\perp }(\tau ,\vec{\eta}(\tau )) +  \frac{\Delta
m}{2m}\, \vec{\rho}(\tau ) \cdot {{\partial}\over {\partial\, \vec
\eta}}\, \Big(\vec{\rho} \cdot \vec{\pi}_{\perp }(\tau
,\vec{\eta}(\tau ))\Big) - \nonumber \\
 &-& \frac{d{\vec{\eta}}(\tau )}{d\tau} \cdot \Big[
 {{\partial}\over {\partial\, \vec \eta}}\, \Big(\vec{\rho}
\cdot \vec{A}_{\perp }(\tau ,\vec{\eta}(\tau ))\Big) -
\vec{\rho}(\tau ) \cdot {{\partial}\over {\partial\, \vec \eta}}\,
\vec{A} _{\perp }(\tau ,\vec{\eta}(\tau ))\Big] +\nonumber \\
 &+& \frac{\Delta m}{2m}\, \frac{d{\vec{\rho} }(\tau )}{d\tau}
 \cdot \Big[\vec{\rho}(\tau ) \cdot {{\partial}\over {\partial\, \vec \eta}}\,
 \vec{A}_{\perp }(\tau ,\vec{\eta}(\tau )) - {{\partial}\over {\partial\,
 \vec \eta}}\, \Big(\vec{\rho}(\tau ) \cdot \vec{A}_{\perp }(\tau
,\vec{\eta}(\tau ))\Big)\Big]\Big) +\nonumber \\
 &&{}\nonumber \\
 &+&{\mathcal{Q}\over c}\, \Big({\vec{\eta}}(\tau ) \cdot \vec{\pi}_{\perp }(\tau
,\vec{\eta}(\tau )) +   \frac{\Delta m}{m}\, {\vec{\rho}}(\tau
)\cdot \vec{\pi}_{\perp }(\tau ,\vec{ \eta}(\tau
)) + \nonumber \\
 &+& \frac{m^{2}-2\mu m}{2m^{2}}\,
\vec{\rho}(\tau ) \cdot \Big(\vec{\rho}(\tau ) \cdot
{{\partial}\over {\partial\, \vec \eta}}\, \vec{\pi}_{\perp }(\tau
,\vec{\eta})\Big)  +\nonumber \\
 &+& \frac{m^{2}-2\mu m}{2m^{2}}\,
 \frac{d{\vec{\rho}}(\tau )}{d\tau}\,
\ \ \cdot \Big[ \vec{\rho}(\tau ) \cdot {{\partial}\over {\partial\,
\vec \eta}}\, \vec{A}_{\perp }(\tau ,{\vec{ \eta}}(\tau )) -
{{\partial}\over {\partial\, \vec \eta}}\, \Big(\vec{\rho}(\tau )
\cdot \vec{A}_{\perp }(\tau ,\vec{\eta}(\tau ))\Big)\Big] -
\nonumber \\
 &-& {\vec{\eta}(\tau ) \cdot
\Big(\frac{d{\vec{\eta}}(\tau )}{d\tau}\, \cdot
{{\partial}\over {\partial\, \vec \eta}}\Big)\,
\vec{A}}_{\perp }(\tau ,\vec{\eta}(\tau )) +  \nonumber \\
 &+& \frac{\Delta
m}{m}\, \frac{d{\vec{\eta}}(\tau )}{d\tau} \cdot \Big[
\vec{\rho}(\tau ) \cdot {{\partial}\over {\partial\, \vec \eta}}\,
\vec{A}_{\perp }(\tau ,{\vec{\eta}}(\tau )) - {{\partial}\over
{\partial\, \vec \eta}}\, \Big(\vec{\rho}(\tau ) \cdot
\vec{A}_{\perp }(\tau ,\vec{\eta}(\tau ))\Big)\Big] - \nonumber \\
 &-& {\vec{\eta}}(\tau ) \cdot
\Big(\frac{d{\vec{\eta}}(\tau )}{d\tau}\, \cdot {{\partial}\over
{\partial\, \vec \eta}}\Big)\, \vec{A} _{\perp }(\tau
,\vec{\eta}(\tau )) \Big) + {{d \tilde S}\over {d\tau}} =\nonumber \\
 &{\buildrel {def}\over =}& L_1(\tau ) + {{d \tilde S}\over {d\tau}}.
 \label{b9}
\end{eqnarray}

\noindent in which we have ignored the second order spatial gradient
terms of the vector potential (dipole approximation) and we have put
$\triangle m = m_2 - m_1$, $\mu = {{m_1\, m_2}\over m}$.
\medskip

The total time derivative terms, not contributing to the equations
of motion,  identify the generating function $\tilde S$:

\begin{eqnarray}
  \tilde S
&=& \frac{Q}{c}\,  \Big[ \vec{\rho}(\tau ) \cdot {\vec{A}}_{\perp
}(\tau ,\vec{\eta }(\tau )) + \frac{\Delta m}{2m}\,
{\vec{\rho}}(\tau )\cdot \Big({\vec{\rho}}(\tau )\cdot
{{\partial}\over {\partial\, \vec \eta}}\, {
\vec{A}}_{\perp }(\tau ,\vec{\eta}(\tau ))\Big)\Big] +  \nonumber \\
 &+&\frac{\mathcal{Q}}{c}\,  \Big[{\vec{\eta}}(\tau ) \cdot
{\vec{A}}_{\perp }(\tau ,\vec{\eta}(\tau )) + \frac{\Delta m}{m}\,
{\vec{\rho}}(\tau ) \cdot {\vec{A}}_{\perp }(\tau ,\vec{\eta}(\tau
)) +\nonumber \\
 &+& \frac{m^{2}-2\mu m}{2m^{2}}\, {\vec{\rho}}(\tau )
\cdot \Big({\vec{\rho }}(\tau )  \cdot {{\partial}\over {\partial\,
\vec \eta}}\, {\vec{A}}_{\perp }(\tau ,\vec{\eta}(\tau ))
\Big)\Big].
 \label{b10}
\end{eqnarray}

Since we are in the radiation gauge we have $\vec{E}_{\perp }(\tau
,\vec{\eta}(\tau )) = -  \frac{\partial \vec{A}_{\perp }{(}\tau
,\vec{\eta}(\tau ))}{\partial \, \tau}$. Then by using

\begin{eqnarray}
\vec{\rho}\times \vec{B} &=& {{\partial}\over {\partial\, \vec
\eta}}\,\Big(\vec{\rho}\cdot \vec{A}_{\perp
}(\tau ,\vec{\eta})\Big) - \vec{\rho}\cdot {{\partial}\over
{\partial\, \vec \eta}}\, \vec{A}_{\perp }(\tau ,\vec{%
\eta}),  \nonumber \\
\vec{\eta}\cdot \Big(\frac{d{\vec{\eta}}}{d\tau}\cdot
\vec{\nabla}\Big)\, \vec{A}_{\perp
}(\tau ,\vec{\eta}) &=&\frac{d{\vec{\eta}}}{d\tau}\cdot
\Big({{\partial}\over {\partial\, \vec \eta}}\, \Big(\vec{\eta}%
\cdot \vec{A}_{\perp }(\tau ,\vec{\eta})\Big) - \vec{A}_{\perp
}(\tau ,\vec{\eta})\Big),\nonumber \\
\vec{\eta}\times \vec{B} &=&\vec{A}_{\perp }(\tau ,\vec{\eta}) -
 {{\partial}\over {\partial\, \vec \eta}}\,%
\vec{\eta}\cdot \ \vec{A}_{\perp }(\tau ,\vec{\eta}) - \vec{\eta}\cdot
{{\partial}\over {\partial\, \vec \eta}}\, \vec{A}_{\perp }(\tau ,\vec{\eta}),
\nonumber \\
\frac{d{\vec{\eta}}}{d\tau}\cdot \vec{\eta}\times \vec{B} &=& - \vec{\eta}\cdot \big(%
\frac{d{\vec{\eta}}}{d\tau}\cdot {{\partial}\over {\partial\, \vec \eta}}\,
\Big)\vec{A}_{\perp }(\tau ,\vec{\eta}%
) - \frac{d{\vec{\eta}}}{d\tau}\cdot \Big(\vec{\eta}\cdot
{{\partial}\over {\partial\, \vec \eta}}\,\Big)\, \vec{A}_{\perp
}(\tau ,\vec{\eta})),\nonumber \\
&&{}
  \label{b11}
\end{eqnarray}

\noindent we obtain

\begin{eqnarray}
L_1(\tau ) &=& - m_{1}\, c\, \sqrt{1 -
\left( \frac{d{\vec{\eta}}(\tau )}{d\tau} + \frac{m_{2}}{m}\,%
\frac{d{\vec{\rho}}(\tau )}{d\tau}\right) ^{2}} - m_{2}\, c\,
\sqrt{1 -  \left( \frac{d{%
\vec{\eta}}(\tau )}{d\tau} - \frac{m_{1}}{m}\,
\frac{d{\vec{\rho}}(\tau )}{d\tau}\right)^{2}} -\nonumber \\
&-&\,\frac{Q_{1}Q_{2}}{4\pi\, c \,|{\vec{\rho}}(\tau )|} + {Q\over
c}\, \Big({\vec{\rho}(\tau ) \cdot }\vec{\pi} _{\perp }(\tau
,\vec{\eta}(\tau )) + \frac{\Delta m}{2m}\, \vec{\rho}(\tau ) \cdot
\Big(\vec{\rho} (\tau ) \cdot {{\partial}\over {\partial\, \vec
\eta}}\, \vec{\pi}_{\perp }(\tau ,\vec{\eta}(\tau ))\Big)  -\nonumber \\
&-& \frac{d{\vec{\eta}}(\tau )}{d\tau} \cdot \vec{\rho} (\tau )
\times \vec{B}(\tau ,{\vec \eta}(\tau )) -  \frac{\Delta m}{2m}\,
\frac{d{\vec{\rho}}(\tau )}{d\tau} \cdot
\vec{\rho}(\tau ) \times \vec{B}(\tau ,\vec \eta (\tau ))\Big) +\nonumber \\
 &&{}\nonumber \\
 &+&{\mathcal{Q}\over c}\, \Big(\frac{\Delta m}{m}\, {\vec{\rho}(\tau )
\cdot }\vec{\pi}_{\perp }(\tau , \vec{\eta}(\tau )) +  \frac{m^{2} -
2\mu m}{2m^{2}}\, \vec{\rho}(\tau ) \cdot \Big(\vec{ \rho}(\tau )
\cdot {{\partial}\over {\partial\, \vec \eta}}\,
 \vec{\pi}_{\perp }(\tau ,\vec{\eta}(\tau ))\Big)  -\nonumber \\
&-& \frac{m^{2} - 2\mu m}{2m^{2}}\, \frac{d{\vec{\rho}}(\tau
)}{d\tau} \cdot \vec{\rho}(\tau ) \times
\vec{B}(\tau ,\vec \eta (\tau )) -\nonumber \\
 &-& \frac{\Delta m}{m}\,
\frac{d{\vec{\eta}}(\tau )}{d\tau} \cdot \vec{\rho}(\tau ) \times
\vec{B}(\tau ,\vec \eta (\tau )) +  \frac{d{\vec{\eta}}(\tau
)}{d\tau} \cdot {\vec{A}}_{\perp }(\tau ,\vec{\eta}(\tau ))\Big).
 \label{b12}
\end{eqnarray}

In the case $e = Q \approx Q_{1} \approx - Q_{2}$, $\mathcal{Q}
\approx 0$, this becomes

\begin{eqnarray}
L_1(\tau ) &=& - m_{1}\, c\, \sqrt{1 -  \left(
\frac{d{\vec{\eta}}(\tau )}{d\tau} + \frac{m_{2}}{m}\,
\frac{d{\vec{\rho}}(\tau )}{d\tau}\right) ^{2}} - m_{2}\, c\,
\sqrt{1 -  \left( \frac{d{%
\vec{\eta}}(\tau )}{d\tau} - \frac{m_{1}}{m}\,
\frac{d{\vec{\rho}}(\tau )}{d\tau}\right) ^{2}} +\nonumber \\
&+&\,\frac{e^{2}}{4\pi\, c \,|{\vec{\rho}}(\tau )|} + {e\over c}\,
{\vec{\rho}}(\tau ) \cdot \vec{\pi}_{\perp }(\tau ,\vec{\eta}(\tau
)) + {e\over c}\, \frac{\Delta m}{2m}\, \vec{\rho}(\tau ) \cdot
\Big(\vec{\rho} (\tau ) \cdot {{\partial}\over {\partial\, \vec
\eta}}\, \vec{\pi}_{\perp }(\tau ,\vec{\eta}(\tau ))\Big)  -\nonumber \\
&-&\frac{e}{c}\, \frac{d{\vec{\eta}}(\tau )}{d\tau} \cdot
\vec{\rho}(\tau ) \times \vec{B}(\tau ,\vec \eta (\tau )) - \frac{e%
}{c}\, \frac{\Delta m}{2m}\, \frac{d{\vec{\rho}}(\tau )}{d\tau}
\cdot \vec{\rho}(\tau ) \times \vec{B}(\tau ,\vec \eta (\tau )).
 \label{b13}
\end{eqnarray}

\noindent and the generating function $\tilde S$ reduces to

\begin{equation}
\tilde S = \frac{e}{c}\, {\vec{\rho}(\tau ) \cdot
\Big[{\vec{A}}_{\perp }(\tau ,\vec{\eta}(\tau ) ) +}\frac{\Delta
m}{2m}\,  \Big({\vec{\rho}}(\tau ) {{\cdot }
 {{\partial}\over {\partial\, \vec \eta}}\Big)\,{
\vec{A}}_{\perp }(\tau ,\vec{\eta}(\tau )))\Big].}
 \label{b14}
\end{equation}

\bigskip
In order to obtain the Hamiltonian from Eq.(\ref{b12}) we revert
from the composite variables back to the constituent variables,
which mean that we need to rewrite $L$ in terms of the individual
velocities using $\vec{\eta} = {\vec \eta}_{12} = \frac{m_{1}\,
\vec{\eta}_{1} + m_{2}\, \vec{\eta}_{2}}{m}$,  $\vec{\rho} = {\vec
\rho}_{12} = {\vec{\eta}}_{1} - \vec{\eta}_{2}$.

We obtain

\begin{eqnarray}
L_1(\tau ) &=&- m_{1}\, c\, \sqrt{1 -  \left(
\frac{d{\vec{\eta}}_{1}(\tau )}{d\tau}\right) ^{2}} - m_{2}\, c\,
\sqrt{1 -  \left(
\frac{d{\vec{\eta}}_{2}(\tau )}{d\tau}\right) ^{2}} +\nonumber \\
 &+& \frac{d{\vec{\eta}}_{1}(\tau )}{d\tau} \cdot \mathcal{\vec{A}}
 _{1}(\tau ,\vec \eta (\tau )) +
\frac{d{\vec{\eta}}_{2}(\tau )}{d\tau} \cdot \mathcal{\vec{A}}
_{2}(\tau ,\vec \eta (\tau )) - \nonumber \\
&-&\,\frac{Q_{1}Q_{2}}{4\pi\, c \,|{\vec{\rho}}(\tau )|} + {Q\over
c}\, {\vec{\rho}(\tau ) \cdot }\vec{\pi} _{\perp }(\tau
,\vec{\eta}(\tau )) + {Q\over c}\, \frac{\Delta m}{2m}\,
\vec{\rho}(\tau ) \cdot \Big(\vec{\rho} (\tau ) \cdot
{{\partial}\over {\partial\, \vec \eta}}\,
 \vec{\pi}_{\perp }(\tau ,\vec{\eta}(\tau ))\Big) +  \nonumber \\
&+&{\mathcal{Q}\over c}\, \frac{ \Delta m}{m}\, {\vec{\rho}(\tau )
\cdot }\vec{\pi}_{\perp }(\tau , \vec{\eta}(\tau )) + {\mathcal{Q
}\over c}\, \frac{m^{2} - 2\mu m}{2m^{2}}\, \vec{\rho}(\tau ) \cdot
\Big(\vec{ \rho}(\tau ) \cdot {{\partial}\over {\partial\, \vec
\eta}}\, \vec{\pi}_{\perp }(\tau ,\vec{\eta}(\tau ))\Big),
 \label{b15}
\end{eqnarray}

\noindent where

\begin{eqnarray}
\mathcal{\vec{A}}_i(\tau ) &=& \frac{\mathcal{Q}m_i}{ cm}\,
{{\vec{A}}_{\perp }(\tau ,\vec{\eta}(\tau ))} -\nonumber \\
 &-& \Big(\frac{Q\, m_i}{mc} + (-)^{i+1}
\frac{Q}{c}\, \frac{\Delta m}{2m} + \frac{\mathcal{Q}}{c}\,
\frac{m^{2} - 2 \mu\, m}{2 m^{2}} + \frac{\mathcal{Q}}{c}\, \frac{
\Delta m\, m_i}{m^{2}}\Big)
\vec{\rho}(\tau ) \times \vec{B}(\tau ,\vec \eta (\tau )).  \nonumber \\
&&{}  \label{b16}
\end{eqnarray}

By evaluating the new momenta

\beq
 \vec{\kappa}_i^{'}(\tau ) =\frac{\partial L_1(\tau )}{\partial
\frac{d{\vec{\eta}}_i(\tau )}{d\tau}} = \frac{m_i\, c\,
\frac{d{\vec{\eta}}_i(\tau )}{d\tau}}{\sqrt{1 - \left(
\frac{d{\vec{\eta}}_i(\tau )}{d\tau}\right) ^{2}}} +
\mathcal{\vec{A}}_i(\tau ).
  \label{b17}
\eeq

\noindent we obtain after some manipulations the Hamiltonian

\bea
 H_1 &=&\vec{\kappa}_{1}^{'}(\tau ) \cdot \frac{d{\vec{\eta}}_{1}(\tau
)}{d\tau} + \vec{\kappa}_{2}^{'}(\tau ) \cdot
\frac{d{\vec{\eta}}_{2}(\tau )}{d\tau} - L_1  =\nonumber \\
&=& \sqrt{m_{1}^{2}\,c^{2} + ({\vec{\kappa}}_{1}^{'}(\tau ) -
\mathcal{\vec{A}}_{1}(\tau )\,)^{2}} +  \sqrt{m_{2}^{2}\,c^{2} +
({\vec{\kappa}}_{2}^{'}(\tau ) - \mathcal{\ \vec{A}}_{2}(\tau
)\,)^{2}} +\nonumber \\
&+&\,\frac{Q_{1}Q_{2}}{4\pi\, c \,|{\vec{\rho}}(\tau )|} - {Q\over
c}\, {\vec{\rho}(\tau ) \cdot }\vec{\pi} _{\perp }(\tau
,\vec{\eta}(\tau )) - {Q\over c}\, \frac{\Delta m}{2m}\,
\vec{\rho}(\tau ) \cdot \Big(\vec{\rho}(\tau ) \cdot
{{\partial}\over {\partial\, \vec \eta}}\Big)\,
\vec{\pi}_{\perp }(\tau ,\vec{\eta}(\tau )) -  \nonumber \\
&-&{{\mathcal{Q}}\over c}\, \frac{\Delta m}{m}\, {\vec{\rho}(\tau )
\cdot }\vec{\pi}_{\perp }(\tau , \vec{\eta}(\tau )) -
{{\mathcal{Q}}\over c}\, \frac{m^{2} - 2 \mu\, m}{2 m^{2}}\,
\vec{\rho}(\tau ) \cdot \Big(\vec{ \rho}(\tau ) \cdot
{{\partial}\over {\partial\, \vec \eta}}\Big)\, \vec{\pi}_{\perp
}(\tau ,\vec{\eta}(\tau )).\nonumber \\
 &&{}
 \label{b18}
\eea

This can be brought to desired form using relative variables by
substituting for $\vec{\kappa}_{i}$ and expanding the vector
potential from the square root by using the Grassmann property. \

Consider first the non-relativistic limit (in which we keep terms of
order $ 1/c$). First we use

\beq
 \sqrt{m_i^{2}\,c^{2} + ({\vec{\kappa}}_i^{'}(\tau ) -
\mathcal{\vec{A}}_i(\tau )\,)^{2}} = \sqrt{m_i^{2}\,c^{2} +
{\vec{\kappa}}_i{}^{{'}\, 2}(\tau )} - \frac{{
\vec{\kappa}}_i^{'}(\tau ) \cdot \mathcal{\vec{A}}_i(\tau
)}{\sqrt{m_i^{2}\,c^{2} + { \vec{\kappa}}_i{}^{{'}\, 2}(\tau )}}.
  \label{b19}
\eeq

Note that we cannot use the Grassmann argument to eliminate the
$\mathcal{\ \vec{A}}^{2}(\tau )$ terms since they involve the
nonzero $Q^{2} = - 2\, Q_{1}\, Q_{2}$ terms even with $ \mathcal{Q}
\approx 0$ terms. We shall instead  ignore them as being too high an
order of $1/c^{2}$. So we get

\bea
 H_1 &=& \sqrt{m_{1}^{2}\,c^{2} + {\vec{\kappa}}_{1}{}^{{'}\, 2}(\tau )} -
\frac{ {\vec{ \kappa}}_{1}^{'}(\tau ) \cdot
\mathcal{\vec{A}}_{1}(\tau )}{\sqrt{m_{1}^{2}\,c^{2} + {\vec{
\kappa}}_{1}{}^{{'}\, 2}(\tau )}} + \sqrt{m_{2}^{2}\,c^{2} +
{\vec{\kappa}} _{2}{}^{{'}\, 2}(\tau )} - \frac{
{\vec{\kappa}}_{2}^{'}(\tau ) \cdot \mathcal{\vec{A}}_{2}(\tau )
}{\sqrt{m_{2}^{2}\,c^{2} + {\vec{\kappa}}_{2}{}^{{'}\, 2}(\tau )}}  +\nonumber \\
&+&\,\frac{Q_{1}Q_{2}}{4\pi\, c \,|{\vec{\rho}}(\tau )|} - {Q\over
c}\, {\vec{\rho}}(\tau ) \cdot \vec{\pi} _{\perp }(\tau
,\vec{\eta}(\tau )) - {Q\over c}\, \frac{\Delta m}{2m}\,
\vec{\rho}(\tau ) \cdot \Big(\vec{\rho}(\tau ) \cdot
{{\partial}\over {\partial\, \vec \eta}}\Big)\,
 \vec{\pi}_{\perp }(\tau ,\vec{\eta}(\tau )) -  \nonumber \\
&-&{{\mathcal{Q}}\over c}\, \frac{ \Delta m}{m}\, {\vec{\rho}(\tau )
\cdot }\vec{\pi}_{\perp }(\tau , \vec{\eta}(\tau )) - {{\mathcal{Q
}}\over c}\, \frac{m^{2} - 2 \mu\, m}{2 m^{2}}\, \vec{\rho}(\tau )
\cdot \Big(\vec{ \rho}(\tau ) \cdot {{\partial}\over {\partial\,
\vec \eta}}\Big)\, \vec{\pi}_{\perp }(\tau ,\vec{\eta}(\tau )).
 \label{b20}
\eea

In terms of the relative and collective variables

\bea
 H_1 &=&  \sqrt{m_{1}^{2}\, c^{2} + (\frac{m_{1}}{m}\,
\vec{\kappa}^{'}(\tau ) + \vec{\pi}^{'}(\tau ))^{2}} -
\frac{(\frac{m_{1}}{m}\, \vec{\kappa}^{'}(\tau ) +
\vec{\pi}^{'}(\tau )) \cdot \mathcal{\vec{A}}_{1}(\tau )}{
\sqrt{m_{1}^{2}\, c^{2} +
(\frac{m_{1}}{m}\, \vec{\kappa}^{'}(\tau ) + \vec{\pi}^{'}(\tau ))^{2}}} +\nonumber \\
&+&  \sqrt{m_{2}^{2}\, c^{2} + (\frac{m_{2}}{m}\,
\vec{\kappa}^{'}(\tau ) - \vec{\pi}^{'}(\tau ))^{2}} - \frac{
(\frac{m_{2}}{m}\, \vec{\kappa}^{'}(\tau ) - \vec{\pi}^{'}(\tau ))
\cdot \mathcal{\vec{A}} _{2}(\tau ) }{\sqrt{ m_{2}^{2}\, c^{2} +
(\frac{m_{2}}{m}\, \vec{\kappa}^{'}(\tau ) - \vec{\pi}^{'}(\tau ))^{2}}}  +\nonumber \\
 &+&\frac{Q_{1}Q_{2}}{4\pi\, c \,|{\vec{\rho}}(\tau )|} - {Q\over c}\, {\vec{\rho}(\tau ) \cdot
}\vec{\pi} _{\perp }(\tau ,\vec{\eta}(\tau )) - {Q\over c}\,
\frac{\Delta m}{2m}\, \vec{\rho}(\tau ) \cdot \Big(\vec{\rho}(\tau )
\cdot {{\partial}\over {\partial\, \vec \eta}}\Big)\,
\vec{\pi}_{\perp }(\tau ,\vec{\eta}(\tau )) -  \nonumber \\
&-&{{\mathcal{Q}}\over c}\, \frac{ \Delta m}{m}\, {\vec{\rho}(\tau )
\cdot }\vec{\pi}_{\perp }(\tau , \vec{\eta}(\tau )) - {{\mathcal{Q
}}\over c}\, \frac{m^{2} - 2 \mu\, m}{2 m^{2}}\, \vec{\rho}(\tau )
\cdot \Big(\vec{ \rho}(\tau ) \cdot {{\partial}\over {\partial\,
\vec \eta}}\Big)\, \vec{\pi}_{\perp }(\tau ,\vec{\eta}(\tau )).
 \label{b21}
\eea

Next, look at the limit as $c\rightarrow \infty $ in which

\bea
 \sqrt{m_i^{2}\,c^{2} + (\frac{m_i}{m}\, \vec{\kappa}^{'}(\tau ) +
(-)^{i+1}\, \vec{\pi}^{'}(\tau ))^{2}} &\rightarrow & m_i\, c +
\frac{(\frac{m_i}{m}\, \vec{\kappa}^{'}(\tau ) + (-)^{i+1}\,
\vec{\pi}^{'}(\tau ))^{2}}{2 m_i\, c},  \nonumber \\
 \frac{1}{\sqrt{m_i^{2}\, c^{2} + (\frac{m_i}{m}\,
\vec{\kappa}^{'}(\tau ) + (-)^{i+1}\, \vec{\pi}^{'}(\tau ))^{2}}}
&\rightarrow &\frac{1}{m_i\, c} - \frac{(\frac{m_i}{m}\,
\vec{\kappa}^{'}(\tau ) + (-)^{i+1}\, \vec{\pi}^{'}(\tau ))^{2}}{2\,
m_i^{3}\, c^{3}},
  \label{b22}
\eea

\noindent and we keep only the terms of $O(1/c)$ and lower obtaining

\bea
 H_1\, c &=& m\, c^{2} +  {{\vec{\kappa}^{{'}\, 2}(\tau )}\over {2 m}} +
\frac{\vec{\pi}^{{'}\, 2}(\tau )}{2 \mu } - \frac{
\vec{\kappa}^{'}(\tau )}{m} \cdot \Big(\,\mathcal{\vec{A}}_{1}(\tau
) + \mathcal{\vec{A}}_{2}(\tau )\Big) - \vec{ \pi}^{'}(\tau ) \cdot
\Big(\frac{\mathcal{\vec{A}}_{1}(\tau )}{m_{1}} -
\frac{\mathcal{\vec{A}}_{2}(\tau )}{m_{2}}\Big)  +\nonumber \\
 &+&\frac{Q_{1}\,Q_{2}}{4\pi \,|{\vec{\rho}}(\tau )|} - Q\, {\vec{\rho}(\tau ) \cdot
}\vec{\pi} _{\perp }(\tau ,\vec{\eta}(\tau )) - Q\, \frac{\Delta
m}{2 m}\, \vec{\rho}(\tau ) \cdot \Big(\vec{\rho}(\tau ) \cdot
{{\partial}\over {\partial\, \vec \eta}}\Big)\,
\vec{\pi}_{\perp }(\tau ,\vec{\eta}(\tau )) - \nonumber \\
 &-& \mathcal{Q}\, \frac{\Delta m}{m}\, {\vec{\rho}(\tau ) \cdot } \vec{\pi}_{\perp
}(\tau , \vec{\eta}(\tau )) - \mathcal{Q}\, \frac{m^{2} - 2 \mu\,
m}{2 m^{2}}\, \vec{\rho}(\tau ) \cdot \Big(\vec{ \rho}(\tau ) \cdot
{{\partial}\over {\partial\, \vec \eta}}\Big)\,
\vec{\pi}_{\perp }(\tau ,\vec{\eta}(\tau )).  \nonumber \\
&&{}  \label{b23}
 \eea

Now using Eq.(\ref{b16}) we obtain in the non-relativistic limit

\bea
 H_1\, c &=& m\, c^{2} +  {{\vec{\kappa}^{{'}\, 2}(\tau )}\over {2 m}} +
\frac{\vec{\pi}^{{'}\, 2}(\tau )}{2\mu } +\nonumber \\
 &+& \frac{\vec{\kappa}^{'}(\tau )}{m} \cdot \Big[ \Big(\frac{Q}{c} +
\frac{\mathcal{Q}}{c}\, \frac{\Delta m}{m}\Big)\, \vec{\rho}(\tau )
\times \vec{B}(\tau ,\vec \eta (\tau )) - \frac{\mathcal{Q}}{c}\,
{{\vec{A}}_{\perp }(\tau ,\vec{\eta}(\tau ))}\Big] +\nonumber \\
&+& \Big(\frac{Q}{c}\, \frac{\Delta m}{2 m\, \mu } +
\frac{\mathcal{Q}}{c}\, \frac{m^{2} - 2 \mu\, m}{2 m^{2}\mu }\Big)\,
\vec{\pi}^{'}(\tau ) \cdot \vec{\rho}(\tau ) \times \vec{B}(\tau ,\vec \eta (\tau )) + \nonumber \\
 &+&\frac{Q_{1}\,Q_{2}}{4\pi \,|{\vec{\rho}}(\tau )|} - Q\, {\vec{\rho}(\tau ) \cdot }
 \vec{\pi}_{\perp }(\tau ,\vec{\eta}(\tau )) - Q\, \frac{\Delta m}{2 m}\, \vec{\rho}(\tau ) \cdot
\Big(\vec{\rho}(\tau ) \cdot {{\partial}\over {\partial\, \vec
\eta}}\Big)\, \vec{\pi}_{\perp }(\tau ,\vec{\eta}(\tau ))  -\nonumber \\
 &-& \mathcal{Q}\, \frac{\Delta m}{m}\, {\vec{\rho}(\tau ) \cdot }
 \vec{\pi}_{\perp}(\tau , \vec{\eta}(\tau )) - \mathcal{Q}\, \frac{m^{2} - 2
 \mu\, m}{2 m^{2}}\, \vec{\rho}(\tau ) \cdot \Big(\vec{ \rho}(\tau ) \cdot {{\partial}\over
{\partial\, \vec \eta}}\Big)\, \vec{\pi}_{\perp }(\tau
,\vec{\eta}(\tau )).
 \label{b24}
 \eea

If we use ${\cal Q} \approx 0$, $e = Q \approx Q_1 \approx - Q_2$,
we get

\bea
 H_1\, c &=& m\, c^{2} + {{ \vec{\kappa}^{{'}\, 2}(\tau )}\over {2 m}}
 - {{e^2}\over {4\pi\, |\vec \rho (\tau )|}} +\nonumber \\
 &+& \frac{\vec{\pi}^{{'}\, 2}(\tau )}{2\mu } + \frac{
\vec{\kappa}^{'}(\tau )}{m} \cdot \frac{e}{c}\, \vec{\rho}(\tau )
\times \vec{B}(\tau ,\vec \eta (\tau )) + \frac{e}{c}\, \frac{
\Delta m}{2 m\mu }\, \vec{\pi}^{'}(\tau ) \cdot \vec{\rho}(\tau )
\times \vec{B}(\tau ,\vec \eta (\tau ))  -\nonumber \\
 &-& e\, {\vec{\rho}(\tau ) \cdot }\vec{\pi}
_{\perp }(\tau ,\vec{\eta}(\tau )) - e\, \frac{\Delta m}{2 m}\,
\vec{\rho}(\tau ) \cdot \Big(\vec{\rho}(\tau ) \cdot
{{\partial}\over {\partial\, \vec \eta}}\Big)
\, \vec{\pi}_{\perp }(\tau ,\vec{\eta}(\tau )),  \nonumber \\
&&{}  \label{b25}
 \eea

\noindent which agrees with Eq.(L.14) and (14.37) of Ref.\cite{8}
with $e^2 = 0$ (one could also find the $e^2$ terms of
Eq.(\ref{7.2})).

\subsection{A Lagrangian for the Internal Motion of the Particles
plus the Electro-Magnetic Field}

Let us consider the isolated system "charged particles with mutual
Coulomb interaction plus the transverse electromagnetic field"
inside the instantaneous Wigner 3-spaces. Its motion is determined
by the Hamiltonian $M c = {1\over c}\, {\cal E}_{(int)}$ given in
Eqs.(\ref{4.1}) and then restricted by the rest-frame conditions
${\vec {\cal P}}_{(int)} \approx 0$ and by their gauge fixing ${\vec
{\cal K}}_{(int)} \approx 0$. Therefore we can look for the
Lagrangian corresponding to the Hamiltonian $M c$.

By using the first half of the Hamilton equations with Hamiltonian
$M c$

\begin{eqnarray}
{\frac{{d\, {\vec \eta}_i(\tau ))}}{{d \tau}}} &{\buildrel \circ
\over {=}} & \{ {\vec \eta}_i(\tau ), M c \} =
\frac{{\vec{\kappa}}_i(\tau ) - {\frac{{Q_i}}{c}{\vec{A}}_{\perp
}(\tau ,{ \vec{\eta}}_i(\tau ))}}{\sqrt{m_i^2\,c^2 + ({\vec{\kappa}}
_i(\tau )-{\frac{{ Q_i}}{c}{\vec{A}}_{\perp }(\tau
,{\vec{\eta}}_i(\tau ))}\,)^2}},\nonumber \\
 { \frac{{\partial\, {\vec A}_{\perp}(\tau ,\vec \sigma
)}}{{\partial\, \tau}}} &\cir& \{ {\vec A}_{\perp}(\tau ,\vec \sigma
), M c \} = - {\vec \pi}_{\perp}(\tau ,\vec \sigma ),\nonumber \\
&&{}  \nonumber \\
&&\Downarrow  \nonumber \\
&&{}  \nonumber \\
{\vec \kappa}_i(\tau) &{\buildrel \circ \over {=}}& \frac{m_i\, c\,
\frac{d{ \vec{\eta}}_i}{d \tau}}{ \sqrt{1-\left(
\frac{d{\vec{\eta}}_i}{d \tau} \right) ^{2}}} + {\frac{{Q_i
}}{c}{\vec{A}}_{\perp }(\tau ,{\vec{\eta}}
_i(\tau ))},  \nonumber \\
&&{}  \nonumber \\
 &&{}\nonumber \\
 \Rightarrow &&\sqrt{m_i^2\,c^2 + ({\vec{\kappa}} _i(\tau
)-{\frac{{Q_i}}{c}{\vec{A}} _{\perp }(\tau ,{\vec{\eta}}_i(\tau
))}\,)^2}\, {\buildrel \circ \over {=}} \, {\frac{{m_i\, c}}{\sqrt{1
- ({\frac{{d {\vec \eta}_i(\tau )}}{{d \tau}}} )^2}}},
 \label{b26}
\end{eqnarray}

\noindent the inverse Legendre transformation leads to the following
Lagrangian

\begin{eqnarray*}
\mathcal{L}(\tau ) &=& \int d^3\sigma\, \Big( \sum_i\, \delta^3(\vec
\sigma - {\vec \eta}_i(\tau))\, \Big[{\vec \kappa}_i(\tau ) \cdot
{\frac{{d\, {\vec \eta}_i(\tau )}}{{d \tau}}}\Big] - {1\over c}\,
\Big[{\vec \pi} _{\perp} \cdot {\frac{{\partial\, {\vec
A}_{\perp}}}{{\partial\, \tau}}}\Big]
(\tau ,\vec \sigma ) \Big) -\nonumber \\
 &-& {1\over c}\,  \mathcal{E}_{(int)} =  \nonumber \\
 &&{}\nonumber \\
 &=& \int d^3\sigma\, \Big( \sum_i\, \delta^3(\vec
\sigma - {\vec \eta} _i(\tau )) \Big[- m_i\, c\, \sqrt{1 -
({\frac{{d\, {\vec \eta}_i(\tau )} }{{d \tau}}})^2} -  \nonumber \\
&-& {\frac{{Q_1\, Q_2}}{{8\pi\, c\, |{\vec \eta}_1(\tau ) - {\vec
\eta}_2(\tau )| }}} + {{Q_i}\over c}\, {\frac{{d\, {\vec
\eta}_i(\tau )}}{{d \tau}}} \cdot {\vec A}
_{\perp}(\tau , \vec \sigma )\Big] +  \nonumber \\
&+& {1\over {2 c}}\, \Big[({\frac{{\partial\, {\vec
A}_{\perp}}}{{\partial\, \tau}}})^2 - { \vec B}^2\Big](\tau ,\vec
\sigma )\Big) =
 \end{eqnarray*}

\bea
 &=& \int d^3\sigma\, \Big( \sum_i\, \delta^3(\vec \sigma - {\vec
\eta} _i(\tau )) \Big[- m_i\, c\, \sqrt{1 - ({\frac{{d\, {\vec
\eta}_i(\tau )} }{ {d \tau}}})^2} - {\frac{{Q_1\, Q_2}}{{8\pi\, c\,
|{\vec \eta}_1(\tau ) - {\vec \eta}_2(\tau )|}}} -  \nonumber \\
 &-& {{Q_i}\over c}\, {\vec \eta}_i(\tau ) \cdot \Big(
{\frac{{\partial\, {\vec A} _{\perp}(\tau ,\vec \sigma
)}}{{\partial\, \tau}}} + {\frac{{d\, {\vec \eta} _i(\tau )}}{{d
\tau}}} \cdot {\frac{{\partial\, {\vec A}_{\perp}(\tau , \vec
\sigma )}}{{\partial\, \vec \sigma}}}\Big)\, \Big]\Big) +  \nonumber \\
 &+& {1\over {2c}}\, \int d^3\sigma\, \Big[({\frac{{\partial\, {\vec
A}_{\perp}}}{{\partial\, \tau}}})^2 - { \vec B}^2\Big](\tau ,\vec
\sigma ) + {{d\, S_2(\tau )}\over   {d \tau}} =\nonumber \\
 &=& {\cal L}_1(\tau ) + {{d\, S_2(\tau )}\over   {d
 \tau}}.
 \label{b27}
\end{eqnarray}

In the last lines we have redefined the Lagrangian by isolating a
total time derivative. This leads to the identification of the
following generating function

\bea
 S_2 &=& {1\over c}\, \sum_i\,\int d^3\sigma\, \delta^3(\vec \sigma - {\vec \eta }_i(\tau
))\, Q_i\, {\vec \eta}_i(\tau ) \cdot {\vec A}_{\perp}(\tau ,\vec
\sigma ) =\nonumber \\
 &=& {\frac{1}{c}}\, \sum_i\, Q_i\, {\vec \eta}_i(\tau ) \cdot {
\vec A}_{\perp}(\tau , {\vec \eta}_i(\tau )) =\nonumber \\
 &&{}\nonumber \\
 &=& {Q\over c}\, \Big[{\vec \rho}_{12}(\tau ) \cdot {\vec A}_{\perp}(\tau
 ,{\vec \eta}_{12}(\tau ) +\nonumber \\
 &+& {{m_2 - m_1}\over m}\, {\vec \rho}_{12}(\tau ) \cdot
 \Big({\vec \rho}_{12}(\tau ) \cdot {{\partial}\over {\partial\,
 {\vec \eta}_{12}}}\Big)\, {\vec A}_{\perp}(\tau ,{\vec \eta}_{12}(\tau ))
 +\nonumber \\
 &+& {\vec \eta}_{12}(\tau ) \cdot \Big({\vec \rho}_{12}(\tau ) \cdot {{\partial}\over {\partial\,
 {\vec \eta}_{12}}}\Big)\, {\vec A}_{\perp}(\tau ,{\vec \eta}_{12}(\tau ))
 \Big] +\nonumber \\
 &+& {{{\cal Q}}\over c}\, \sum_{i=1}^2\,
 {\vec \eta}_i(\tau ) \cdot {\vec A}_{\perp}(\tau ,{\vec
 \eta}_i(\tau )) + O\Big(({\vec \rho}_{12} \cdot {{\partial}\over
 {\partial\, {\vec \eta}_{12}}})^2\, {\vec A}_{\perp}\Big)\nonumber \\
 &&{}\nonumber \\
 &\rightarrow_{{\cal Q} \approx 0}&
{Q\over c}\, \Big[{\vec \rho}_{12}(\tau ) \cdot {\vec
A}_{\perp}(\tau ,{\vec \eta}_{12}(\tau )) +\nonumber \\
 &+& {{m_2 - m_1}\over m}\, {\vec \rho}_{12}(\tau ) \cdot
 \Big({\vec \rho}_{12}(\tau ) \cdot {{\partial}\over {\partial\,
 {\vec \eta}_{12}}}\Big)\, {\vec A}_{\perp}(\tau ,{\vec \eta}_{12}(\tau ))
 +\nonumber \\
 &+& {\vec \eta}_{12}(\tau ) \cdot \Big({\vec \rho}_{12}(\tau ) \cdot {{\partial}\over {\partial\,
 {\vec \eta}_{12}}}\Big)\, {\vec A}_{\perp}(\tau ,{\vec \eta}_{12}(\tau ))
 \Big] + O\Big(({\vec \rho}_{12} \cdot {{\partial}\over
 {\partial\, {\vec \eta}_{12}}})^2\, {\vec A}_{\perp}\Big)
 =\nonumber \\
 &=& {1\over c}\, \vec d(\tau ) \cdot \Big[{\vec
A}_{\perp}(\tau ,{\vec \eta}_{12}(\tau )) + {{m_2 - m_1}\over m}\,
\Big({\vec \rho}_{12}(\tau ) \cdot {{\partial}\over {\partial\,
 {\vec \eta}_{12}}}\Big)\, {\vec A}_{\perp}(\tau ,{\vec \eta}_{12}(\tau ))
\Big] +\nonumber \\
 &+& {Q\over c}\, {\vec \eta}_{12}(\tau ) \cdot
\Big({\vec \rho}_{12}(\tau ) \cdot {{\partial}\over {\partial\,
 {\vec \eta}_{12}}}\Big)\, {\vec A}_{\perp}(\tau ,{\vec \eta}_{12}(\tau ))
 + O\Big(({\vec \rho}_{12} \cdot {{\partial}\over
 {\partial\, {\vec \eta}_{12}}})^2\, {\vec A}_{\perp}\Big).
 \label{b28}
 \eea

\noindent In Eq.(\ref{b28}) we have shown the dipole approximation
of $S_2$. It contains an electric dipole term similar to the
function $S$ of Eqs.(\ref{b3}) - (\ref{7.3}) but with ${{m_2 -
m_1}\over {2m}}$ replaced by ${{m_2 - m_1}\over m}$. Moreover it
contains an extra term connected to the motion of the collective
variable ${\vec \eta}_{12}(\tau )$ of the 2-particle system.

\vfill\eject

\end{document}